\documentclass[a4paper, 11pt]{article}

\usepackage{fullpage}
\usepackage{my_package}
\usepackage[numbers,sort&compress]{natbib}
\usepackage{enumerate}
\usepackage{authblk}
\usepackage{commands}

\numberwithin{equation}{section}

\title{Time-dependent modelling of thin poroelastic films drying on deformable plates}

\date{}

\author[1]{Matthew~G.~Hennessy\thanks{matthew.hennessy@bristol.ac.uk}}
\author[2]{Richard~V.~Craster}
\author[3]{Omar~K.~Matar}

\affil[1]{Department of Engineering Mathematics, University of Bristol, Ada Lovelace Building, Bristol, BS8 1TW, UK}
\affil[2]{Department of Mathematics, Imperial College London, South Kensington Campus, London, SW7 2AZ, UK}
\affil[3]{Department of Chemical Engineering, Imperial College London, South Kensington Campus, London, SW7 2AZ, UK}

\begin{document}

\maketitle

\begin{abstract}
Understanding the generation of mechanical stress in drying,
particle-laden films 
is important for a wide range of industrial processes. 
The cantilever experiment allows the stress in a 
drying film that has been deposited onto a thin plate to be quantified.
Mechanical stresses in the film are transmitted to the plate and drive
bending.
Mathematical modelling enables the film stress to be inferred
from measurements of the plate deflection.  The aim of this paper is
to present simplified models of the cantilever experiment that have been
derived from the time-dependent
equations of continuum mechanics using asymptotic methods. 
The film is described using nonlinear poroelasticity and the plate using nonlinear elasticity.
In contrast to Stoney-like
formulae, the simplified models
account for films with non-uniform thickness and stress.  The film
model reduces to a single differential equation that can be solved
independently of
the plate equations.
The plate model reduces to an extended form of the Foppl-von Karman (FvK) equations that accounts for gradients in the
longitudinal traction acting on the plate surface.  Consistent boundary conditions for the FvK equations are derived by resolving the Saint-Venant boundary layers at the free edges of the plate.  The asymptotically
reduced models are in excellent agreement with finite element solutions of the full governing equations.
As the P\'eclet number
increases, the time evolution of the plate deflection changes from $t$ to $t^{1/2}$, in
agreement with experiments.  

\end{abstract}

\section{Introduction}

The drying of thin films that consist of a volatile solvent and
a particulate phase 
is relevant to a wide range of industrial applications~\cite{kolegov2020},
such as the fabrication of electrodes for lithium-ion batteries~\cite{zhang2022}
and flexible conductive coatings~\cite{shimoni2014}.  As solvent is removed from
the mixture by evaporation, the particulate phase aggregates to form a
porous solid.  The
evaporation-driven flow of solvent through the pores will generate
a pressure gradient that, in turn, can deform the solid. The
removal of solvent from the pore space will also cause the solid to contract.
If the film has been placed on a rigid substrate, then the adhesion
of the solid to the substrate will restrict contractions in the longitudinal
plane.
As a result, mechanical stresses will develop within the
film~\cite{croll1979}.  The  film stress can be relieved
through a myriad of mechanical instabilities~\cite{giorgiutti2018}, including
fracture~\cite{dufresne2003, bourrianne2021}, 
buckling~\cite{pauchard2003}, 
and delamination~\cite{giorgiutti2015, osman2020}. 
Although drying-induced
instabilities have traditionally been viewed as detrimental, there 
is increasing interest in understanding how these can be harnessed in
applications such as medical 
diagnostics~\cite{sefiane2021} and lithography~\cite{kim2016}. 

Avoiding or triggering
instabilities during film drying requires 
quantitative knowledge about the evolution of the film stress. However,
directly measuring the mechanical stress during drying
is difficult.  A number of innovative experiments have therefore
been designed around the aim of indirectly measuring the film 
stress~\cite{bouchaudy2019, xu2010}.
The cantilever technique is one such experiment, whereby a thin film 
(or drop) is deposited onto the surface of a thin plate
and left to dry~\cite{evans2006}.  Due to the adhesion between
the film and the plate, the mechanical stresses that are generated in the
film are transmitted to the plate.  In response, the plate will bend, causing
a deflection that can be measured.  The film stress can then be inferred
from the deflection using mathematical modelling. 

In the seminal work of Stoney~\cite{stoney1909}, a simple model is presented
for the cantilever experiment.  Using
classical beam theory, Stoney was able to determine a relationship 
between the film stress and the beam deflection.
Stoney's derivation relies on a number of key assumptions,
namely: the film is thin compared to the
beam, the film has a uniform and constant thickness, and bending is driven by a uniform
longitudinal film stress.  In the context of drying, the latter point implies that 
the film has a homogeneous composition.  Moreover, Stoney's derivation 
requires the explicit definition of a neutral axis, 
which is a longitudinal axis in the beam where no strain occurs.
A number of authors have extended Stoney's model by utilising more sophisticated
beam theories~\cite{francis2002}.  
For example, Petersen \etal\cite{petersen1999} accounted the finite thickness of the film.  
However, Chiu~\cite{chiu1990} has criticised several of these models and argues that
Stoney's choice of neutral axis is only correct when the film and beam have the 
same Young's modulus.  As discussed by Francis \etal\cite{francis2002}, 
the Young's modulus of the film is typically
one or two orders of magnitude smaller than that of the plate.  
At the other end of the modelling spectrum, Lei \etal\cite{lei2002} used 
the finite element method to simulate the drying of an elasto-viscoplastic
film on a cantilever beam.  
To the best of our knowledge, no comparison has been made 
between simple models based on beam theory and finite-element 
solutions of the full equations of continuum mechanics.

By analysing a cantilever experiment with a Stoney-like model, Croll~\cite{croll1978}
observed that the stress in the film is independent of the film height.
Similar observations were recently made for drying polymer films by 
Tomar \etal\cite{tomar2020}, who also noted that the kinetics of stress generation
are sensitive to the P\'eclet number.  
The film stress was found to increase linearly with time for small P\'eclet numbers 
and nonlinearly for larger P\'eclet numbers.    
The nonlinear evolution of the stress was attributed to the formation of a 
polymer-rich skin at the film surface caused by a large rate of evaporation. 
When a skin forms, the longitudinal stress across the film will be highly non-uniform
and the applicability of Stoney's model 
to this scenario comes into question. 

The aim of this paper is to revisit the derivation of Stoney-like formulae
using  asymptotic methods.  
Starting from the time-dependent equations of three-dimensional
continuum mechanics, a hierarchy of simplified models for the
cantilever experiment will be  derived.  More specifically, the film will be 
modelled using nonlinear poroelasticity~\cite{biot1941, macminn2016}
and the plate using nonlinear elasticity~\cite{howell2009}.  
The asymptotic analysis is based on the 
small aspect ratios of the film and the plate.  By using asymptotic methods
to simplify the equations 
of elasticity rather than directly appealing to beam theory, the explicit definition of a 
neutral axis can be avoided.  The reduced models that are presented here extend 
Stoney-like models by capturing the evolution of the film thickness, 
composition, and stress, all of which can be highly non-uniform.  Moreover, the analysis 
reveals parameter regimes where the plate deflection is driven by both the longitudinal and 
transverse stress in the film.  The reduced models are validated by comparing them against
finite element solutions of the full system of equations and found to be very accurate.

A key feature of the asymptotic reduction is that it leads to a 
one-way decoupling of the models, thus allowing the film and plate equations 
to be solved sequentially.  Moreover, the dimensionality of the film
and plate models are reduced.  The combined result is a
computationally efficient model that is able to resolve the 
spatio-temporal dynamics of drying and bending.  
In asymptotically reducing the equations for the film, we invoke
a poroelastic version of lubrication theory~\cite{jensen1994, hewitt2015, etzold2022}.  
Furthermore,
we extend our previous work on drying-induced stresses in poroelastic
drops~\cite{hennessy2022} 
by accounting for a wide range of drying regimes that are determined 
by the magnitude of the P\'eclet number.  In all cases, the film model 
reduces to either a nonlinear ordinary differential equation or a 
nonlinear diffusion equation for the solvent concentration.  
The deformation of the plate can be described by an extended form of the
Foppl-von Karman (FvK) equations that accounts for gradients in the longitudinal 
traction exerted on the upper plate surface by the film.  In order to derive a consistent set
of boundary conditions at the free edges of the plate, a careful analysis of the
Saint-Venant boundary layers is required.  If considering a beam rather than
a plate, we show that the FvK equations reduce to a linear second-order
differential equation.

The governing equations for the film and the plate are presented in 
Sec.~\ref{sec:model}.  A scaling analysis and preliminary
reduction of the plate equations are carried out in Sec.~\ref{sec:plate_scaling}.  
The equations are non-dimensionalised in Sec.~\ref{sec:non_dim}.  A detailed
asymptotic reduction is then carried out in Sec.~\ref{sec:reduction}.  
In Sec.~\ref{sec:fem}, the solutions of the asymptotically reduced models are
compared against finite-element solutions of the full model.  A parametric
study is also conducted.  The paper concludes in Sec.~\ref{sec:conclusion}.

\section{Modelling}
\label{sec:model}

The formation of a coating is a complex process that involves
the transformation of a liquid-like mixture into
a solid.  The generation of mechanical
stresses occurs after the gel point of the mixture
has been crossed and the mechanical response of the film
becomes more solid-like than liquid-like.  
In cantilever experiments, the crossing of the 
gel point can be identified as the time at which the plate begins to
bend.  Before this time, film drying takes place but
the deflection of the plate is negligible.  

\begin{figure}
  \centering
  \includegraphics[width=\textwidth]{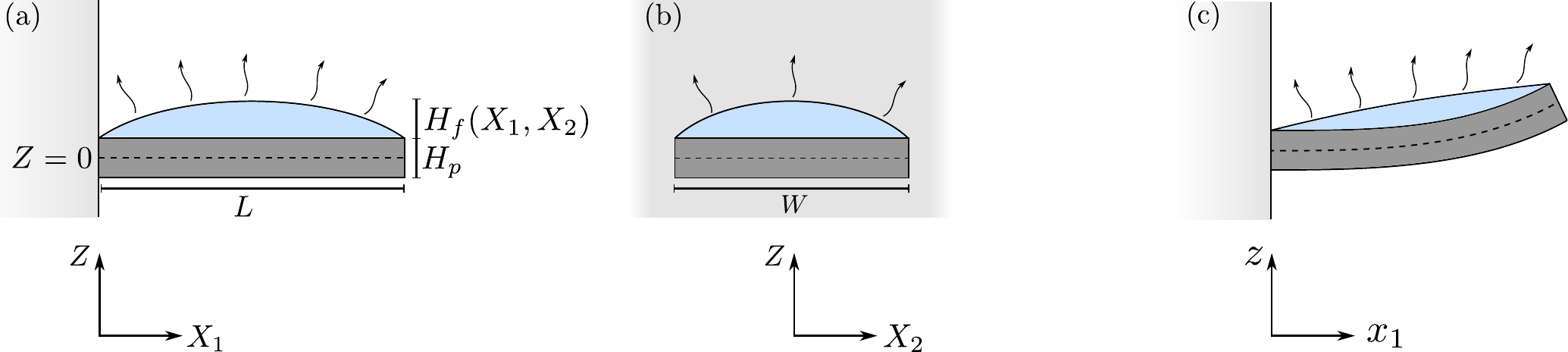}
  \caption{Bending of a plate during during film drying.
    The plate has length $L$, width $W$, and height $H_p$.
    The initial film thickness is given by $H_f(X_1, X_2)$,
    where $X_1$ and $X_2$ are Lagrangian coordinates that lie
    in the plane spanned by the plate centreline (dashed line).
    The plate is clamped to a wall at $X_1 = 0$. The origin of the vertical
    Lagrangian coordinate $Z$ coincides with the plate centreline.
    Panels (a) and (b): Cross-sections of the Lagrangian (undeformed)
    configuration. Panel (c):
    Cross-section of the Eulerian
    (deformed) configuration.}
  \label{fig:drop_cantilever}
\end{figure}

Given that we are interested in modelling the deflection of the
plate, we will assume that the gel point has just been crossed
so that a poroelastic matrix has formed throughout the film.
One-dimensional models that couple solidification and poromechanics
have been considered by Style and Peppin~\cite{style2011}
and Punati and Tirumkudulu~\cite{punati2022}.
We therefore consider a poroelastic film that is drying on a 
thin, flexible plate, as illustrated in Fig.~\ref{fig:drop_cantilever}.  
The plate is assumed to be rectangular
with length $L$, width $W$, and thickness $H_p$.  The length and width
are assumed to be comparable in size.  However, the thickness of the
plate is assumed to be much smaller than the length
and width. 
The small aspect ratio of the plate is denoted by $\delta = H_p / L \ll 1$. 

We let $X_1$, $X_2$, and $X_3 = Z$ denote
Cartesian coordinates associated with the initial (Lagrangian) 
configuration of the system.  The coordinates $X_1$ and $X_2$ are chosen
to lie in the plane that is parallel to the upper
and lower surfaces of the plate.  
The vertical (or transverse) Lagrangian
coordinate, $Z$, is chosen such that $Z = 0$ corresponds to the centre surface of the
plate.  Consequently, $Z = \pm H_p/2$ denotes the upper and lower surfaces. 
The plate is envisioned as being clamped to a wall at $X_1 = 0$ and
having free edges at $X_1 = L$ and $X_2 = \pm W / 2$.

The film has a non-uniform thickness that evolves in time.
The initial film thickness is denoted by $H_f(X_1, X_2)$ and has a 
maximum value given by $\hf = \max(H_f)$.  The film is also assumed to be
thin relative to the length of the plate so that 
$\epsilon = \hf / L \ll 1$.  We assume that the film thickness tends to
zero as the edges of the plate are approached.  That is, the contact line of
the film is pinned to the edges of the plate.  This setup is similar to the
experiments of Tomar \etal\cite{tomar2020}, where a drop of polymer solution
was deposited into the middle of a cantilever plate and allowed to dry.

The film is assumed to remain bonded to the plate during drying. 
The upper surface of the plate, originally located at $Z = H_p/2$,
becomes vertically displaced to the Eulerian position $z = H_p/2 + w(x_1, x_2, t)$,
where $x_1$ and $x_2$ are in-plane coordinates associated with the
deformed (Eulerian) configuration and $t$ is time.  
The Eulerian thickness of the film is denoted by
$h_f(x_1, x_2,t)$. Hence, the Eulerian
position of the free surface of the film
is given by $z = H_p / 2 + w(x_1, x_2,t) + h(x_1, x_2,t)$.

\subsection{Notation}
\label{sec:notation}

Occasionally, vector and tensor quantities will be decomposed into in-plane and
vertical components that are parallel and perpendicular to the undeformed
plate, 
respectively. We let $\vec{e}_1$, $\vec{e}_2$, and $\vec{e}_3 = \ez$ denote
Cartesian basis vectors for the $X_1$, $X_2$, and $X_3 = Z$ directions,
as well as the $x_1$, $x_2$, and $z = x_3$ directions. 
If $\vec{a} = a_i \vec{e}_i$ denotes an arbitrary vector, then we can write
$\vec{a} = \vec{a}_{\parallel} + a_z \ez$, where
$\vec{a}_{\parallel} = a_\alpha \vec{e}_\alpha$ is defined as the vector of 
in-plane components and $a_z = a_3$ is the vertical
component. Einstein summation notation is used and we adopt the convention that Greek indices are equal to 1 or 2.  We let $\nabla$ denote the material gradient
taken with respect to the Lagrangian coordinates $\vec{X} = X_i \vec{e}_i$.  The in-plane
gradient operator is defined as $\nx = \nabla - \ez\,\pdf{}{Z}$. Tensors as written as $\tens{T} = \Txx + \tens{T}_{\perp} \otimes \ez + {\sf T}_{z\alpha} \ez \otimes \vec{e}_\alpha + {\sf T}_{zz} \ez \otimes \ez$ where $\Txx = {\sf T}_{\alpha \beta} \vec{e}_\alpha \otimes \vec{e}_\beta$ and $\tens{T}_\perp = {\sf T}_{\alpha z} \vec{e}_\alpha$.

\subsection{Kinematics}

The governing equations are formulated in terms of Lagrangian
coordinates $\vec{X} = X_i \vec{e}_i$
associated with the initial (undeformed) configuration of the film and plate.
We let $\vec{x} = x_i(\vec{X},t) \vec{e}_i$ denote
Eulerian coordinates associated with the current (deformed) configuration.
During drying, the solid element originally located
at $\vec{X}$ is displaced to $\vec{x}$, thereby generating a displacement
$\as = \vec{x}(\vec{X},t) - \vec{X}$. The deformation gradient
tensor $\tens{F}$ describes the distortion of material elements
and is given by
\begin{align}
  \tens{F} = \nabla \vec{x} = \tens{I} + \nabla \as,
  \label{eqn:F_inv}
\end{align}
where $\tens{I}$ is the identity tensor.
We adopt the convention that
the gradient of a vector $\vec{a} = a_i \vec{e}_i$
is given by $\nabla \vec{a} = (\pdf{a_{i}}{X_j})\,\vec{e}_i\otimes\vec{e}_j$.
The determinant of $\tens{F}$, denoted by $J = \det \tens{F}$,
accounts for volumetric changes of material elements.

\subsection{A model for a poroelastic film}
\label{sec:model_film}

Poroelastic materials consist of a porous and deformable
solid matrix that is filled with fluid.  The theory of 
poroelasticity was first developed by 
Biot~\cite{biot1941} in
the context of soil mechanics.  
Coussy~\cite{coussy2004} 
provides a comprehensive overview of the theory.
We treat the drying film as a poroelastic material
and describe it using a Lagrangian version of the
model proposed by MacMinn \etal\cite{macminn2016}.
In essence, the model couples  the equations of nonlinear
elasticity to those for flow in a porous medium.

Conservation of liquid in the porous film leads to
\subeq{
  \label{eqn:phi}
\begin{align}
  \pd{\Phi}{t} + \nabla \cdot \Q &= 0,
\end{align}
}
where $\Phi$ is the nominal volume fraction of fluid
and $\Q$ is the nominal fluid flux.
The Eulerian volume fraction of fluid, which is equivalent to the
film porosity, is given by
$\phif = \Phi / J$.  
The transport of fluid within the pore space is
governed by Darcy's law, which can be written
in terms of the reference configuration as
\begin{align}
  \Q &= -\tens{K} \nabla p, \label{eqn:darcy}
\end{align}
where $\tens{K}$ is a permeability tensor and
$p$ is the fluid pressure.  The permeability 
tensor can be written in terms of the 
scalar permeability $k(\phif)$ as
$\tens{K} = (k(\phif) / \mu_f) J \tens{C}^{-1}$,
where $\mu_f$ is the fluid viscosity and
$\tens{C} = \tens{F}^T \tens{F}$ is the 
right Cauchy--Green tensor.  
The factor of $J \tens{C}^{-1}$ in $\tens{K}$ 
is the result of mapping Darcy's
law in the current configuration to the reference configuration.
We let $k_0 = k(\phif_0)$ denote the initial permeability, where
$\phi_0$ is the initial volume fraction (or porosity) of the film.
For simplicity, we assume
that the initial fluid fraction is spatially uniform.

At the microscopic level, 
the solid and fluid phases are assumed to be incompressible.  However, at the macroscopic
level, the film is compressible, with volumetric changes being accommodated by
rearrangements of the pore geometry
and the removal (or addition) of fluid from material elements.
The connection between macroscopic volume changes
and the amount of fluid in the pore space of a material element
is captured through the incompressibility condition
\begin{align}
  J = 1 + \Phi - \phif_{0}.  \label{eqn:J}
\end{align}
Evaluating \eqref{eqn:J} at $t = 0$ gives $J = 1$. Since drying leads to the
removal of fluid from the pore space, $\Phi$ decreases relative to
$\phif_0$, and we expect that $J \leq 1$.  If the composition of the film remains
uniform during drying, then $J$ can be written in terms of the total 
film volume $V(t)$
as $J = V(t) / V(0)$.  In light of this, we will refer to $J$ as the 
contraction ratio. 

Conservation of linear and angular momentum in the film leads to
\begin{align}
\nabla \cdot \tens{S} &= \vec{0}, \label{eqn:lin_mom_film} \\
\tens{S} \tens{F}^T &= \tens{F} \tens{S}^T. 
\end{align}
The first Piola--Kirchhoff (PK1) stress tensor $\tens{S}$ is
decomposed as $\tens{S} = \tens{\Sigma} - p J \tens{F}^{-T}$,
where the first component, $\tens{\Sigma}$, represents the
effective (Terzaghi) elastic stress of the solid.
The second contribution to $\tens{S}$ accounts for the
stress exerted by the fluid.
The solid matrix is assumed to be isotropic and obey a neo-Hookean
equation of state. The elastic component of the stress tensor can be written as
\begin{align}
  \tens{\Sigma} = \frac{\nu E_f}{(1 + \nu_f)(1-2\nu_f)} J (J - 1)\tens{F}^{-T} + \frac{E_f}{2(1+\nu_f)}\,(\tens{F} - \tens{F}^{-T}),
  \label{eqn:sigma}
\end{align}
where $E_f$ and $\nu_f$ are the Young's modulus and Poisson's ratio of the
film, respectively.  Both of these parameters are assumed to remain constant
during the drying process.
In the limit of small deformations, $\nabla \as \ll 1$, we find that 
$\tens{F} \sim \tens{I} + \nabla \as$, 
$\tens{F}^{-T} \sim \tens{I} - (\nabla \as)^T$,
and $J = \det \tens{F} \sim 1 + \nabla \cdot \as$.
Hence, the stress-strain relation \eqref{eqn:sigma} reduces
to
\begin{align}
    \vec{\Sigma} \sim \frac{\nu E_f}{(1 + \nu_f)(1-2\nu_f)}(\nabla \cdot \as)\tens{I} + \frac{E_f}{2(1+\nu_f)}\,\left(\nabla \as + (\nabla \as)^T\right),
  \label{eqn:sigma_lin}
\end{align}
thus recovering linear elasticity.  When carrying out a scaling analysis in 
Sec.~\ref{sec:plate_scaling}, it will be convenient to write the stress
balance \eqref{eqn:lin_mom_film} in component form as
\subeq{\label{eqn:lin_mom_film_comp}
\begin{align}
\nx \cdot \Sxx + \pd{\Sxz}{Z} = 0, \\
\pd{\Sza}{X_\alpha} + \pd{\Szz}{Z} = 0,
\end{align}}
where the definitions of $\Sxx$ and $\Sxz$ can be found in Sec.~\ref{sec:notation}.

\subsection{A model for a deformable plate}
\label{sec:model_plate}

Conservation of linear and angular momentum for the plate leads to
\subeq{
\begin{align}
\nabla \cdot \tens{S} &= \vec{0}, \\
\tens{S} \tens{F}^T &= \tens{F} \tens{S}^T,
\end{align}}
where $\tens{S}$ is the PK1 stress tensor, 
$\tens{F} = \tens{I} + \nabla \vec{u}$ is the deformation gradient tensor,
and $\vec{u}$ is the plate displacement. 
The mechanical response of the plate is described using the
Saint Venant--Kirchhoff constitutive relation given by
\begin{align}
  \tens{S} = \tens{F} \tens{P}, \quad
  \tens{P} = \frac{\nub E_p}{(1+\nub)(1-2\nub)}\tr( \tens{E}) \tens{I} + \frac{E_p}{1+\nub}\tens{E},
  \label{eqn:SFP}
\end{align}
where $\nub$ and $E_p$ are the Poisson's ratio and Young's modulus
of the plate, respectively;
$\tens{P}$ is the second Piola--Kirchhoff
stress tensor, and $\tens{E} = (1/2) (\tens{F}^T \tens{F} - \tens{I})$ is
the strain tensor.

\subsection{Boundary and initial conditions}

In the reference configuration, the film thickness is constant
in time and given by $H_f(\XX)$. The position of the film surface
is then $Z =H_f + H_p / 2$. At the film surface, we assume
that fluid evaporates with volumetric flux $V_e$ 
that depends on the Eulerian fluid fraction. 
We therefore impose
\begin{align}
  \Q \cdot \vec{N} = V_e(\phif) \mathcal{S}(\tens{F}),
  \quad Z =H_f(\XX) + H_p / 2,
\end{align}
where $\mathcal{S}(\tens{F}) =   J |\tens{F}^{-T} \cdot \vec{N}|$
is a dimensionless function that accounts for how the differential
area element differs between the reference and deformed configurations.
The unit normal
vector can be written as $\vec{N} = \mathcal{N}^{-1}(-\nx H_f + \ez)$
where $\mathcal{N} = (1 + |\nx H_f|^2)^{1/2}$. 
In addition,
we assume that the film surface is stress free:
\begin{align}
  \tens{S} \cdot \vec{N} = \vec{0}, \quad Z = H_f(\XX) + H_p / 2.
\end{align}

The adhesion between the film and the plate is assumed to be perfect;
consequently, neither slip nor delamination can occur.
The
assumption of perfect adhesion manifests as continuity of displacement,
\begin{align}
  \left.\vec{u}\right|_{-} = \left.\vec{u}\right|_{+},
  \quad Z = H_p/2,
\end{align}
where $-$ means approaching $Z = H_p/2$ from the plate and $+$ means approaching
$Z = H_p / 2$ from the film.
The traction exerted by the film on the plate is denoted by
$\vec{\tau}$. 
Continuity of stress across the plate-film interface can therefore
be expressed as
\begin{align}
  \left.\tens{S}\cdot \ez \right|_{-} = \vec{\tau} = \left.\tens{S}\cdot \ez\right|_{+}, \quad Z = H_p/2.
\end{align}
The plate is assumed to be impermeable and hence the following no-flux
condition is imposed:
\begin{align}
  \Q \cdot \ez = 0, \quad Z = H_p / 2.
\end{align}

The lower surface of the plate is assumed to be stress free, resulting
in
\begin{align}
  \tens{S} \cdot \ez = \vec{0}, \quad Z = -H_p / 2.
\end{align}
At the contact surface between the plate and the wall, zero-displacement
conditions are imposed:
\begin{align}
\vec{u} = \vec{0}, \quad X_1 = 0.
\end{align}
The free edges of the plate are stress free; thus,
\subeq{
\begin{align}
\tens{S} \cdot \vec{e}_1 &= \vec{0}, \quad X_1 = L, \\
\tens{S} \cdot \vec{e}_2 &= \vec{0}, \quad X_2 = \pm W/2.
\end{align}}
The initial condition for the nominal fluid fraction is given by
 $\Phi(\XX,Z,0) = \phif_0$.  

\section{Scaling analysis and reduction of the plate equations}
\label{sec:plate_scaling}

The amplitude of the plate deflection can be estimated
by carrying out a scaling analysis of the governing equations.  
This estimate will guide the asymptotic reduction of the plate model.
To facilitate the scaling analysis, the film and plate are assumed to behave
as linear elastic materials with stress-strain relations that are,
respectively, identical
or analogous
to \eqref{eqn:sigma_lin}.  In addition, it will be assumed that the
deflection of the plate is driven by the in-plane traction, in line
with Stoney-like models~\cite{stoney1909, evans2006}.  We will
show below that these two assumptions can be broken in the following
ways.  The vertical traction can play a comparable role to the
in-plane traction.  The shear stresses in the
film, $\Saz$ and $\Sza$, can have different orders of
magnitude.
These cases will be specifically addressed when constructing
asymptotic solutions to the governing equations.
The strategy behind the scaling analysis is to estimate the size of
the in-plane traction from the film equations. Then, the magnitude of the
plate displacements will be estimated by examining the plate equations.

To begin, we consider the stress balance in the film \eqref{eqn:lin_mom_film_comp}. 
The size of the shear stress in the film can be found
by noting that 
vertical gradients in the shear stress
$\pdf{\Sxz}{Z} \sim \Sxz / \hf$ must balance in-plane gradients of the
in-plane stress $\nx \cdot \Sxx \sim E_f / L$.
Consequently, the shear stress and in-plane traction scale like
$\Sxz \sim \epsilon E_f$ and $\TTx \sim \epsilon E_f$.
Applying similar arguments to the vertical component
of the stress balance \eqref{eqn:lin_mom_film_comp} and assuming
$\Sza \sim \Saz \sim \epsilon E_f$ shows that
$\Szz \sim \epsilon^2 E_f$ and hence
$\TTz \sim \epsilon^2 E_f$.  Thus, under these
assumptions, the
vertical traction is much smaller than the
in-plane traction. 

The in-plane traction exerted on the surface of the plate
generates an internal shear stress; thus,
$\Sxz \sim \TTx \sim \epsilon E_f$.  As in the film,
the vertical gradients in the shear stress,
$\pdf{\Sxz}{Z} \sim \Sxz / H_p$,
are balanced by the in-plane gradients of the 
in-plane stresses, $\nx \cdot \Sxx \sim \Sxx / L$, leading to
$\Sxx \sim \delta^{-1} \epsilon E_f$. A linear stress-strain relation
implies that $\Sxx \sim E_p \ubx / L$. Thus,
we find that the in-plane displacement in the plate scales as
$\ubx \sim \delta^{-1} (E_f / E_p) \hf$. A scale for the vertical
displacement of the plate is obtained by assuming that bending
is the primary mode of deformation and 
balancing the components of the
shear strains, $\ubx / H_p \sim \ubz / L$, which leads to
$\ubz \sim \delta^{-2} (E_f / E_p) \hf$.  Since the displacements
in the film and plate must match at the film-plate surface,
these displacement scales also apply to the film.

The scaling analysis motivates introducing the
non-dimensional parameter $\E \equiv \delta^{-2} E_f / E_p$,
which compares the film modulus $E_f$ to a reduced modulus
for the plate, $\delta^2 E_p$.  Thus, $\E$ characterises how
stiff the film is relative to the plate.  
The displacements will therefore scale as 
$\ubx \sim \delta \E \hf$ and $\ubz \sim \E \hf$.
The size of $\E$, therefore, plays a key role in determining the 
magnitudes of the
displacements and controls the mechanics of the film and plate.  
For films that are thin relative to the plate, $\epsilon \ll \delta$,
three regimes
of plate mechanics can be identified:
\begin{enumerate}[(i)]
\item{\emph{Soft films}: $\E = O(1)$.
    The deflection of the plate scales like the film thickness,
    which is much smaller than the plate thickness.  
	Linear plate theory can be applied.}
\item{\emph{Stiff films}: $\E = O(\delta \epsilon^{-1}) \gg 1$.
    The plate deflection is proportional to its
    thickness. Foppl-von Karman theory applies to the plate.}
\item{\emph{Very stiff films}: $\E \gg O(\delta \epsilon^{-1}) \gg 1$.
    The deflection
    of the plate greatly exceeds its thickness. This regime
    is beyond the validity of Foppl-von Karman theory.}
\end{enumerate}

The model reduction presented below will focus on the first two of these
regimes.  
An analysis of Regime (iii), in which the film is very stiff, is left 
as an area of future work.  The case when the film and the plate
have similar thicknesses, i.e.\ $\delta = O(\epsilon)$ as $\epsilon \to 0$,
is a distinguished limit.    
In this case, Regimes (i) and (ii) coincide
and the plate can be modelled using a modified form of the 
Foppl-von Karman (FvK)
equations.
The reduced equations for the plate are formulated by assuming that
$\delta = O(\epsilon)$ and are then specialised to the first
two regimes when $\epsilon \ll \delta$ in Sec.~\ref{sec:reduction}.

\subsection{Reduced equations for the plate}

The mechanics of the plate are described using a modified form of the
Foppl-von Karman (FvK) equations that accounts for non-uniform in-plane
tractions. These equations are
systematically derived from the equations of nonlinear elasticity in
Appendix \ref{app:fvk}.  
The leading-order contribution to the vertical displacement of the
plate, which we denote by $\wb$, is independent of $Z$ and
satisfies
\subeq{
\begin{align}
  -B \nx^4 \wb + \nx \cdot(\Sxxbar \nx \wb)
  &= -\frac{H_p}{2}\nx \cdot \TTx - \TTz,
    \label{plate:fvk}
\end{align}
where
$B = \Eb H_p^3  / 12$ is the bending modulus and
$\Eb = E_p / (1 - \nu_p^2)$ is the effective Young's modulus.
The first term on the right-hand side of \eqref{plate:fvk} is absent from
traditional formulations of the FvK equations;
see, e.g.\ Landau and Lifshitz~\cite{landau1986} or
Howell \etal\cite{howell2009}. 
The mean in-plane displacement $\ubar$ can be obtained by solving the
mean in-plane stress balance given by
\begin{align} 
  \nx \cdot \Sxxbar &= -\TTx,
\end{align}
where the mean in-plane stress $\Sxxbar$ and strain $\ee$ are defined as
\begin{align}
  \Sxxbar &= H_p \Eb\left[(1 - \nu_p) \ee + \nu_p \tr(\ee) \Ixx\right], \\
  \ee &= \frac{1}{2}\left[\nx \ubar + (\nx \ubar)^T + \nx \wb \otimes \nx \wb\right].
\end{align}}
The in-plane displacements are linked to the mean in-plane and vertical
displacements via
\begin{align}
  \vec{u}_\parallel = \ubar -Z \nx \wb.
  \label{plate:u}
\end{align}

\subsection{Boundary conditions at the edges of the plate}
\label{sec:beam_bc}

At the clamped edge of the plate, located at $X_1 = 0$, we impose
\subeq{
\begin{align}
  \ubar = 0,
  \quad 
  \wb = 0,
  \quad
  \pd{\wb}{X_1} = 0; \quad X_1 = 0.
  \label{plate:bc:clamp}
  \end{align}
}
The boundary conditions at a free edge can be derived by resolving the mechanics
of the plate in a thin boundary layer where the stresses become large, details
of which are provided in Appendix \ref{app:fvk_bc}. The boundary conditions
at $X_1 = L$ are 
\subeq{
  \begin{alignat}{2}
    \bar{\mathsf{S}}_{\alpha 1} &= 0, &\quad X_1 &= L;
    \label{plate:bc:S11}
    \\
    \pdd{\wb}{X_1} + \nub \pdd{\wb}{X_2} &= 0, &\quad X_1 &= L;
    \label{plate:bc:wxx}
    \\
    B \left[\frac{\p^3\wb}{\p X_1^3} + (2-\nub)\frac{\p^3\wb}{\p X_1 \p X_2^2}\right] &= \frac{H_p}{2}\TTi{1}, &\quad X_1 &= L.
    \label{plate:bc:wxxx}
  \end{alignat}}
Similarly, the boundary conditions at $X_2 = \pm W / 2$ are
\subeq{
  \begin{alignat}{2}
    \bar{\mathsf{S}}_{\alpha 2} &= 0, &\quad X_2 &= \pm W/2; \\
    \nub \pdd{\wb}{X_1} + \pdd{\wb}{X_2} &= 0, &\quad X_2 &= \pm W/2; \\
  \quad
  B\left[(2-\nub)\frac{\p^3\wb}{\p X_1^2 \p X_2} + \frac{\p^3\wb}{\p X_2^3}\right] &= \frac{H_p}{2}\TTi{2}, &\quad X_2 &= \pm W/2.
\end{alignat}}

\section{Non-dimensionalisation}
\label{sec:non_dim}

The governing equations for the film and plate
are non-dimensionalised by
scaling the in-plane coordinates as $\xx \sim L$.  Time is scaled as
$t \sim \tau_{\text{pe}}$, where the poroelastic time scale
$\tau_{\text{pe}} = \mu_f L^2 / (k_0 E_f)$
describes the time required for
a pressure gradient of magnitude $E_f / L$ to transport fluid a distance
$L$ through a porous medium with permeability $k_0$.

\subsection{Film model}

In the film, the vertical coordinate
is scaled according to $Z \sim \hf$.  The components
of the fluxes are scaled as
$\Qx \sim (k_0 / \mu_f) E_f / L$ and $\Qz \sim \epsilon (k_0 / \mu_f) E_f / L$.
The film displacements and stresses are scaled according to 
the estimates from Sec.~\ref{sec:plate_scaling}; thus,
$\asx \sim \delta \E \hf$, $\asz \sim \E \hf$, 
$\Sxx \sim E_f$, $\Sxz \sim \epsilon E_f$, 
$\Sza \sim \epsilon E_f$, and $\Szz \sim \epsilon^2 E_f$.
The pressure
and the elastic stress tensor are scaled
as $p \sim E_f$ and $\tens{\Sigma} \sim E_f$.

Under this choice of scales, the conservation law for the
nominal fluid fraction becomes
\begin{align}
  \pd{\Phi}{t} + \nx \cdot \Qx + \pd{\Qz}{Z} = 0.
  \label{nd:f:Phi}
\end{align}
By writing the symmetric permeability tensor as
$\tens{K} = \Mxx + \Mxz \otimes \ez + \ez \otimes \Mxz + \Mzz \ez \otimes \ez$,
the components of the fluid flux are
\subeq{
  \label{nd:f:Q}
\begin{align}
  \Qx &= -\Mxx \nx p - \epsilon^{-1} \Mxz \pd{p}{Z}, \label{nd:f:Qx}\\
  \Qz &= -\epsilon^{-1} \Mxz \cdot \nx p - \epsilon^{-2} \Mzz \pd{p}{Z}.
        \label{nd:f:Qz}
\end{align}}
The stress balances representing conservation of linear momentum are
\subeq{
\label{nd:f:lin_mom}
\begin{align}
  \nx \cdot \Sxx + \pd{\Sxz}{Z} = 0, \label{nd:f:lin_mom_x} \\
  \pd{S_{z\alpha}}{X_\alpha} + \pd{\Szz}{Z} = 0, \label{nd:f:lin_mom_z}
\end{align}}
where the PK1 stress tensor is
given by
\begin{align}
  \Sxx + \epsilon \left(\Sxz \otimes \ez + \Sza \ez \otimes \vec{e}_\alpha \right) + \epsilon^2 \Szz \ez \otimes \ez = \tens{\Sigma} - p J \tens{F}^{-T}.
  \label{nd:f:S}
\end{align}
The elastic stress tensor is written as
\begin{align}
  \tens{\Sigma} = a(\nu_f) (J - 1) J \tens{F}^{-T} + b(\nu_f)(\tens{F} - \tens{F}^{-T}),
  \label{nd:f:Sigma}
\end{align}
where the functions $a$ and $b$ are defined, for convenience, as
\begin{align}
a(\nu) \equiv \frac{\nu}{(1 + \nu)(1 - 2 \nu)}, \quad b(\nu) \equiv \frac{1}{2(1 + \nu)}.
\end{align}
The deformation gradient tensor is given by
\begin{align}
  \tens{F} = \tens{I} + \E \left(\delta \epsilon \nx \asx + \delta \pd{\asx}{Z} \otimes \ez + \epsilon \ez \otimes \nx \asz + \pd{\asz}{Z} \ez \otimes \ez\right). \label{nd:f:F}
\end{align}
The appearance of $\E$, $\epsilon$, and $\delta$ in the deformation gradient tensor 
lead to distinct asymptotic regimes
that are defined by the relative sizes of these parameters.  
Finally, the incompressibility condition is given by \eqref{eqn:J}.

\subsection{Plate model}

In the plate, the displacements are scaled according to
$\ubx \sim \delta \E \hf$, $\ubar \sim \delta \E \hf$, and
$\wb \sim \E \hf$.  The mean in-plane strains and stresses
are scaled as $\ee \sim \delta \epsilon \E$ and
$\Sxxbar \sim \delta \epsilon \E H_p E_p$.
The tractions are scaled as $\TTx \sim \epsilon E_f$
and $\TTz \sim \epsilon^2 E_f$; see Sec.~\ref{sec:plate_scaling}.

The rescaled FvK equations are then given by
\subeq{
  \label{nd:p:fvk}
\begin{align}
  -\B \nx^4 \wb + \epsilon \delta^{-1} \E \nx \cdot(\Sxxbar \nx \wb)
  &= -\frac{1}{2}\nx \cdot \TTx - \epsilon \delta^{-1} \TTz,
    \label{nd:p:w}
  \\
  \nx \cdot \Sxxbar &= -\TTx,
                      \label{nd:p:lin_mom_xx}
\end{align}
where $\B = 1/[12(1-\nu_b^2)]$ is a non-dimensional bending modulus.  The
rescaled mean stresses and strains are
\begin{align}
  \Sxxbar &= \frac{1}{1-\nu_p^2}\left[(1 - \nu_p) \ee + \nu_p \tr(\ee) \Ixx\right], \label{nd:p:Sxx}
  \\
  \ee &= \frac{1}{2}\left[\nx \ubar + (\nx \ubar)^T + \epsilon \delta^{-1} \E \nx \wb \otimes \nx \wb\right]. \label{nd:p:Exx}
\end{align}}
By scaling $Z \sim H_p$ in the plate, the leading-order contribution to the in-plane
displacement is given by
\begin{align}
  \ubx = \ubar - Z \nx w.
  \label{nd:p:uxx}
\end{align}

\subsection{Boundary conditions}

In the context of the film model, in which $\hf$ has been used to non-dimensionalise
the vertical coordinate $Z$, the plate has a non-dimensional thickness of
$\hp = \delta \epsilon^{-1}$.  The upper surface of the plate is therefore
located at $Z = \hp / 2$. 

The boundary conditions at the free surface of the film, $Z = H_f + \hp /2$, are given by
\subeq{
\begin{align}
  -\nx H_f \cdot \Qx + \Qz &= \Pe\, \mathcal{V}(\phif)\, \mathcal{S}(\tens{F})\,\mathcal{N}, \label{ndbc:f:top_Q} \\
  \Sxz - \Sxx \nx H_f &= 0, \label{ndbc:f:top_Sxz} \\
  \Szz - \Sza \pd{H_f}{X_\alpha} &= 0, \label{ndbc:f:top_Szz}
\end{align}}
where the P\'eclet number is defined as $\Pe = V_e^0 \hf \mu_f / (\epsilon^2 k_0 E_f)$,
where $V_e^0$ is a characteristic evaporative flux.
Moreover, $\mathcal{V} = V_e / V_e^0$ denotes the composition-dependent
non-dimensional
evaporative flux and $\mathcal{N} = (1 + \epsilon^2 |\nx H_f|^2)^{1/2}$.  
At the upper plate surface, $Z = \hp / 2$, the no-flux condition on the fluid
can be written as
\begin{align}
Q_z|_{+} = 0. \label{ndbc:f:bot_Q}
\end{align}
Using \eqref{nd:p:uxx}, continuity of displacement can be written as
\subeq{\label{ndbc:f:cont_u}
\begin{align}
\left.\asx\right|_{+} &= \ubar - (1/2) \nx \wb, \label{ndbc:f:cont_u_xx}\\
\left.\asz\right|_{+} &= \wb. \label{ndbc:f:cont_uz}
\end{align}}

The non-dimensionalised boundary conditions at the clamped and free edges
of the plate are identical to those written in Sec.~\ref{sec:beam_bc}.
However, the free edges are now located at $x_1 = 1$ and $x_2 = \pm \mathcal{W}/2$,
where $\mathcal{W} = W / L$; $B$ can be replaced with $\B$; and
$H_p$ can be set to one.

The tractions can be found by evaluating the normal components of the PK1
stress in the film at $Z = \hp / 2$ to give $\TTx = \Sxz \cdot \ez|_{+}$ and $\TTz = \Szz|_{+}$.
However, equivalent expressions that are more convenient to use can be found by
integrating the stress balances \eqref{nd:f:lin_mom}
across the film thickness, resulting in
\subeq{\label{nd:tau}
\begin{align}
  \TTx &= \nx \cdot \int_{\hp/2}^{H_f + \hp / 2} \Sxx\, \d Z, \label{nd:tau_x} \\
  \TTz &= \pd{}{X_\alpha}\int_{\hp/2}^{H_f + \hp / 2} \Sza \, \d Z. \label{nd:tau_z}
\end{align}
}

\section{Asymptotic reductions of the coupled film-plate model}
\label{sec:reduction}

The coupled system of equations for the film and the plate can be
asymptotically reduced by taking the limit $\epsilon \to 0$.
Substantial simplifications occur through a decoupling
of the models.  In particular, the equations for the film can be
solved independently of those for the plate. 
The problem for the film also simplifies through
a decoupling of the mechanical and fluid-transport
problems.  This decoupling occurs because 
the deformation gradient tensor, to leading order, is simply
$\tens{F} =\mathrm{diag}(1, 1, J)$.  Thus, the mechanical problem
can be readily solved in terms of $J = 1 + \Phi - \phif_0$
and used to formulate a closed-form problem for $\Phi$.
Using the asymptotically reduced model in practice therefore
involves three steps:
\begin{enumerate}
\item Solve a reduced model for the nominal fluid fraction $\Phi$ and compute
	the local volumetric contraction $J$.
\item Compute the tractions, which can be expressed in terms of $J$ and $\nx w$, and substitute them into the FvK equations.
\item Solve the FvK equations for the plate displacements.
\end{enumerate}

Despite the simplified pathway to obtaining a solution, there are
several asymptotic regimes in both the mechanical and fluid-transport
problems for the film that must be delicately handled.  These regimes arise 
from the various sizes that $\E$ and $\Pe$
can take.  The former alters the structure of the mechanical problem
whereas the latter alters the fluid-transport problem.  In terms
of the mechanical problem, we consider three cases that can be summarised
as follows:
\begin{itemize}
\item{$\delta = O(\epsilon)$ with $\E = O(1)$: the deformation of the film is
    driven by a combination of volumetric contraction and plate bending. Both
	the in-plane and vertical tractions drive bending.}
\item{$\delta \gg \epsilon$ with $\E = O(1)$: in-plane film deformation
    is dominated by plate bending. The vertical film deformation driven by
    volumetric contraction and plate bending.  Only the in-plane traction
	drives bending.}
\item{$\delta \gg \epsilon$ with $\E = O(\delta \epsilon^{-1})$: the
    deformation of the film is dominated by plate bending. Both the
	in-plane and vertical traction drive bending.}
\end{itemize}
In terms of the drying dynamics, there are three cases to consider:
\begin{itemize}
\item{$\Pe = O(1)$: the nominal fluid fraction is uniform in the vertical direction. The flow of fluid predominantly occurs in the in-plane directions. 
    The case $\Pe \ll 1$ is a sub-limit in which the film remains homogeneous
	during drying; that is, the fluid fraction is spatially uniform.}
\item{$\Pe = O(\epsilon^{-1})$: the nominal fluid fraction is uniform in the vertical direction. Fluid flow is two dimensional, with the in-plane and
  vertical flux components having similar orders of magnitude.}
\item{$\Pe = O(\epsilon^{-2})$: the nominal fluid fraction is non-uniform in the vertical direction. Fluid flow mainly occurs in the vertical direction.}
\end{itemize}
In Sec.~\ref{sec:asy_mech}, asymptotic solutions to the mechanical problems
are detailed for the three cases described above. In Sec.~\ref{sec:asy_dry},
the fluid transport problem is asymptotically reduced according to the
size of the P\'eclet number.

\subsection{Reduction of the mechanical problems}
\label{sec:asy_mech}

Asymptotic expansions are used to solve the governing equations for
the film mechanics. The goal is to obtain expressions for
the traction. As the plate model has already been reduced, we
do not asymptotically expand the solutions to the FvK equations.
Instead, we comment on which terms can be neglected depending on
the sizes of $\delta$ and $\E$ relative to $\epsilon$.

\subsubsection{Films and plates of similar thicknesses}
\label{sec:am:c1}

We first consider the distinguished limit in which the film and the
plate have similar thicknesses by letting $\delta = \delta_0 \epsilon$,
where $\delta_0 = O(1)$ as $\epsilon \to 0$.  
To ensure that the FvK limit remains valid, we also assume that
$\E = O(1)$ as $\epsilon \to 0$.  With this choice of $\delta$, 
the deformation gradient tensor for the film \eqref{nd:f:F}
can be asymptotically expanded as
$\tens{F} = \tens{F}^{(0)} + \epsilon \tens{F}^{(1)} + O(\epsilon^2)$,
where
\subeq{
\begin{align}
\tens{F}^{(0)} &= \Ixx + \left(1 + \E \pd{\asz^{(0)}}{Z}\right) \ez \otimes \ez, \label{am:c1:F0} \\
\tens{F}^{(1)} &= \E \left(\delta_0 \pd{\asx^{(0)}}{Z}\otimes \ez + \ez \otimes \nx \asz^{(0)}\right)
+ \Fzz^{(1)} \ez \otimes \ez. \label{am:c1:F1}
\end{align}}
Importantly, the leading-order contribution to $\tens{F}$ is
diagonal and can be expressed as $\tens{F}^{(0)} = \mathrm{diag}(1,1, J^{(0)})$,
where the local contraction ratio is given by
$J^{(0)} = 1 + \Phi^{(0)} - \phif_0$ from the incompressibility
condition \eqref{eqn:J}. Thus, removal of fluid from the pore space drives
a uniaxial contraction of the solid matrix along the vertical direction.

The components of the PK1 stress tensor for the film are expanded as
$\Sxx = \Sxx^{(0)} + O(\epsilon)$, $\Sxz = \Sxz^{(1)} + O(\epsilon)$,
$\Sza = \Sza^{(1)} + O(\epsilon)$,
and $\Szz = \Szz^{(2)} + O(\epsilon)$.  Non-zero superscripts are used when a 
component has already been scaled by a power of $\epsilon$ when non-dimensionalising
the model. The elastic stress tensor 
and the pressure are written as $\tens{\Sigma} = \tens{\Sigma}^{(0)} + \epsilon \tens{\Sigma}^{(1)} + O(\epsilon^2)$
and $p = p^{(0)} + O(\epsilon)$.  From \eqref{am:c1:F0} and
\eqref{nd:f:Sigma}, we can deduce that the leading-order
contribution to the elastic stress tensor is diagonal and given by
$\tens{\Sigma}^{(0)} = \sxx^{(0)} + \szz^{(0)} \ez \otimes \ez$,
where
\subeq{\label{am:c1:Sigma_0}
\begin{align}
\sxx^{(0)} &= a(\nu_f)(J^{(0)} - 1) J^{(0)} \Ixx, \label{am:c1:Sigma_xx_0}\\
\szz^{(0)} &= a(\nu_f)(J^{(0)} - 1) + b(\nu_f)\left(J^{(0)} - \frac{1}{J^{(0)}}\right). \label{am:c1:Sigma_zz_0}
\end{align}}
In addition, by considering the $O(1)$ contributions of the PK1 stress tensor
\eqref{nd:f:S}, we see that
\subeq{\label{am:c1:S_0}
\begin{align}
\Sxx^{(0)} &= \sxx^{(0)} - p^{(0)} J^{(0)} \Ixx, \\
0 &= \szz^{(0)} - p^{(0)}.
\end{align}}
Therefore, by substituting \eqref{am:c1:Sigma_0} into \eqref{am:c1:S_0}, 
we can deduce that the pressure and the in-plane PK1 stress are given by
\subeq{
\begin{align}
  p^{(0)} &=a(\nu_f)(J^{(0)} - 1) + b(\nu_f)\left(J^{(0)} - \frac{1}{J^{(0)}}\right), \label{am:c1:p_0}\\
  \Sxx^{(0)} &=  b(\nu_f)\left[1 - (J^{(0)})^2\right]\Ixx. \label{am:c1:Sxx_0}
\end{align}}
Since $J^{(0)} < 1$, the pressure will be negative. In dimensional terms,
this means the fluid pressure will be below atmospheric pressure (which has
been set to zero). The in-plane elastic stresses will be compressive, while the total (PK1)
in-plane stresses will be tensile. 
For convenience, we define 
\begin{align}
\Sxxx(J) \equiv b(\nu_f)\left(1 - J^2\right)
\label{eqn:scalar_Sxx}
\end{align}
so that the in-plane PK1 
stress tensor can be written as $\Sxx^{(0)} = \Sxxx(J^{(0)}) \Ixx$. 
From the $O(\epsilon)$ contributions to the PK1 stress tensor \eqref{nd:f:S},
we find, after simplification, that
\subeq{
\begin{align}
\Sxz^{(1)} = \E b(\nu_f) \left(\delta_0 \pd{\asx^{(0)}}{Z} + J^{(0)} \nx \asz^{(0)}\right), \label{am:c1:Saz_1}\\
\Sza^{(1)} \vec{e}_\alpha = \E b(\nu_f) \left(\delta_0 J^{(0)} \pd{\asx^{(0)}}{Z} + \nx\asz^{(0)}\right). \label{am:c1:Sza_1}
\end{align}}
Equation \eqref{am:c1:Saz_1} will enable the the in-plane displacement
to be determined, whereas \eqref{am:c1:Sza_1} will enable
the vertical traction to be computed via \eqref{nd:tau_z}.  

The shear stress $\Sxz^{(1)}$ in the film can be determined by 
integrating the
in-plane stress balance \eqref{nd:f:lin_mom_x} and imposing the stress-free 
boundary condition \eqref{ndbc:f:top_Sxz} to obtain, after simplification,
\begin{align}
  \Sxz^{(1)} = \nx \int_{Z}^{H_f + \hp / 2}\Sxxx(J^{(0)})\, \d Z.
  \label{am:c1:Saz_1_v2}
\end{align}
A differential equation for the
vertical component of the film displacement $\asz^{(0)}$
can be obtained by calculating $J^{(0)} = \det \tens{F}^{(0)}$
using \eqref{am:c1:F0} to obtain
\begin{align}
  1 + \E \pd{\asz^{(0)}}{Z} = J^{(0)}.
  \label{am:c1:ode_w}
\end{align}
Having determined $\Sxz^{(1)}$ and $\pdf{\asz^{(0)}}{Z}$, we can use
\eqref{am:c1:Saz_1} to formulate a differential equation for  $\asx^{(0)}$.
After solving \eqref{am:c1:Saz_1} and \eqref{am:c1:ode_w}
and imposing continuity of
displacement \eqref{ndbc:f:cont_u}, we
find that the film displacements can be written as
\subeq{\label{am:c1:u_0}
  \begin{align}
    \asx^{(0)} &= \ubar - (1/ 2)\nx\wb  \nonumber \\ 
    & \quad + (\delta_0 \E b(\nu_f))^{-1}\Bigg\{
      \int_{\hp/2}^{Z}\bigg[\Sxz^{(1)} - \E b(\nu_f) \nx \wb
    - b(\nu_f) J^{(0)} \int_{\hp/2}^{Z} \nx J^{(0)}\,\d Z'\bigg]\,\d Z \Bigg\},
    \\
    \asz^{(0)} &= \E^{-1} \int_{\hp / 2}^{Z}\left(J^{(0)} - 1\right)\,\d Z + w. \label{am:c1:uz_0}
\end{align}}
The Eulerian film thickness 
can be obtained from \eqref{am:c1:uz_0} as
\begin{align}
  h_f = \int_{\hp/2}^{H_f + \hp/2} J^{(0)}(\XX, Z, t)\,\d Z.
  \label{am:c1:h}
\end{align}

We are now in a position to calculate the tractions. Substituting
\eqref{am:c1:Sxx_0} into \eqref{nd:tau_x} and using \eqref{eqn:scalar_Sxx}
leads to an expression for the in-plane traction given by
\subeq{\label{am:c1:tau}
  \begin{align}
    \TTx = \nx \int_{\hp/2}^{H_f + \hp / 2}\Sxxx(J^{(0)})\, \d Z. \label{am:c1:tau_x}
  \end{align}
  The vertical traction can be obtained by first substituting the expressions
  for the displacements \eqref{am:c1:u_0} into \eqref{am:c1:Sza_1} and
  then inserting the result into \eqref{nd:tau_z} to find
  \begin{align}
    \TTz &= \nx^2 \int_{\hp/2}^{H_f + \hp/2} J^{(0)} \left(\int_{Z}^{H_f + \hp/2} \Sxxx\,\d Z'\right)\,\d Z \nonumber \\
         &- \nx \cdot \int_{\hp / 2}^{H_f + \hp / 2}\left\{\left(\int_{Z}^{H_f + \hp/2} \Sxxx\, \d Z'\right)\nx J^{(0)} - \Sxxx \left(\int_{\hp/2}^{Z} \nx J^{(0)}\, \d Z'\right)\, \d Z\right\} \nonumber \\
         &+ \nx \cdot \int_{\hp/2}^{H_f + \hp/2} \E \Sxxx \nx w\, \d Z. \label{am:c1:tau_z}
\end{align}}
At this point, the leading-order mechanical problem for the film has
been solved in terms $J^{(0)} = 1 + \Phi^{(0)} - \phi_0$. Thus, all that
remains is to solve the transport problem for the nominal
solvent fraction $\Phi^{(0)}$.

The expressions for the traction
\eqref{am:c1:tau} can be inserted into the FvK equations \eqref{nd:p:fvk}.
In the distinguished limit $\delta = O(\epsilon)$ with $\E = O(1)$
as $\epsilon \to 0$, the full set of FvK equations must be solved;
no further simplifications to the plate model are possible. In particular,
both the in-plane and vertical tractions drive plate bending despite
the differences in their orders of magnitude.

\subsubsection{Soft films on thick plates}
\label{sec:am:c2}

We now consider the limit in which the film is thin
relative to the plate, $\epsilon \ll \delta$.  We further assume
that the film is soft so that $\E = O(1)$. In this case, 
the deflection of the plate will be proportional to the
film thickness.  The film displacements are expanded
as $\asx = \asx^{(0)} + o(1)$ and 
$\asz = \asz^{(0)} + o(1)$. The deformation gradient
tensor $\tens{F}$ remains diagonal to leading order; see
\eqref{nd:f:F}.  Therefore, many
of the results obtained in Sec.~\ref{sec:am:c1} for the case
$\delta = O(\epsilon)$ still apply. 
In particular, the leading-order contributions to
pressure, in-plane stresses, shear stresses, and vertical displacement
are given by \eqref{am:c1:p_0},
\eqref{am:c1:Sxx_0}, \eqref{am:c1:Saz_1_v2}, and
\eqref{am:c1:uz_0}, respectively.
The main
difference occurs in the form of the in-plane displacement,
which can be found from examining the 
$\alpha z$ component of the stress-strain relation \eqref{nd:f:S},
which reads as
\begin{align}
  \epsilon \Saz = \vec{e}_\alpha \tens{\Sigma} \ez - p J \vec{e}_\alpha \tens{F}^{-T} \ez
  = \E b(\nu_f)\left(\delta \pd{u_\alpha^{(0)}}{Z} + \epsilon J^{(0)} \pd{\asz^{(0)}}{X_\alpha}
  \right)
  + o(\epsilon),
  \label{am:c2:Saz}
\end{align}
where $J^{(0)} = 1 + \E \pdf{\asz^{(0)}}{Z} = 1 + \Phi^{(0)} - \phif_0$.  The leading-order
contributions to \eqref{am:c2:Saz}, which are $O(\delta)$ in size,
imply that $\pdf{u_\alpha^{(0)}}{Z} = 0$. Solving this equation and 
imposing continuity of in-plane displacements \eqref{ndbc:f:cont_u_xx}
then shows that
$\asx^{(0)} = \ubar - (1/2) \nx w$.  
Therefore, the in-plane displacement of the film simply matches
that of the upper plate surface.  In-plane deformation due to volumetric contraction is a higher-order effect.


The leading-order problem for the plate can be obtained by taking
$\epsilon \to 0$ with $\epsilon \ll \delta$ and $\E = O(1)$ in the FvK
equations given by \eqref{nd:p:fvk}. 
The vertical traction $\TTz$ drops out of \eqref{nd:p:w} and
is not required. Thus, plate bending is dominated by the in-plane traction
$\TTx$, which can be obtained from \eqref{am:c1:tau_x}.
The nonlinear terms from $\nx \cdot (\Sxxbar \nx w)$ also drop
out of \eqref{nd:p:w}; thus the
vertical plate displacement satisfies
a linear plate equation.  The mean in-plane plate displacements $\ubar$ 
can be obtained by solving \eqref{nd:p:lin_mom_xx}--\eqref{nd:p:Exx}
after neglecting the nonlinear terms in the strain tensor \eqref{nd:p:Exx}.

\subsubsection{Stiff films on thick plates}

The final case to consider is when the film is thin relative to
the plate, $\epsilon \ll \delta$, and stiff, so that
$\E = O(\delta \epsilon^{-1})$. We therefore write
the effective film stiffness as $\E = \delta \epsilon^{-1} \E_1$.  
The deformation gradient
tensor for the film \eqref{nd:f:F} is given by
\begin{align}
  \tens{F} = \tens{I} + \E_1 \left(\delta^2 \nx \asx + \delta^2 \epsilon^{-1} \pd{\asx}{Z} \otimes \ez + \delta \ez \otimes \nx \asz + \delta \epsilon^{-1}\pd{\asz}{Z} \ez \otimes \ez\right).
  \label{am:c3:FF}
\end{align}
The requirement that $\tens{F}$ is invertible implies
that the large $O(\delta \epsilon^{-1})$ term in \eqref{am:c3:FF} must
be smaller in magnitude;
consequently, the vertical displacement must have the expansion
\begin{align}
\asz(\XX, Z, t) = \asz^{(0)}(\XX,t) + \epsilon \delta^{-1}\aszt(\XX, Z, t)
+ o(\epsilon \delta^{-1}).
\end{align}
The expansion for the in-plane film displacement can be found using a
similar argument to that in Sec.~\ref{sec:am:c2}, leading to
\begin{align}
\asx(\XX, Z, t) = \asx^{(0)}(\XX,t) + \epsilon \delta^{-1}\asxt(\XX, Z, t)
+ o(\epsilon \delta^{-1}).
\end{align}
By imposing continuity of displacement \eqref{ndbc:f:cont_u}, we conclude that
$\asx^{(0)} = \ubar - (1/2) \nx \wb$ and $\asz^{(0)} = \wb$. Thus, the
leading-order film displacement matches the plate displacement, i.e.\
the film simply bends with the plate. Boundary conditions for
$\asxt$ and $\aszt$ could be obtained by calculating a higher-order solution
to the plate model and imposing continuity of displacement; however, we do
not do this here.
The deformation gradient tensor for the film therefore reduces to
\begin{align}
  \tens{F} = \tens{I} + \E_1 \pd{\aszt}{Z} \ez \otimes \ez + \delta \E_1 \left(\pd{\asxt}{Z} \otimes \ez + \ez \otimes \nx \wb\right) + O(\epsilon).
  \label{am:c3:F}
\end{align}
Although $\tens{F}$ is diagonal to leading order, the $O(\delta)$
corrections result in substantially different film mechanics compared to
when the film is soft ($\E = O(1)$). To see this,
the
deformation gradient tensor \eqref{am:c3:F}
can be substituted into the stress-strain relation \eqref{nd:f:S}. 
Collecting the $O(\delta)$ contributions to the $\alpha z$ and
$z \alpha$ components leads to the following equations, respectively,
\subeq{\label{am:c3:u}
\begin{align}
  \pd{\asxt}{Z} + J^{(0)} \nx \wb &= \vec{0},
                                                    \label{am:c3:eq1}
  \\
  J^{(0)}\pd{\asxt}{Z} + \nx \wb &= \vec{0}, \label{am:c3:eq2}
\end{align}}
where $J^{(0)} = 1 + \E_1 \pdf{\aszt}{Z}$. 
Solving \eqref{am:c3:u} shows that $\nx \wb = \vec{0}$, which implies
that the plate remains flat. This unphysical result stems from the assumption
that the shear stresses $\Saz$ and $\Sza$ in the film have the same
order of magnitude, which would be the case in linear elasticity.
Thus, a resolution can be found by elevating the asymptotic order of 
$\Sza$ and writing $\Sza = \delta \epsilon^{-1} \tSza$. From the vertical
stress balance \eqref{nd:f:lin_mom_z}, we must also elevate the  order
of $\Szz$ by writing $\Szz = \delta \epsilon^{-1} \tSzz$.
Importantly, the elevated order of the
vertical stress implies that the non-dimensional vertical traction must also be
rescaled according to $\TTz = \delta \epsilon^{-1} \tTTz$.
In terms of dimensional quantities,
these rescalings imply that $\Sza \sim \delta E_f$, 
$\Szz \sim \epsilon \delta E_f$, and
$\TTz \sim \epsilon \delta E_f$.

With these rescalings,
we proceed by expanding $\Sxz = \Sxz^{(1)} + o(1)$ and 
$\tSza$ as $\tSza = \tSza^{(1,1)} + o(1)$. The $\alpha z$
component of the stress-strain relation still leads to \eqref{am:c3:eq1};
however, the $z \alpha$ component now gives
\begin{align}
  \tSza^{(1,1)} \vec{e}_\alpha =  \E_1 b(\nu_f) \left(J^{(0)}\pd{\asxt}{Z} + \nx \wb \right) = \E_1 \Sxxx(J^{(0)}) \nx \wb,
  \label{am:c3:Sza}
\end{align}
which replaces \eqref{am:c3:eq2}. The second equality arises from using
\eqref{am:c3:eq1} and the definition of
$\Sxxx$ in \eqref{eqn:scalar_Sxx}.
The vertical traction can now be obtained by substituting \eqref{am:c3:Sza}
into \eqref{nd:tau_z} to find that
\begin{align}
  \tTTz = \nx \cdot \int_{\hp / 2}^{H_f + \hp/2} \E_1 \Sxxx(J^{(0)}) \nx \wb\,\d Z.
  \label{am:c3:tau_z}
\end{align}

Despite the elevated asymptotic order of the vertical stress
$\Szz$, the $zz$ component
of the stress-strain relation \eqref{nd:f:S}
leads to an expression for the pressure
that is identical to \eqref{am:c1:p_0}. Consequently, the in-plane
stresses and traction are given by \eqref{am:c1:Sxx_0} and \eqref{am:c1:tau_x},
respectively. 

In the regime of stiff films on thick plates, $\epsilon \ll \delta$
with $\E = O(\delta \epsilon^{-1})$, the full set of FvK equations given by
\eqref{nd:p:fvk} must be solved. The small $O(\epsilon \delta^{-1})$
prefactor
of $\TTz$
in \eqref{nd:p:w} is exactly offset by the large
$O(\delta \epsilon^{-1})$ magnitude of $\TTz$. Consequently, the
vertical traction must be considered despite the film being much
thinner than the plate.

\subsection{Film drying}
\label{sec:asy_dry}

Having solved the mechanical problem in the film, the fluid-transport
problem is asymptotically reduced by taking the limit $\epsilon \to 0$.
The goal is to derive a hierarchy of evolution equations for the nominal
solvent fraction that capture different drying regimes depending on the
magnitude of the P\'eclet number, $\Pe$.

Physically, the different drying regimes arise from differences in the
relative rates at which fluid is removed from the pore space by evaporation
and replenished by bulk transport.  When $\Pe = O(1)$, the time scales
of evaporation and in-plane fluid transport are commensurate.  If
$\Pe \ll 1$, evaporation is slow and a homogeneous drying process can be
expected. When $\Pe \gg 1$, evaporation is fast relative to in-plane transport
and a highly non-uniform film composition is likely to arise.

In reducing the fluid-transport problem, we work within the distinguished
limit of $\delta = O(\epsilon)$ as $\epsilon \to 0$ by assuming the
film and plate have comparable thicknesses. In this case, the
deformation gradient tensor has the asymptotic expansion
$\tens{F} = \mathrm{diag}(1, 1, J)
+ O(\epsilon)$. Thus, the off-diagonal components of the
permeability tensor are small, specifically, $\Mxz = O(\epsilon)$.
We therefore expand the pressure, nominal fluid fraction, and
components of the permeability tensor
as $p = p^{(0)} + O(\epsilon)$, $\Phi = \Phi^{(0)} + O(\epsilon)$,
$\Mxx = \Mxx^{(0)} + O(\epsilon)$, $\Mxz = \epsilon \Mxz^{(1)} + O(\epsilon^2)$,
and $\Mzz = \Mzz^{(0)} + O(\epsilon)$. The expansions of the flux will
depend on the size of $\Pe$. 

Although the reduction of the fluid-transport problem will be carried out
in the limit that $\delta = O(\epsilon)$, the results also apply
when the film is thin relative to the
plate and $\delta \gg \epsilon$.

\subsubsection{Slow evaporation: $\Pe = O(1)$}
\label{sec:ad:c1}

We first consider the drying problem when the evaporation is
slow, as characterised by $\Pe = O(1)$ as $\epsilon \to 0$.
The components of the flux are expanded as
$\Qx = \Qx^{(0)} + O(\epsilon)$ and $\Qz = \Qz^{(0)} + O(\epsilon)$.
Collecting the $O(\epsilon^{-2})$ terms in \eqref{nd:f:Qz} lead to
the conclusion that the pressure $p^{(0)}$, and hence $J^{(0)}$ and $\Phi^{(0)}$, are independent of $Z$.  Therefore, the film composition is uniform along
the vertical direction. 
The $O(1)$ terms in \eqref{nd:f:Qx} provide an
expression for the in-plane flux,
$\Qx^{(0)} = -\Mxx^{(0)} \nx p^{(0)}$,
where $\Mxx^{(0)} = k(\phif^{(0)}) J^{(0)} \Ixx$. The Eulerian fluid fraction
is given by $\phif^{(0)} = \Phi^{(0)} / J^{(0)}$.

By integrating the solvent conservation equation \eqref{nd:f:Phi} 
from $Z = \hp/2$ to $Z = H_f(\XX) + \hp/2$
and using the boundary conditions \eqref{ndbc:f:top_Q} and \eqref{ndbc:f:bot_Q},
we find that the nominal solvent fraction evolves according to
\begin{align}
  H_f \pd{\Phi^{(0)}}{t} = \nx \cdot \left(H_f J^{(0)} k(\phi^{(0)}) \nx p^{(0)}\right) - \Pe\,\mathcal{V}(\phi^{(0)}),
  \label{ad:c1:Phi}
\end{align}
where $J^{(0)} = 1 + \Phi^{(0)} - \phi_0$. Upon substituting the expression
for the pressure $p^{(0)}$ given by \eqref{am:c1:p_0} into \eqref{ad:c1:Phi},
a nonlinear diffusion equation for $\Phi^{(0)}$ is obtained. 
Equation \eqref{ad:c1:Phi}
can be solved with the boundary conditions $\Qx^{(0)} \cdot \vec{N} = 0$ at
the contact line, $X_1 = 1$ or $X_2 = \pm \mathcal{W}$, along with the
initial condition $\Phi^{(0)} = \phi_0$.

Upon solving for $\Phi^{(0)}$ and computing $J^{(0)}$,
the Eulerian film thickness can be obtained
from \eqref{am:c1:h} as $h_f = J^{(0)} H_f$. Moreover, the
traction can be determined from \eqref{am:c1:tau} as
\subeq{\label{ad:c1:tau}
\begin{align}
  \TTx &= \nx (H_f \Sxxx), \label{ad:c1:tau_x} \\
  \TTz &= \frac{1}{2}\nx^2 \left(J^{(0)} H_f^2  \Sxxx\right) + \E \nx \cdot \left(H_f \Sxxx \nx \wb\right), \label{ad:c1:tau_z}
\end{align}}
where $\Sxxx = \Sxxx(J^{(0)})$ is given by \eqref{eqn:scalar_Sxx}.
In the thin-film regime $\delta \gg \epsilon$ with $\E = O(1)$, the
in-plane traction is given by \eqref{ad:c1:tau_x} and the vertical traction
\eqref{ad:c1:tau_z} is not required.  When $\delta \gg \epsilon$ with
$\E = O(\delta \epsilon^{-1})$, the in-plane traction is given by \eqref{ad:c1:tau_x}
and the vertical traction is
\begin{align}
\TTz = \E \nx \cdot \left(H_f \Sxxx \nx \wb\right). \label{ad:c1:tau_z_r3}
\end{align}

If evaporation is very slow, $\Pe \ll 1$, then time can be rescaled
as $t \sim \Pe^{-1}$.  The leading-order (in $\Pe$) part of \eqref{ad:c1:Phi} 
then implies that $p^{(0)}$ and hence $\Phi^{(0)}$ are uniform in space.
Integrating \eqref{ad:c1:Phi} over the plate surface leads to
a simple differential equation for the nominal volume fraction,
which can be expressed in terms of the original non-dimensional time as
\begin{align}
  \td{\Phi^{(0)}}{t} = -\frac{\Pe \mathcal{W} \mathcal{V}(\phi^{(0)})}{V_0},
  \label{ad:c1:Phi_slow}
\end{align}
where $V_0$ is the initial volume of the film. By taking
$J^{(0)}$ and hence $\Sxxx$ to be constant in \eqref{ad:c1:tau_x},
the in-plane traction reduces to $\TTx = \Sxxx \nx H_f$. Given that
$\Sxxx > 0$, the direction of the traction follows the gradient of the film
thickness. Near the contact line, where $H_f$ tends to zero,
the film will always pull the plate inwards, that is, towards the
centre of the plate.

\subsubsection{Moderate evaporation: $\Pe = O(\epsilon^{-1})$}
\label{sec:ad:c2}

The limit of moderate evaporation is characterised by 
$\Pe = O(\epsilon^{-1})$.  Thus, we write
$\Pe = \epsilon^{-1} \Pe_0$.  In order to obtain a balance in the
boundary condition \eqref{ndbc:f:top_Q}, we must rescale
the vertical component of the flux as $\Qz = \epsilon^{-1} \tQz$.
Both flux components now have the same order of magnitude.
Despite the elevated asymptotic order of $\Qz$, the
vertical component of Darcy's law \eqref{nd:f:Qz} 
implies that $p^{(0)}$, $J^{(0)}$, and $\Phi^{(0)}$ are
still independent of $Z$.  
Balancing terms in
the conservation law for the nominal fluid fraction
\eqref{nd:f:Phi} requires rescaling time as
$t = \epsilon \tilde{t}$.  Integrating
the $O(\epsilon^{-1})$ terms in the conservation
law across the film thickness leads to 
\begin{align}
  H_f \pd{\Phi^{(0)}}{\tilde{t}} = - \Pe_0 \,\mathcal{V}(\phi^{(0)}).
  \label{ad:c2:Phi}
\end{align}
Equation \eqref{ad:c2:Phi} can be treated as an ordinary differential
equation for $\Phi^{(0)}$, with the in-plane coordinates $\XX$ acting
as parameters, and solved with the
initial condition $\Phi^{(0)} = \phi_0$ at $\tilde{t} = 0$.  
The expressions for the traction that are provided
in Sec.~\ref{sec:ad:c1} also apply to this drying regime.

The reduced transport problem defined by \eqref{ad:c2:Phi},  
does not capture the in-plane components of the fluid flux.
However, near the contact line, where $H_f$ is small, 
\eqref{ad:c2:Phi} predicts that the nominal fluid fraction
will undergo a rapid decrease.  Consequently, large in-plane
gradients in the fluid fraction are expected near the contact line, 
and these will locally amplify the in-plane flux.
From an asymptotics perspective, near the contact line,
the order of the in-plane flux must be elevated,
resulting in additional terms that must be considered
in the reduced model.  The dynamics near the contact line
can therefore be resolved by carrying out a boundary-layer analysis
using matched asymptotic expansions.  However,   
we do not resolve these boundary layers because they are
unimportant in terms of plate bending; see 
Sec.~\ref{sec:fem_dynamic} for more details.

\subsubsection{Fast evaporation: $\Pe = O(\epsilon^{-2})$}
\label{sec:ad:c3}

The limit of fast evaporation occurs when
$\Pe = O(\epsilon^{-2})$.  Thus, we write
$\Pe = \epsilon^{-2} \Pe_1$.  
By following the same strategy as in
Sec.~\ref{sec:ad:c2}, we deduce that capturing the dynamics
requires rescaling time as $t = \epsilon^2 \hat{t}$ and
the vertical component of the flux as $\Qz = \epsilon^{-2} \hQz$.
By writing $\hQz = \hQz^{(-2)} + O(\epsilon)$ and 
equating the $O(\epsilon^{-2})$ components of 
\eqref{nd:f:Qz}, we find that $\hQz^{(-2)} = -\Mzz^{(0)} \p_Z p^{(0)}$,
with $\Mzz^{(0)} = k(\phif^{(0)}) / J^{(0)}$.  
Therefore, the rate of evaporation is now sufficiently large that
vertical gradients will occur in the
pressure $p^{(0)}$, the volumetric contraction ratio $J^{(0)}$,
and the nominal fluid fraction $\Phi^{(0)}$.  
Equating the $O(\epsilon^{-2})$ contributions of the
conservation law \eqref{nd:f:Phi} leads to
a nonlinear diffusion equation for the fluid fraction
given by
\begin{align}
\pd{\Phi^{(0)}}{\hat{t}} + \pd{\hQz^{(-2)}}{Z} = 0.
\label{ad:c3:Phi}
\end{align}
The boundary conditions and initial for \eqref{ad:c3:Phi} are
\subeq{
\begin{alignat}{2}
\hQz^{(-2)} &= 0, &\quad Z &= \hp/2; \\
\hQz^{(-2)} &= \Pe_1\,\mathcal{V}(\phi^{(0)}), &\quad Z  &= H_f + \hp/2; \\
\Phi^{(0)} &= \phif_0, &\quad \hat{t} &= 0.
\end{alignat}}
Again, the in-plane coordinates $\XX$ play the role of parameters.
After solving for $\Phi^{(0)}$ and hence $J^{(0)}$, the
tractions can be evaluated by computing the integrals in
\eqref{am:c1:tau_x} and \eqref{am:c1:tau_z} or \eqref{am:c3:tau_z}. 

The reduced transport problem \eqref{ad:c3:Phi} does not capture
in-plane fluid transport.  However, in contrast to the case
$\Pe = O(\epsilon^{-1})$, the contributions from the 
in-plane fluid flux are always asymptotically smaller than those
from the large vertical flux.  Therefore,
there is no need to resolve the dynamics near the contact line
using a separate boundary-layer analysis.

\section{Validation and parametric studies}
\label{sec:fem}

The asymptotic reductions
are validated by comparing them against finite element (FE)
solutions of the fully coupled film-plate model proposed
in Sec.~\ref{sec:model}.  The FE method is
implemented in Python with FEniCS~\cite{logg2012, alnaes2015}. 
The multiphenics~\cite{ballarin2019}
package is used to couple the film and plate problems.  
The traction $\TT$ is treated as a Lagrange multiplier that
enables the continuity of displacement at the film-plate
interface to be imposed.  
The film and plate displacements are represented using
P2 elements.  The pressure and nominal fluid fraction
 are represented using P1 elements.  Finally,
DGT1 elements are used for the traction.  Time is
discretised using a fully implicit Euler method with
a fixed time step.  The film and plate problems
are simultaneously solved as a monolithic nonlinear system.
We employ the PETSc
SNES nonlinear solver, using MUMPS to
solve the linear systems during each Newton iteration.

The FE simulations are carried out in a two-dimensional
Cartesian geometry, which is valid when the width of
the plate is small, $\mathcal{W} \ll 1$. In this
case, the plate reduces to a beam.
The film is assumed to have a 
parabolic initial profile that can be written in 
dimensional form as $H_f(X) = 4 \epsilon X (1 - X / L)$,
where we define $X \equiv X_1$ for simplicity.  The
plate is assumed to be made of steel with a 
Poisson's ratio of $\nu_p = 0.27$.  
The Poisson's ratio of the film is set to
$\nu_f = 0.2$, corresponding to a porous 
matrix formed from a dried colloidal 
dispersion~\cite{style2011}.

\begin{figure}
  \centering
  \includegraphics[width=\textwidth]{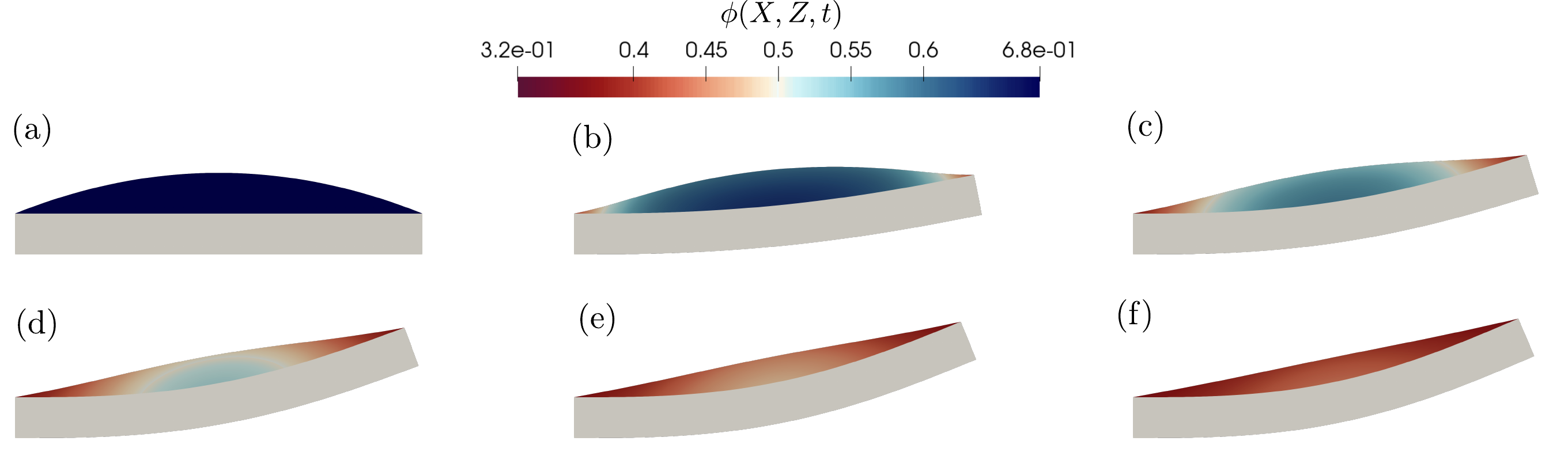}
  \caption{Finite element simulations of the film-plate model
    presented in Sec.~\ref{sec:model}. The heat map represents the
    Eulerian fluid fraction (porosity) of the film, $\phif = \Phi / J$.
    Simulation snapshots are shown at times
	(a) $t = 0$, (b) $t = 0.005$, (c) $t = 0.010$, (d) $t = 0.015$,
	(e) $t = 0.020$, and (f) $t = 0.025$.  The parameter values are
	$\epsilon = \delta = 0.1$, $\Pe = 100$, $\E = 3$ with
	$\phif_0 = 0.68$, and $k(\phi) \equiv 1$. The non-dimensional
        evaporative flux is $\mathcal{V}(\phif) = \phif - \phif_\infty$, where
        $\phif_\infty = 0.36$.}
  \label{fig:sim}
\end{figure}

Snapshots from a typical FE simulation are shown
in Fig.~\ref{fig:sim}.  The Eulerian fluid fraction
in the film $\phif = \Phi / J$ is plotted as a heat map.
It has been assumed that the film and plate have the same 
thickness, $\epsilon = \delta = 0.1$.
The P\'eclet number has been set to $\Pe = 100$.  The
effective film stiffness has been set to $\E = 3$.  
The other parameters are discussed in Sec.~\ref{sec:fem_dynamic}.
Due to P\'eclet number being $O(\epsilon^{-2})$ in size,
film drying is highly non-uniform.  Drying first occurs
at the contact line, where the film is the thinnest, and
then proceeds inwards.   Due to
the high rate of evaporation, fluid that is removed
from the pore space near the free surface cannot be
replenished by fluid in the bulk; hence, vertical
gradients in the composition arise.  Despite
drying initially being confined to the contact line,
the plate undergoes an appreciable deflection for small
times.  The deflection then monotonically increases as the
film approaches its homogeneous steady state.

\subsection{Comparison of the steady states}
\label{sec:fem_static}

Calculating the steady states of the film-plate system
enables the accuracy of the
asymptotic solutions of the mechanical problem to be established.
In particular, the steady states satisfy a purely mechanical problem
and can be obtained without solving the fluid-transport problem. 
Thus, in this
section, it is assumed that the film has dried
to a homogeneous steady state with
uniform contraction ratio $J < 1$. 
In this case, the poroelastic model for the film
developed in Sec.~\ref{sec:model_film} reduces to the
equilibrium equations of incompressible nonlinear elasticity,
$\nabla \cdot \tens{S} = \vec{0}$
with $\det \tens{F} = J$, where $J$ is now treated as a
prescribed constant. 

When the full problem is posed in two Cartesian dimensions, the
FvK equations \eqref{nd:p:fvk} become one-dimensional.  
Moreover, when $J$ is uniform in the vertical direction
so that the expressions for the tractions in \eqref{ad:c1:tau}
apply, then the mean in-plane plate stress can be obtained
from \eqref{nd:p:lin_mom_xx} as $\bar{\mathsf{S}}_{11} = -H_f \Sxxx(J)$,
where $\Sxxx(J)$ is given by \eqref{eqn:scalar_Sxx}.
Inserting this result along with the tractions \eqref{ad:c1:tau}
into \eqref{nd:p:w}
leads to cancellations that allow two integrations to
be performed.  After imposing the boundary conditions at
the free edge $X = 1$, the equation for the vertical plate 
displacement \eqref{nd:p:w} simplifies to
\begin{align}
-\B \tdd{w}{X} = -\frac{1}{2} H_f \Sxxx(J) - \frac{1}{2}\epsilon \delta^{-1} J H_f^2 \Sxxx(J).
\label{nd:p:w_ode}
\end{align} 
The boundary conditions for \eqref{nd:p:w_ode} are 
 $w(0) = 0$ and $\p_X w(0) = 0$.
The first term on the right-hand side of \eqref{nd:p:w_ode} captures
the influence of the in-plane traction, whereas the second term
captures the vertical traction. 
When $J$ is a constant, \eqref{nd:p:w_ode} can be solved to find
\begin{align}
      w(X) = \frac{1 }{30 \B}\, \Sxxx(J) X^3 \left[10 - 5X + 4 \epsilon \delta^{-1} J (2X^2 - 6X + 5) X \right].
\label{nd:p:w_sol}
\end{align}
Equation \eqref{nd:p:w_ode} and its solution \eqref{nd:p:w_sol} apply to all
of the mechanical regimes considered in Sec.~\ref{sec:asy_mech}; however, for asymptotic consistency, the terms that are proportional to $\epsilon\delta^{-1}$ should be dropped when $\epsilon \ll \delta$.  

\begin{figure}
  \centering
  \subfigure[]{\includegraphics[width = 0.49\textwidth]{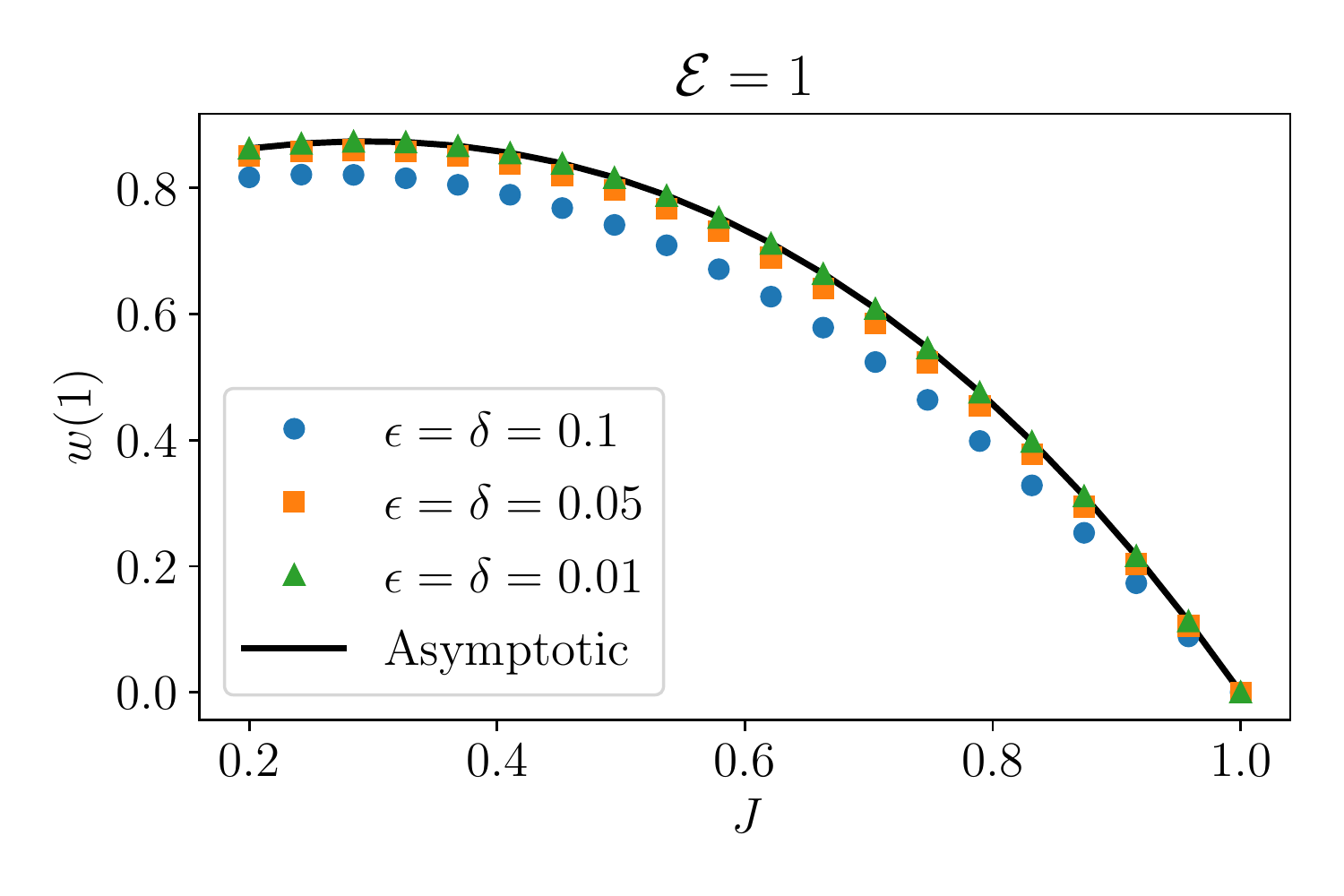}}
  \subfigure[]{\includegraphics[width = 0.49\textwidth]{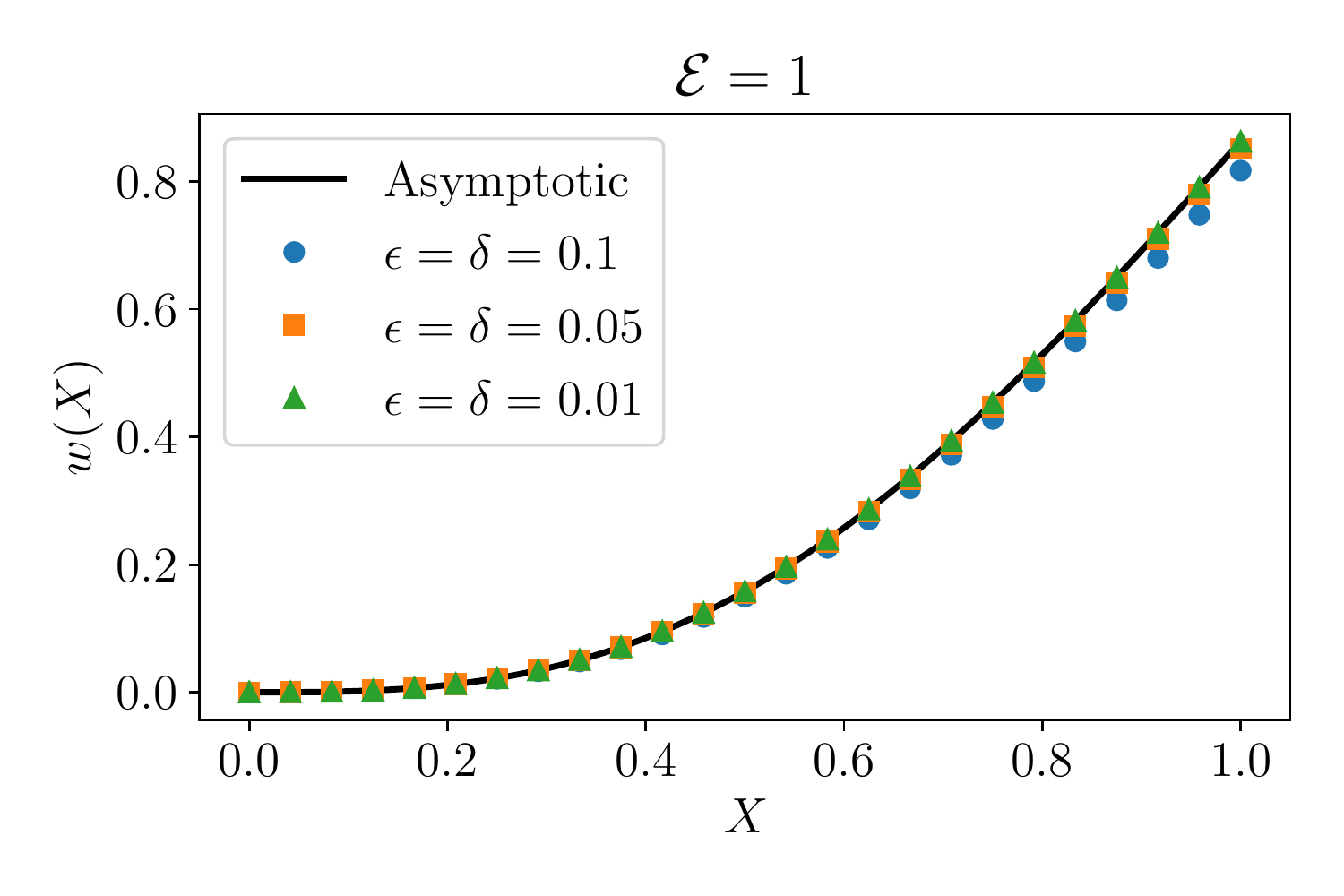}}
  \caption{Steady-state solutions of the
    vertical plate displacement when the
    film and plate have similar thicknesses, $\delta = O(\epsilon)$ with
    $\E = O(1)$. Asymptotic solutions are shown as lines and obtained
    from \eqref{nd:p:w_sol}; finite element solutions are
    shown as circles.
    (a) The deflection at the end of the plate as a function of the
    imposed (uniform) contraction ratio $J$ in the film.  (b) The 
    vertical displacement of the plate as a function of space when $J = 0.2$.}
  \label{fig:static_compare}
\end{figure}

We first compare asymptotic and FE solutions when the film and plate
have similar thicknesses.  In particular, we set $\delta = \epsilon$
and $\E = 1$.  A comparison of the deflection of the end of the plate,
$\wb(X = 1)$, shows that the asymptotic and FE solutions are in
good agreement across a wide range of $J$ values, 
especially when $\epsilon = \delta \leq 0.05$; see Fig.~\ref{fig:static_compare}~(a).
The deflection is seen to increase as $J$ is decreased from one,
corresponding to greater volumetric contraction of the film.
However, the deflection reaches a maximum at $J \simeq 0.30$,
after which it decreases with further decreases in $J$. 
The non-monotonic behaviour results from a competition between the
in-plane stress in the film $\Sxxx$ and the contribution to
\eqref{nd:p:w_ode} from the vertical traction.  The former
increases as $J$ decreases, whereas the latter decreases. 

Profiles of the vertical plate displacement are found to
compare favourably in Fig.~\ref{fig:static_compare}~(b).  As expected,
the plate deflection monotonically increases as the distance
from the wall increases.

\begin{figure}
  \centering
  \subfigure[]{\includegraphics[width = 0.49\textwidth]{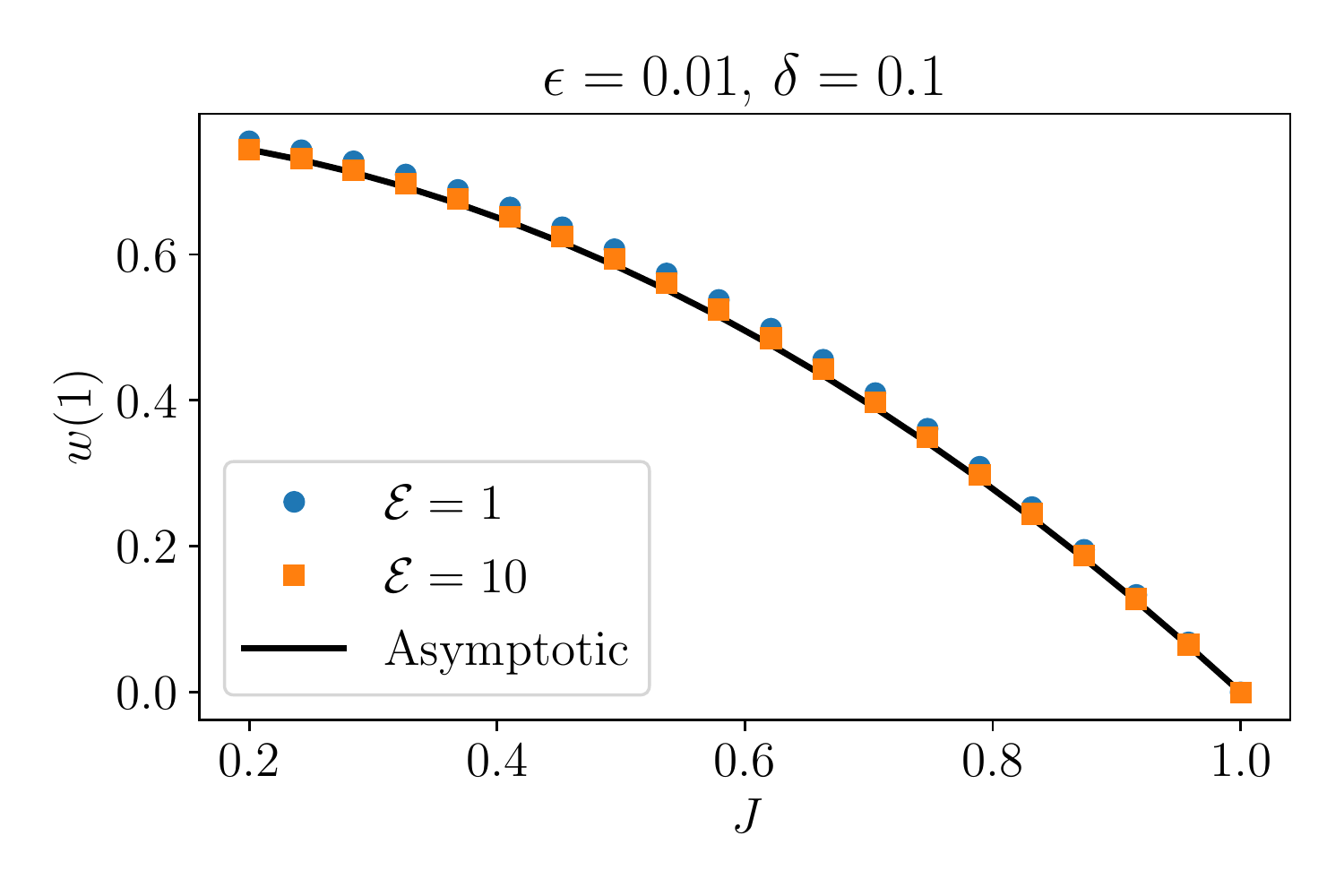}}
  \subfigure[]{\includegraphics[width = 0.49\textwidth]{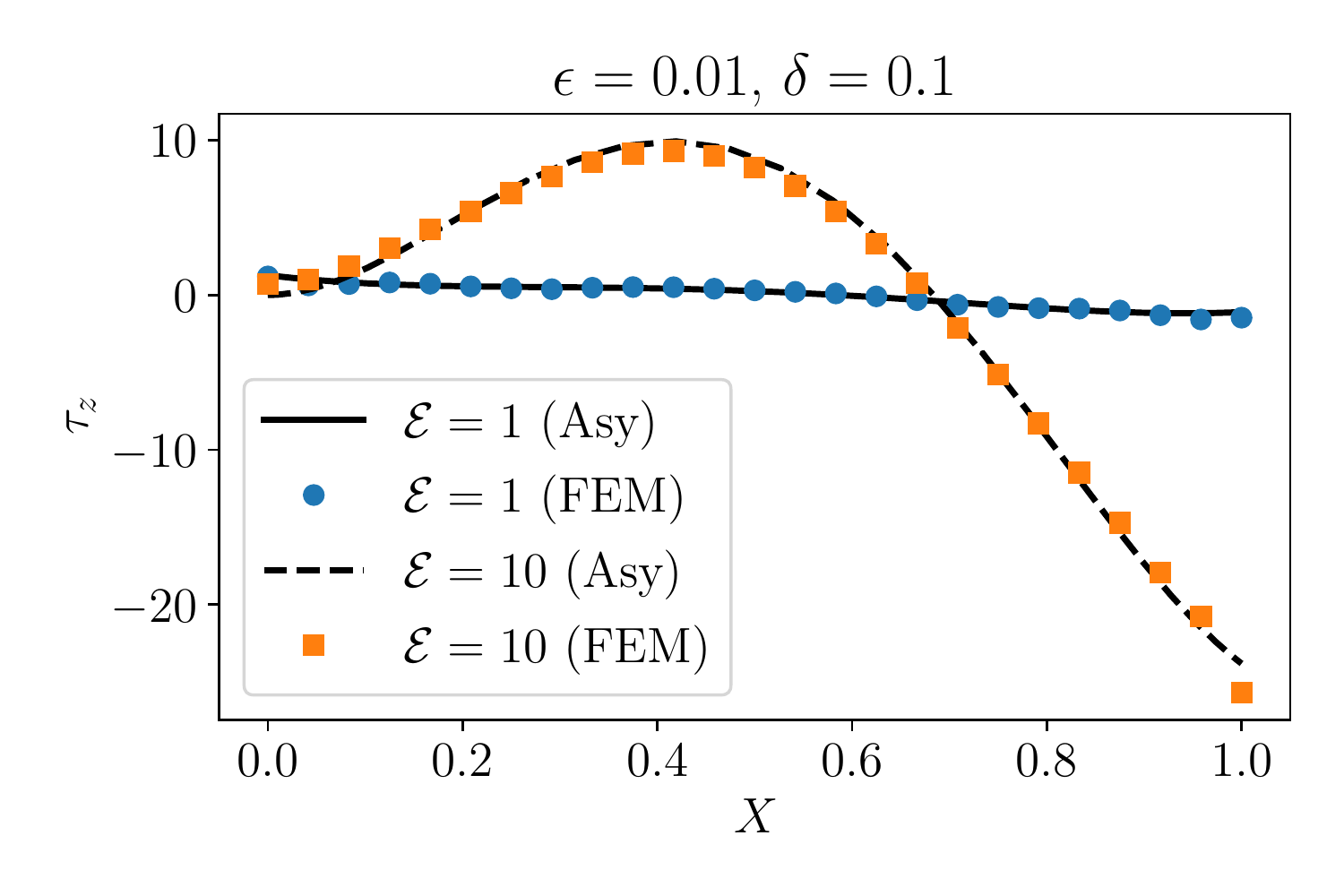}}
  \caption{Steady-state solutions when the
    film is thin relative to the plate, $\epsilon \ll \delta$.
    Asymptotic solutions are shown as lines; finite element solutions are
    shown as circles.
    (a) The deflection at the end of the plate as a function of the imposed
    (uniform) contraction ratio $J$ in the film.  The
    asymptotic solution for $\wb$ is obtained from \eqref{nd:p:w_sol}
    with $\epsilon \delta^{-1} = 0$. 
    (b) The
    vertical traction $\tau_z$ as a function of space.
    The asymptotic solutions for the traction are given by
    \eqref{ad:c1:tau_z} when $\E = 1$ and \eqref{ad:c1:tau_z_r3} when $\E = 10$.}
  \label{fig:static_compare_thick}
\end{figure}

To validate the asymptotic reduction of the mechanical problem when the
film is thin relative to the plate, we set $\epsilon = 0.01$ and
$\delta = \epsilon^{1/2} = 0.1$.  Solutions are computed when
$\E = 1$ and $\E = \delta \epsilon^{-1} = 10$ across a range of $J$ values.
By computing the deflection at the
end of the plate, we again find excellent agreement between the
asymptotic and FE solutions; see Fig.~\ref{fig:static_compare_thick}~(a).
The strong 
agreement when $\E = 10$ is remarkable given that the film undergoes extremely 
large deformations, with the vertical displacement being much greater
than the film thickness.  

A distinguishing feature of the thin-film regime ($\epsilon \ll \delta$)
is that the
asymptotic order of the vertical traction increases when the
non-dimensional stiffness $\E$ increases. To observe this increase,
the vertical traction is computed from the FE and asymptotic solutions
by fixing $\epsilon = 0.01$, $\delta = 0.1$, and $J = 0.2$ while
considering $\E = 1$ and $\E = 10$.
When $J$ is independent of $Z$, the asymptotic solution for the vertical
traction is obtained from \eqref{ad:c1:tau_z} when $\E = 1$ and 
\eqref{ad:c1:tau_z_r3} when $\E = 10$. 
The solutions for the traction are plotted as a function of space in
Fig.~\ref{fig:static_compare_thick}~(b). There is clearly a large difference
in the magnitude of the traction in the two cases, with the traction
remaining $O(1)$ in size when $\E = 1$ and becoming $O(\delta \epsilon^{-1})$
in size  when $\E = 10$.
Again, excellent agreement between the asymptotic and FE solutions is
found.

\subsection{Dynamic simulations}
\label{sec:fem_dynamic}

Time-dependent simulations are used to explore the dynamics of drying
and bending. 
The non-dimensional evaporative flux is written in terms of
a simple phenomenological law given by
$\mathcal{V}(\phif) = \phif - \phif_\infty$. The parameter $\phif_\infty$
can have different interpretations depending on the modelling context.
For instance, $\phif_\infty$ can represent
a constant amount of fluid vapour 
in the surrounding environment. In any case, 
with this model for $\mathcal{V}$, 
the film will dry until its porosity (fluid fraction)
reaches $\phif_\infty$. For simplicity, we set the
scalar permeability of the film, $k(\phif)$,
to be a constant. Hennessy \etal\cite{hennessy2022}
showed that a porosity-dependent permeability can
lead to drying fronts propagating into the bulk of the film
from the contact line. These fronts separate wet and
completely dry solid and hence require the porosity to
become very small. By taking $\phif_\infty$ to be sufficiently
large, the formation of drying fronts will be 
suppressed and the role of a porosity-dependent
permeability will be negligible.

Our time-dependent simulations are based on the work of
Bouchaudy and Salmon~\cite{bouchaudy2019}, who measured
drying-induced stresses in poroelastic discs formed from
nanoparticle suspensions. Gelation was estimated to
occur at a solid fraction of 0.32. Hence, we take
$\phif_0 = 1 - 0.32 = 0.68$. We imagine that $1 - \phif_\infty$
represents the closest packing fraction of the colloidal dispersion.
For a poroelastic
film composed of monodisperse nanoparticles,
$\phif_\infty \simeq 1 - 0.64 = 0.36$. 
The drying that occurs after the closest packing fraction is reached
can lead to pore invasion by air~\cite{lilin2022}, a phenomenon that
our model does not capture. For this set of parameters, the film will
shed half of its volume during drying, which can be seen by
calculating $J = (1 - \phif_0) / (1 - \phif_\infty) = 0.5$.

\subsubsection{Films drying on plates with similar thickness}

The first set of time-dependent simulations captures the dynamics when the
film and plate have similar thicknesses, $\delta = O(\epsilon)$ with
$\E = O(1)$. The evaporation is assumed to be slow, $\Pe = O(1)$,
or moderate, $\Pe = O(\epsilon^{-1})$. 
We set $\delta = \epsilon = 0.01$ with
$\E = 1$, and consider $\Pe = 1$ or $\Pe = 100$. For both of these cases,
the asymptotic solution to the plate displacement can be obtained from
\eqref{nd:p:w_ode}. The asymptotic solutions for the
in-plane film stress, $\Sxxx$, and local contraction
ratio, $J$, can be obtained by numerically solving \eqref{ad:c1:Phi}
when $\Pe = O(1)$ and \eqref{ad:c2:Phi} when $\Pe = O(\epsilon^{-1})$.

\begin{figure}
  \centering
  \subfigure[]{\includegraphics[width = 0.49\textwidth]{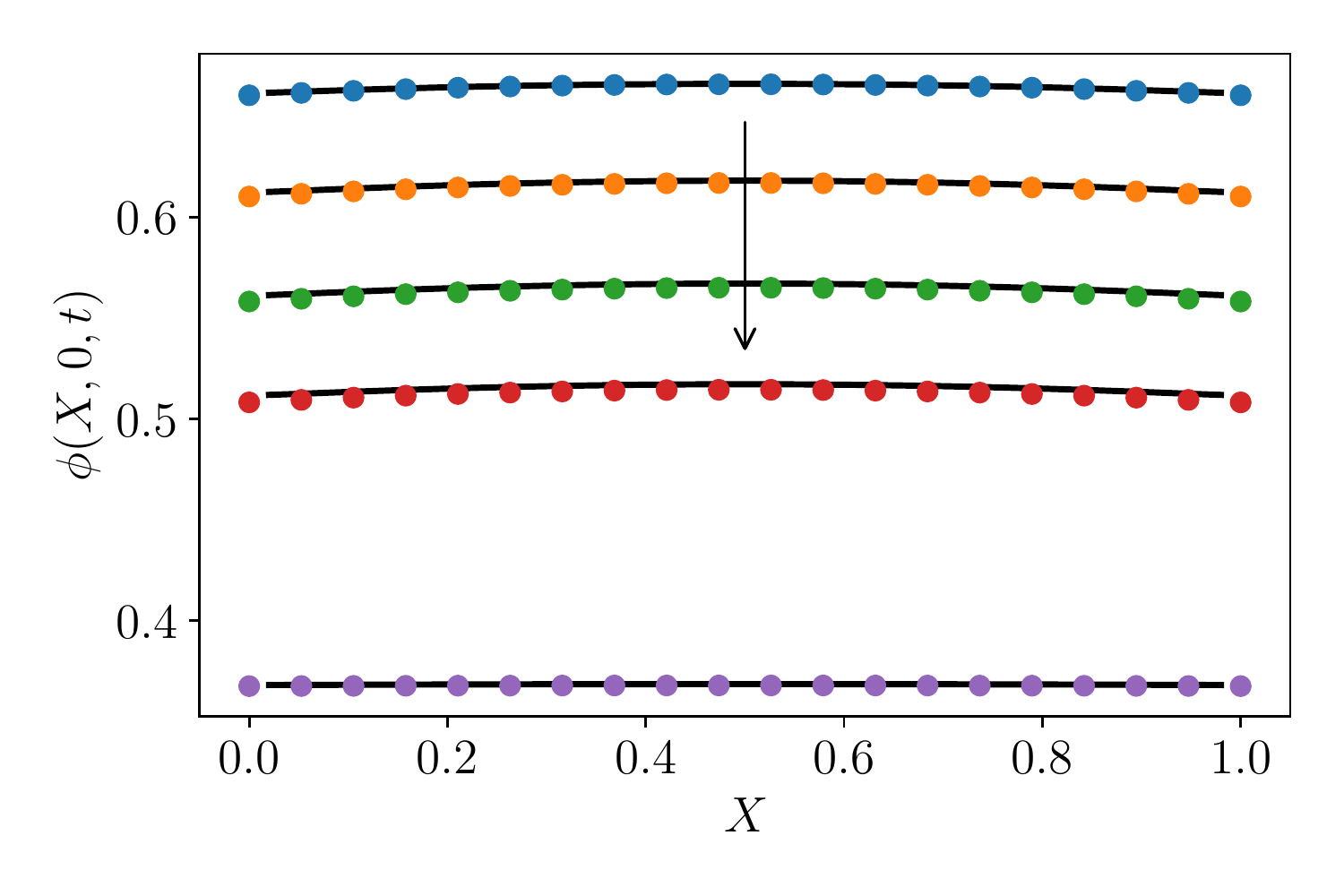}}
  \subfigure[]{\includegraphics[width = 0.49\textwidth]{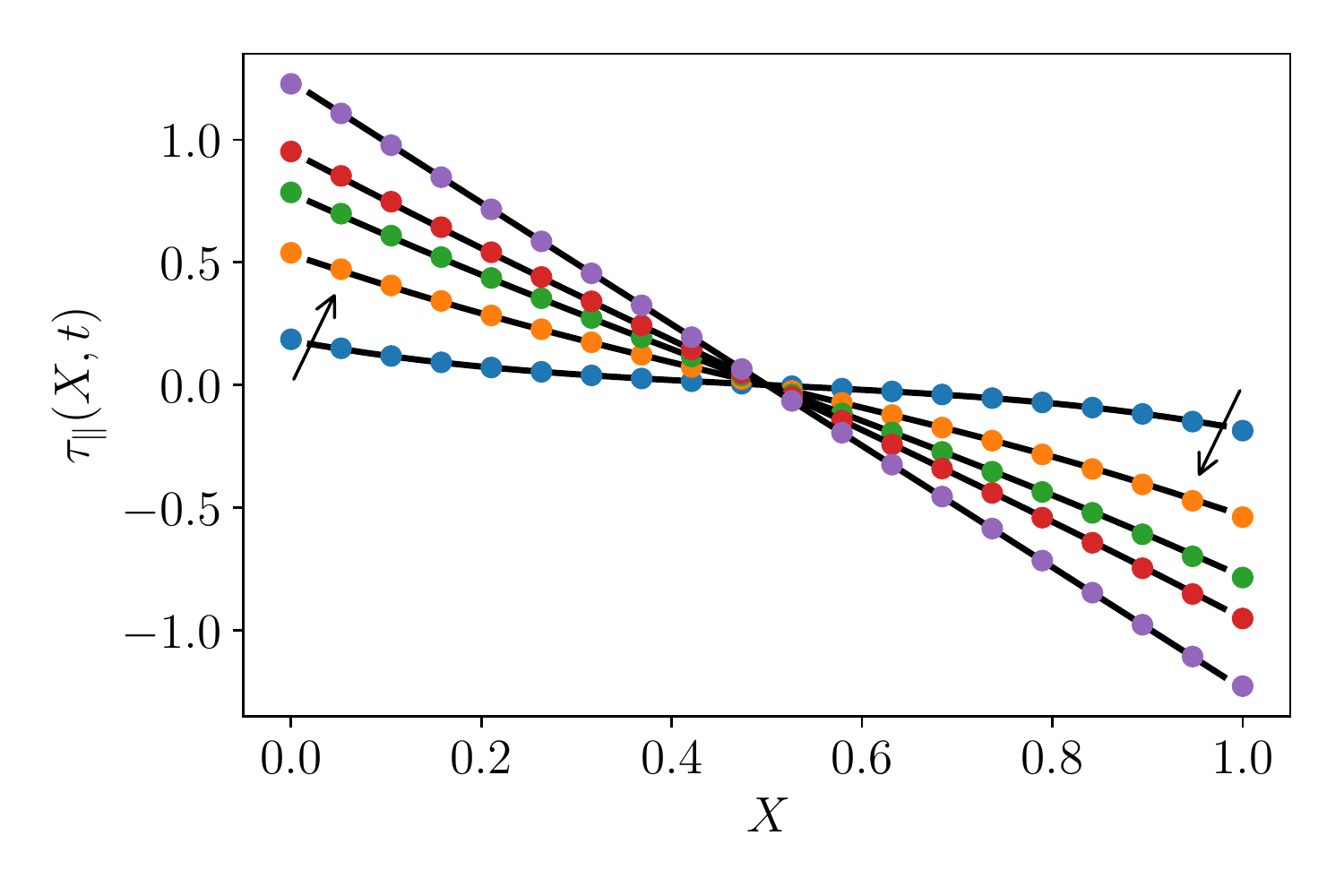}}
  \\
  \subfigure[]{\includegraphics[width = 0.49\textwidth]{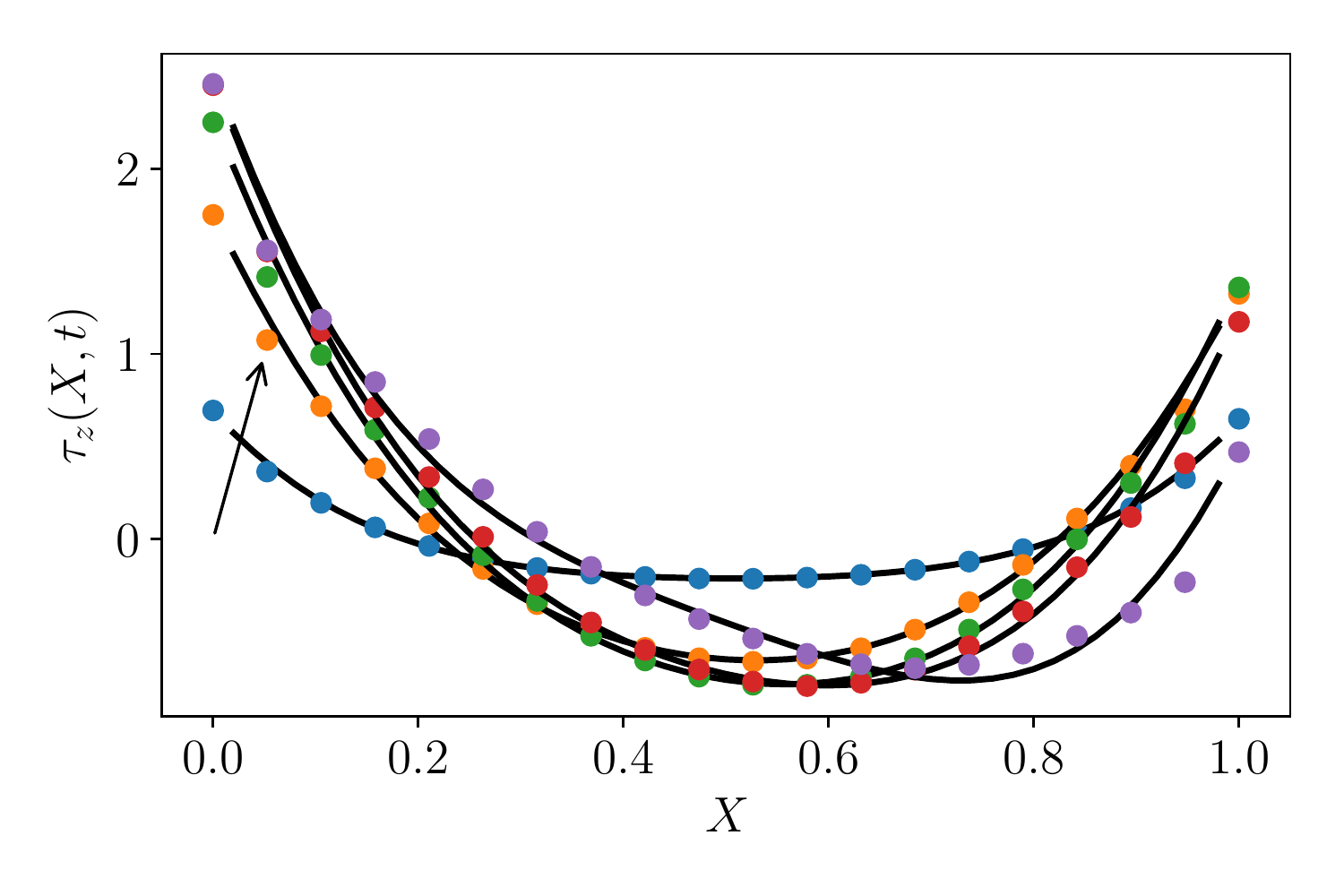}}
  \subfigure[]{\includegraphics[width = 0.49\textwidth]{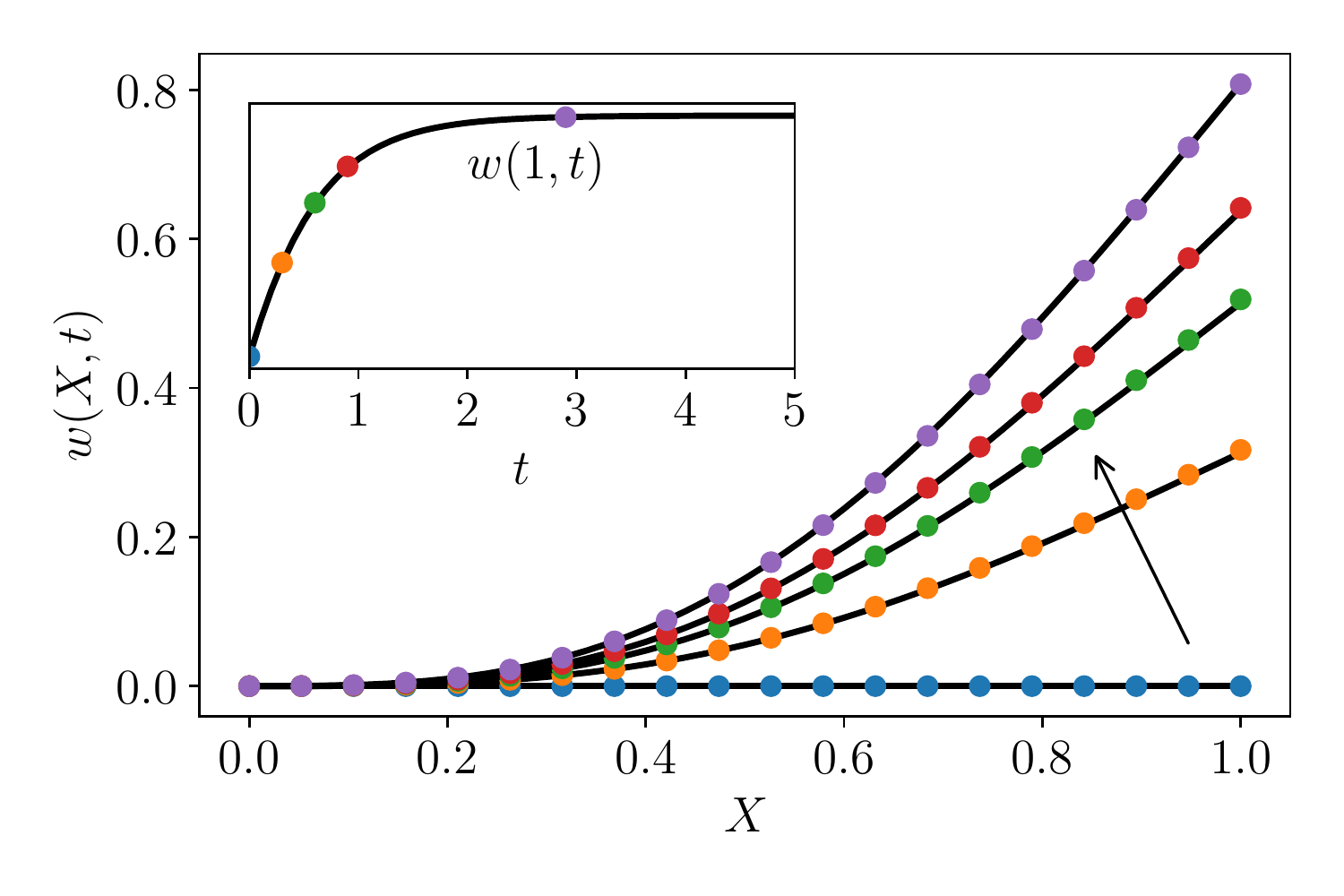}}

  \caption{Drying dynamics for small evaporation rates, $\Pe = O(1)$,
    when the film and plate have similar thicknesses, $\delta = O(\epsilon)$.
    Asymptotic solutions are shown as lines;
    finite element solutions are shown as circles. (a) The Eulerian fluid fraction
    at the bottom of the film. (b) and (c) The in-plane and vertical
    traction. (d) The vertical plate displacement. The inset shows the
    evolution of the deflection of the end of the plate. Solutions are shown
    at times $t = 0.1$, $0.4$, $0.7$, $1$, and $3$. The arrows show the
    direction of increasing time.
    The parameter values are $\epsilon = \delta = 0.01$,
    $\Pe = 1$, and $\E = 1$.}    
  \label{fig:sim_dis_Pe_one}
\end{figure}

When evaporation is slow, $\Pe = 1$, the Eulerian fluid
fraction (porosity) of the film
remains relatively uniform during drying. That is, the in-plane composition
gradients, in addition to the vertical composition gradients, are weak.
The weak in-plane gradients can be seen in Fig.~\ref{fig:sim_dis_Pe_one}~(a),
where the fluid fraction at the bottom of the film ($Z = 0$) is plotted
as a function of the in-plane coordinate $X$ at various times.
Black lines represent numerical solutions of the asymptotically reduced
transport problem defined by \eqref{ad:c1:Phi}; circles
represent the FE solution. The uniform loss of fluid across the film induces
a homogeneous contraction of the film that predominately occurs in the
vertical direction. The in-plane stresses in the film, $\Sxxx$, will also
be approximately homogeneous. Since the initial
film profile is quadratic, the in-plane traction
is expected to vary linearly in space, which can be deduced from
\eqref{ad:c1:tau_x} and is shown in Fig.~\ref{fig:sim_dis_Pe_one}~(b).
The evolution of the
vertical traction is more complicated, as seen from
Fig.~\ref{fig:sim_dis_Pe_one}~(c).
The vertical traction is typically positive near the contact line, implying
the film pulls upwards on the plate. However, near the centre of the plate,
the traction becomes negative, implying the film pushes downwards. 
The vertical displacement of the plate is shown
in Fig.~\ref{fig:sim_dis_Pe_one}~(d); the inset depicts
the time evolution of the deflection at the end of the plate, $w(1,t)$.
The plate deflection monotonically increases in space and time. The deflection
at the end of the beam initially grows linearly with time and then saturates
as the film approaches its steady state. The initial linear increase in the
beam deflection is a result of the evaporation rate being approximately
constant for small times, with an evaporative flux given by
$\mathcal{V} \sim \phif_0 - \phif_\infty$. In all of the panels of
Fig.~\ref{fig:sim_dis_Pe_one}, excellent agreement between the
asymptotic and FE solutions can be seen.

\begin{figure}
  \centering
  \subfigure[]{\includegraphics[width = 0.49\textwidth]{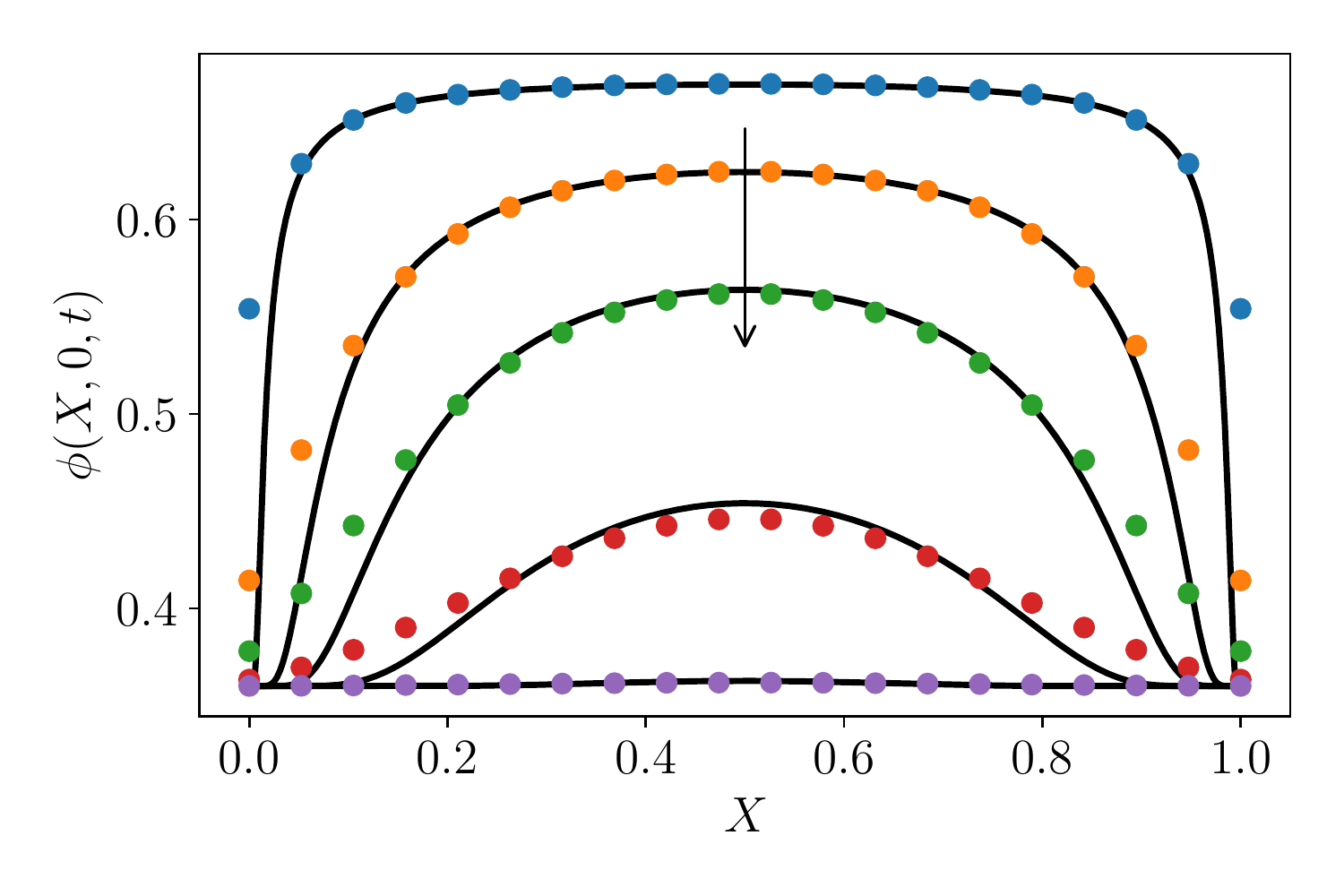}}
  \subfigure[]{\includegraphics[width = 0.49\textwidth]{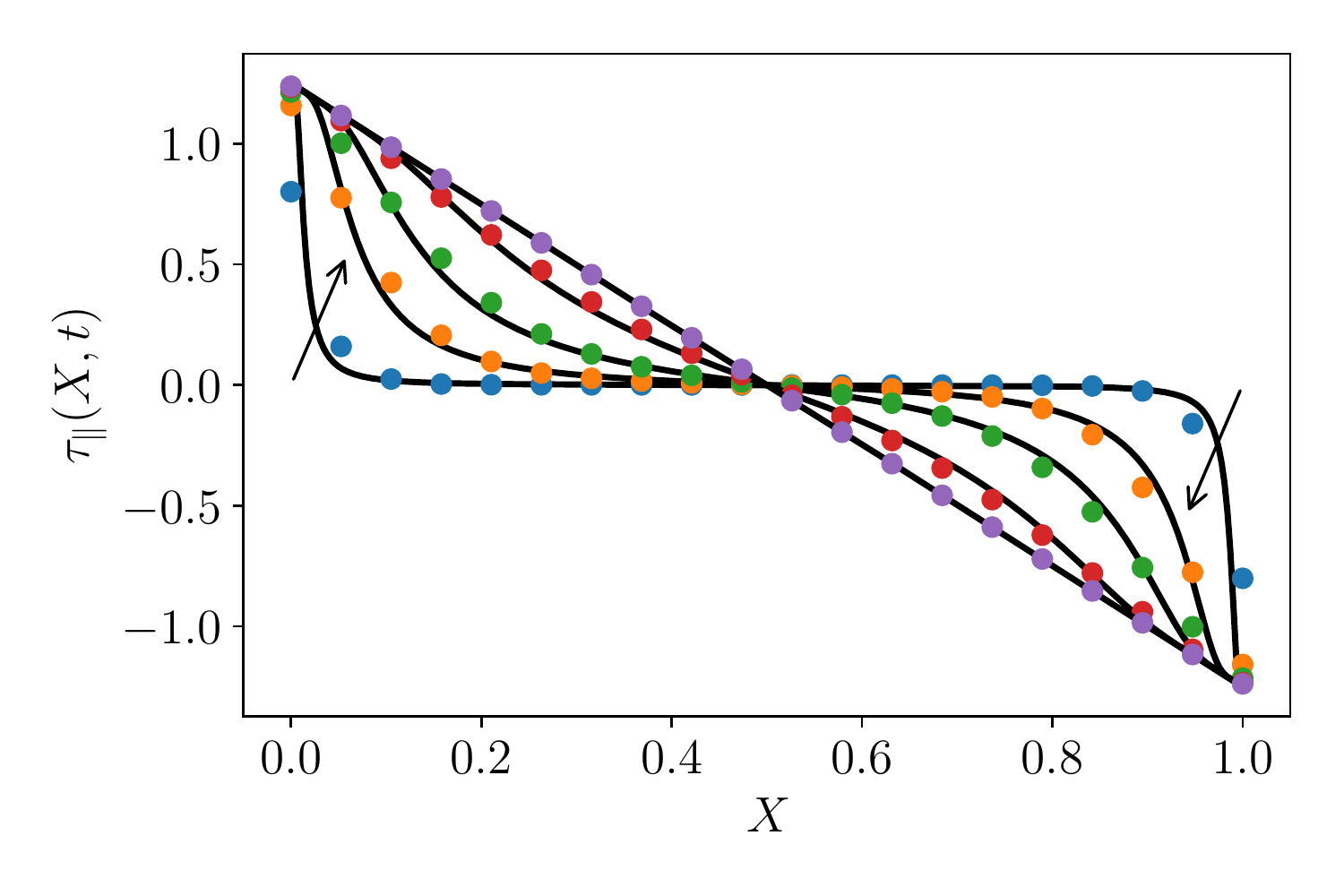}}
  \\
  \subfigure[]{\includegraphics[width = 0.49\textwidth]{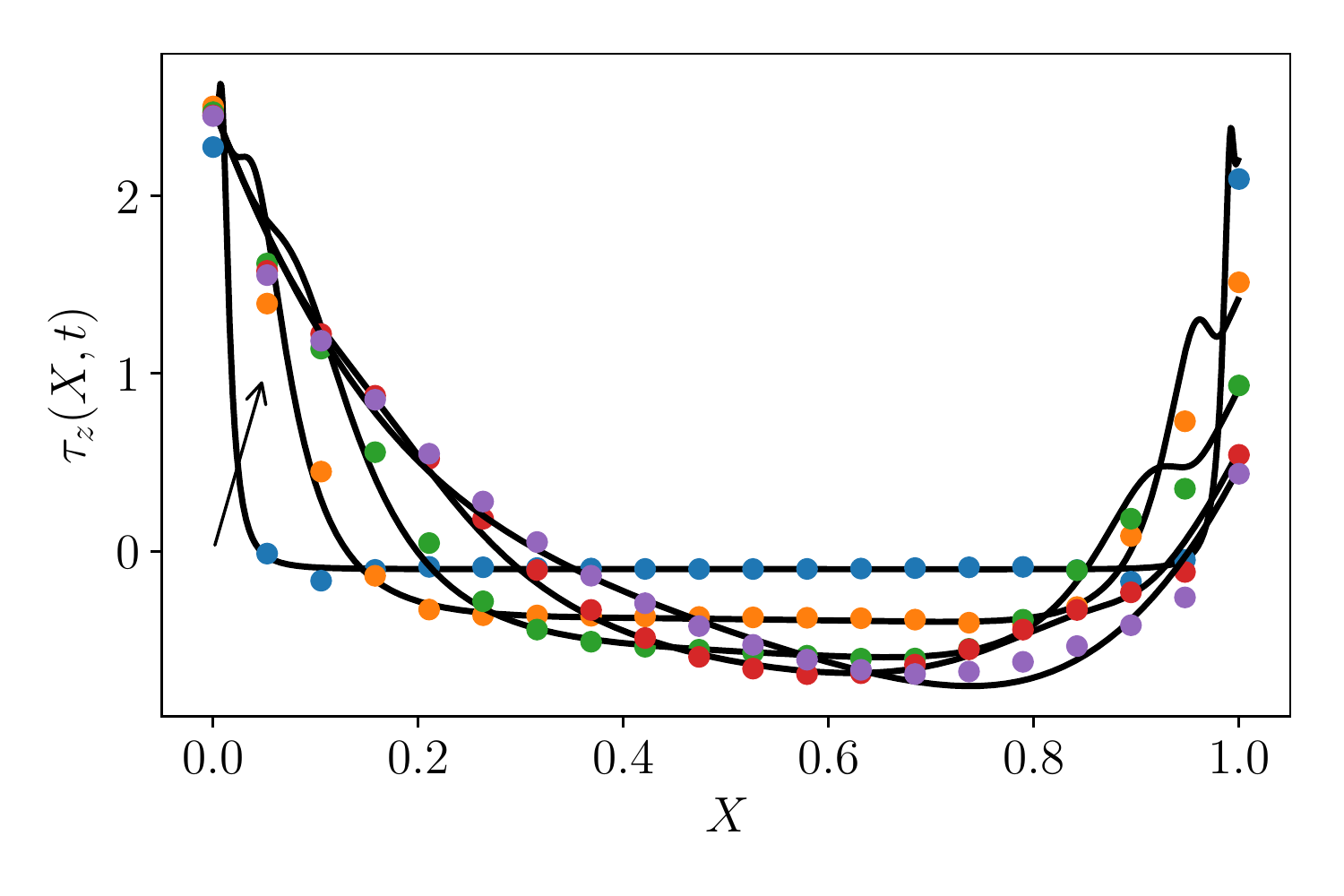}}
  \subfigure[]{\includegraphics[width = 0.49\textwidth]{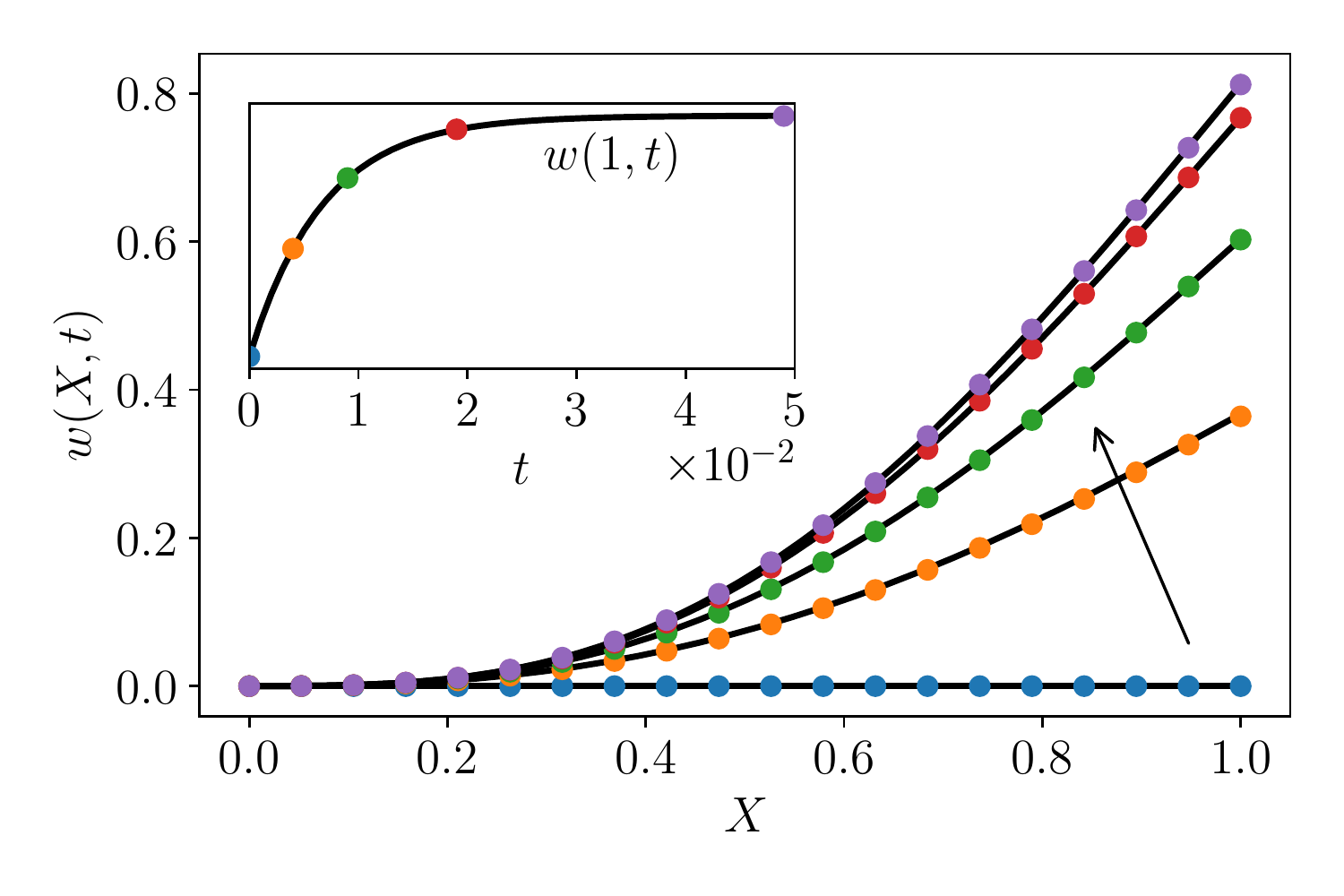}}
  \caption{Drying dynamics for moderate evaporation rates
    , $\Pe = O(\epsilon^{-1})$,
    when the film and plate have similar thicknesses, $\delta = O(\epsilon)$.
    Asymptotic solutions are shown as lines;
    finite element solutions are shown as circles. (a) The Eulerian fluid fraction
    at the bottom of the film. (b) and (c) The in-plane and vertical
    traction. (d) The vertical plate displacement. The inset shows the
    evolution of the deflection of the end of the plate. Solutions are shown
    at times $t = 0.001$, $0.005$, $0.01$, $0.02$, and $0.05$.
    The arrows show the
    direction of increasing time.
	The parameter values are $\epsilon = \delta = 0.01$,
    $\Pe = 100$, and $\E = 1$.}    
  \label{fig:sim_dis_Pe_big}
\end{figure}

Increasing the P\'eclet number to $\Pe = 100$, corresponding to moderate
evaporation,
leads to a non-uniform drying process with significant in-plane
gradients; see Fig.~\ref{fig:sim_dis_Pe_big}. The film rapidly dries
near the contact line, as seen in Fig.~\ref{fig:sim_dis_Pe_big}~(a), which
leads to a localisation of both the in-plane and vertical components of
the traction; see Figs~\ref{fig:sim_dis_Pe_big}~(b) and (c), respectively.
Despite the strongly non-uniform tractions, the evolution of the plate
deflection, shown in Fig.~\ref{fig:sim_dis_Pe_big}~(d), is remarkably similar
to the case when $\Pe = O(1)$. A more in-depth comparison of the plate
deflection for different evaporation rates will be provided in
Sec.~\ref{sec:Pe_compare}. The asymptotic solutions (lines) are in good
agreement with the FE simulations (circles). The largest discrepancy
occurs in the fluid fraction near the contact line; see Fig.~\ref{fig:sim_dis_Pe_big}~(a). 
The asymptotically
reduced model \eqref{ad:c2:Phi} does not capture the influence of
in-plane pressure gradients, which provide a mechanism to replenish
the fluid that evaporates at the contact line. 
Hence,
the asymptotically reduced model leads to faster drying compared to the
FE solutions. Despite the asymptotically reduced model not correctly
capturing the dynamics near the contact line, it is still able to provide a
highly accurate prediction of the plate deflection; see Fig.~\ref{fig:sim_dis_Pe_big}~(d).

\subsubsection{Films drying on thick plates}

The final comparison that we make between the asymptotic and FE solutions
considers the case of films that are thin relative to the plate,
$\epsilon \ll \delta$. Moreover, the film is assumed to be soft,
$\E = O(1)$, and evaporation is  fast,
$\Pe = O(\epsilon^{-2})$. In particular, we set 
$\epsilon = 10^{-2}$, $\delta = \epsilon^{1/2} = 10^{-1}$, 
$\E = 1$, and $\Pe = 5\,\epsilon^{-2} = 5\times 10^{4}$.
In this regime, the vertical displacement
of the plate can be obtained by solving
\begin{align}
  -\B \pdd{w}{X} = -\frac{1}{2} H_f \langle\Sxxx(J)\rangle,
  \quad
  \langle\Sxxx(J)\rangle = \frac{1}{H_f} \int_{\hp/2}^{H_f + \hp/2} \Sxxx(J)\,
  \d Z,
\label{nd:p:w_ode_thick}
\end{align}
subject to $w(0,t) = 0$ and $\p_X w(0,t) = 0$,
where $\langle \Sxxx \rangle$ denotes the vertically averaged in-plane
film stress. The contraction ratio $J$ and the in-plane film stress
$\Sxxx$ can be obtained by numerically solving \eqref{ad:c3:Phi}.

\begin{figure}
  \centering
  \subfigure[]{\includegraphics[width = 0.49\textwidth]{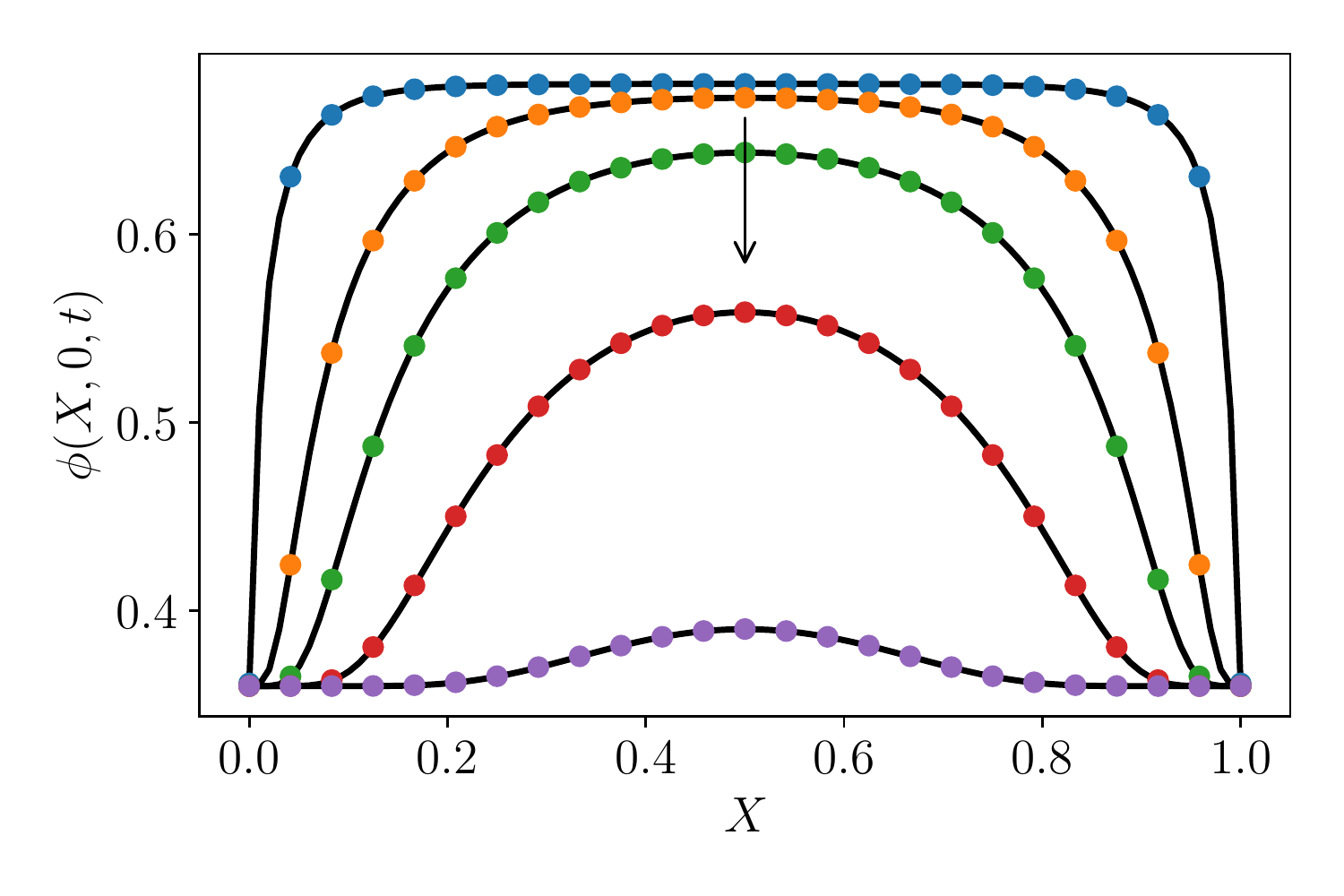}}
  \subfigure[]{\includegraphics[width = 0.49\textwidth]{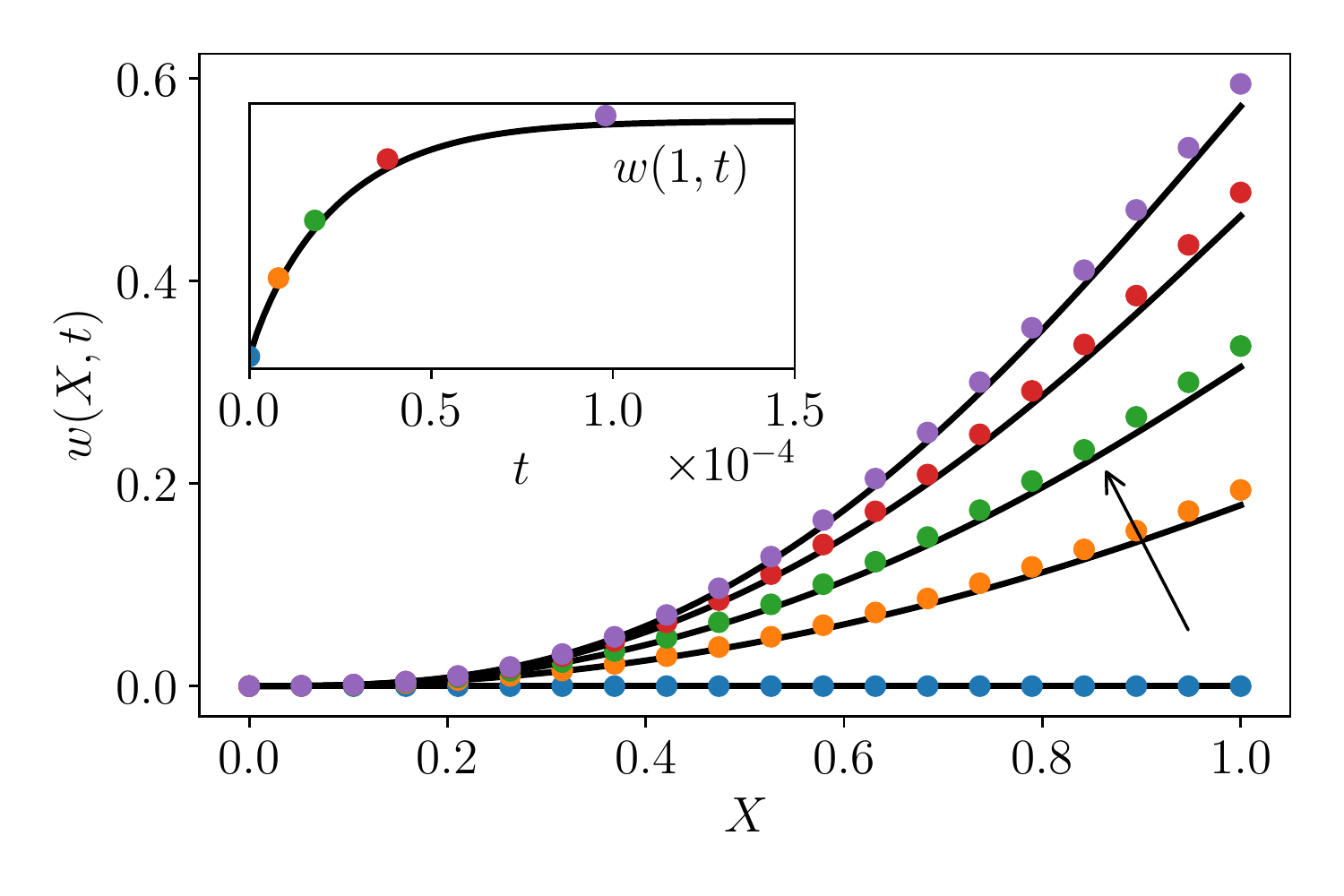}}
  \\
  \subfigure[]{\includegraphics[width = 0.49\textwidth]{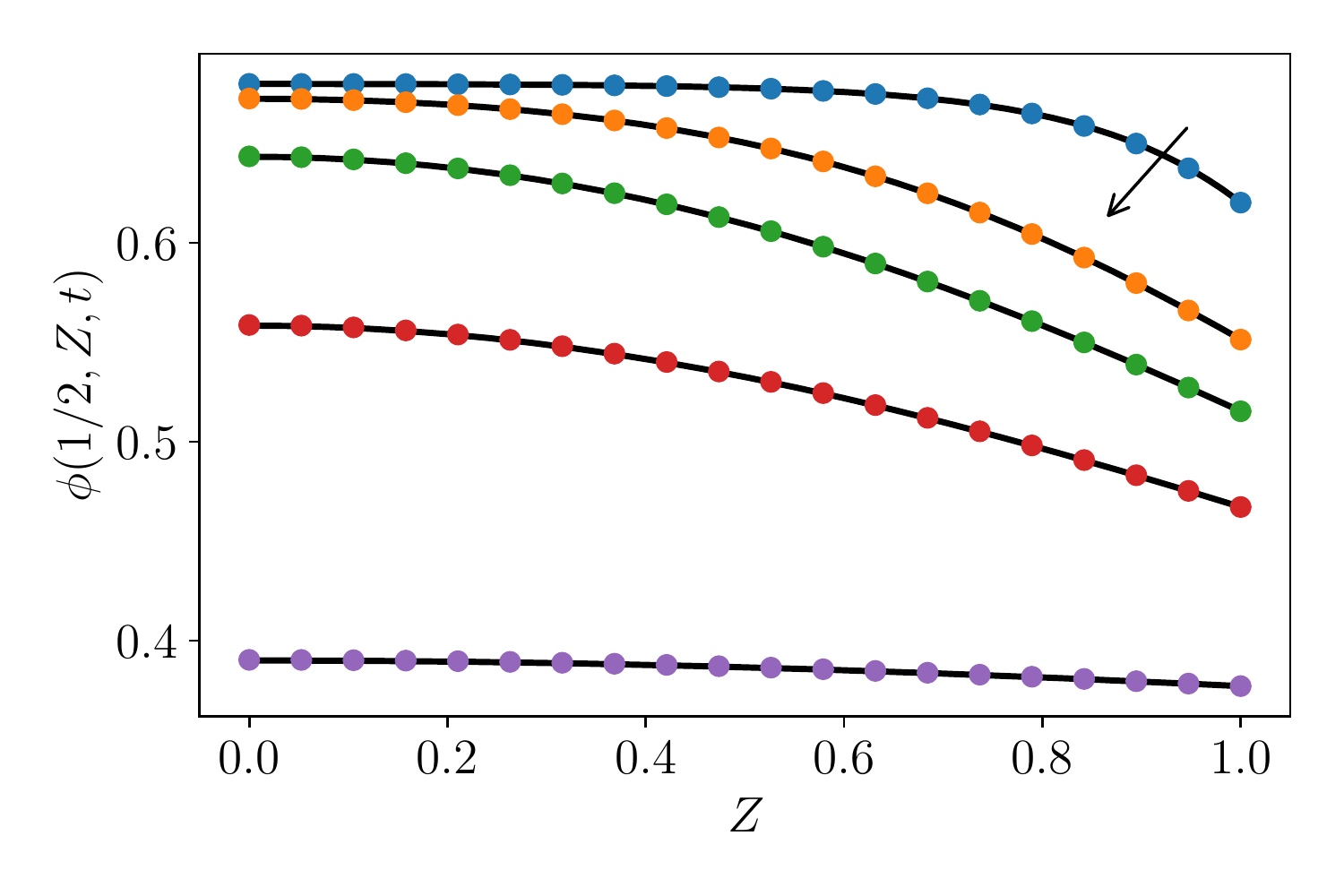}}
  \subfigure[]{\includegraphics[width = 0.49\textwidth]{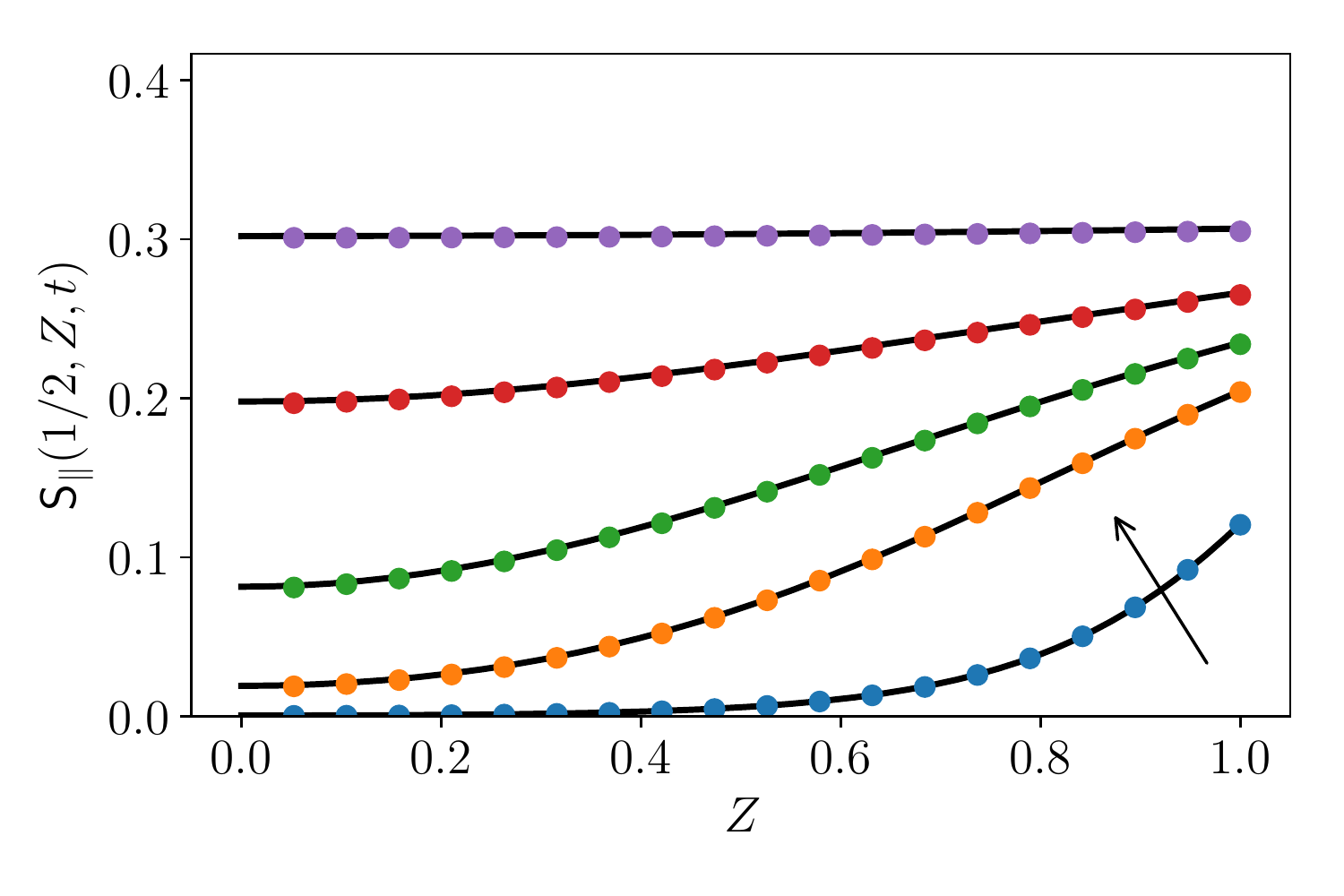}}
  \caption{Drying dynamics for large evaporation rates, $\Pe = O(\epsilon^{-2})$,
    when the film is much thinner than the plate, $\epsilon \ll \delta$. 
    Asymptotic solutions are shown as lines;
    finite element solutions are shown as circles. (a) The Eulerian fluid fraction
    at the bottom of the film. (b) The vertical plate displacement.
    The inset shows the
    evolution of the deflection of the end of the plate.
    (c) and (d) The Eulerian fluid fraction and in-plane film stress
    at the centre of the plate. 
    Solutions are shown
    at times $t = 2 \times 10^{-6}$, $10^{-5}$, $2 \times 10^{-5}$, $4 \times 10^{-5}$, 
    and $10^{-4}$.
    The arrows show the direction of increasing time.
    The parameter values are $\epsilon = 0.01$,
    $\delta = 0.1$,
    $\Pe = 5 \times 10^{4}$, and $\E = 1$.}
  \label{fig:sim_thick}
\end{figure}

The evolution of the fluid fraction at the bottom of the film and
the plate deflection are shown in Figs~\ref{fig:sim_thick}~(a)
and \ref{fig:sim_thick}~(b), respectively. Qualitatively, the
solutions appear very similar to the $\Pe = O(\epsilon^{-1})$ case
shown in Fig.~\ref{fig:sim_dis_Pe_big}. There is a rapid depletion of fluid
near the contact line, which leads to large in-plane gradients. However,
a key difference is that vertical composition gradients occur when
$\Pe = O(\epsilon^{-2})$, which can be seen in Fig.~\ref{fig:sim_thick}~(c).
As expected, the fluid fraction is the greatest at the bottom of the film
and the smallest at the free surface. Consequently, the in-plane film stress
$\Sxxx$ monotonically increases with $Z$, as shown in
Fig.~\ref{fig:sim_thick}~(d).

The asymptotically reduced model for the fluid-transport problem
\eqref{ad:c3:Phi} provides an excellent approximation to the
FE solutions of the nominal fluid fraction and the in-plane film
stress; see Figs.~\ref{fig:sim_thick}~(a), (c), and (d).  In particular,
the fluid fraction near the contact line is well captured.
The reduced model for the plate deflection provides a close
approximation to the FE solution, as shown in Fig.~\ref{fig:sim_thick}~(b).
Since the reduced model does not account for the vertical traction when
$\epsilon \ll \delta$ and $\E = O(1)$, the deflection is slightly smaller
than that obtained using the FE method.

We have also compared the asymptotic and FE solutions when $\E$ is increased to
$\E = \delta \epsilon^{-1} = 10$, keeping the other parameters the same
(not shown).
The solutions are virtually the same as those  in Fig.~\ref{fig:sim_thick}.
The agreement between the fluid fraction and in-plane stresses remains
excellent. However, there is much stronger agreement in the plate
deflection. This is because the
asymptotically reduced model for the plate now captures
terms associated with the in-plane (plate) stress and the vertical traction.
However, these terms
serendipitously cancel out so that \eqref{nd:p:w_ode_thick} still applies.

\subsection{Impact of the evaporation rate on bending}
\label{sec:Pe_compare}

The insets of Fig.~\ref{fig:sim_dis_Pe_one}~(d) and
Fig.~\ref{fig:sim_dis_Pe_big}~(d) show that the deflection of the plate can
undergo similar evolutions
despite the evaporation rate differing by two
orders of magnitude. We now explore the dependence of the deflection
on the evaporation rate in more detail.
The asymptotically reduced models for the
fluid-transport problem are numerically solved when
$\Pe \ll 1$, $\Pe = 1$, $\Pe = \epsilon^{-1}$, and
$\Pe = \epsilon^{-2}$. The  equations for the reduced models
are given by \eqref{ad:c1:Phi_slow}, \eqref{ad:c1:Phi},
\eqref{ad:c2:Phi}, and \eqref{ad:c3:Phi}, respectively.
The plate deflection is then computed by assuming that
the film is thin and solving \eqref{nd:p:w_ode_thick}. In doing so, we take
$\E = 1$.

\begin{figure}
  \centering
  \subfigure[]{\includegraphics[width=0.48\textwidth]{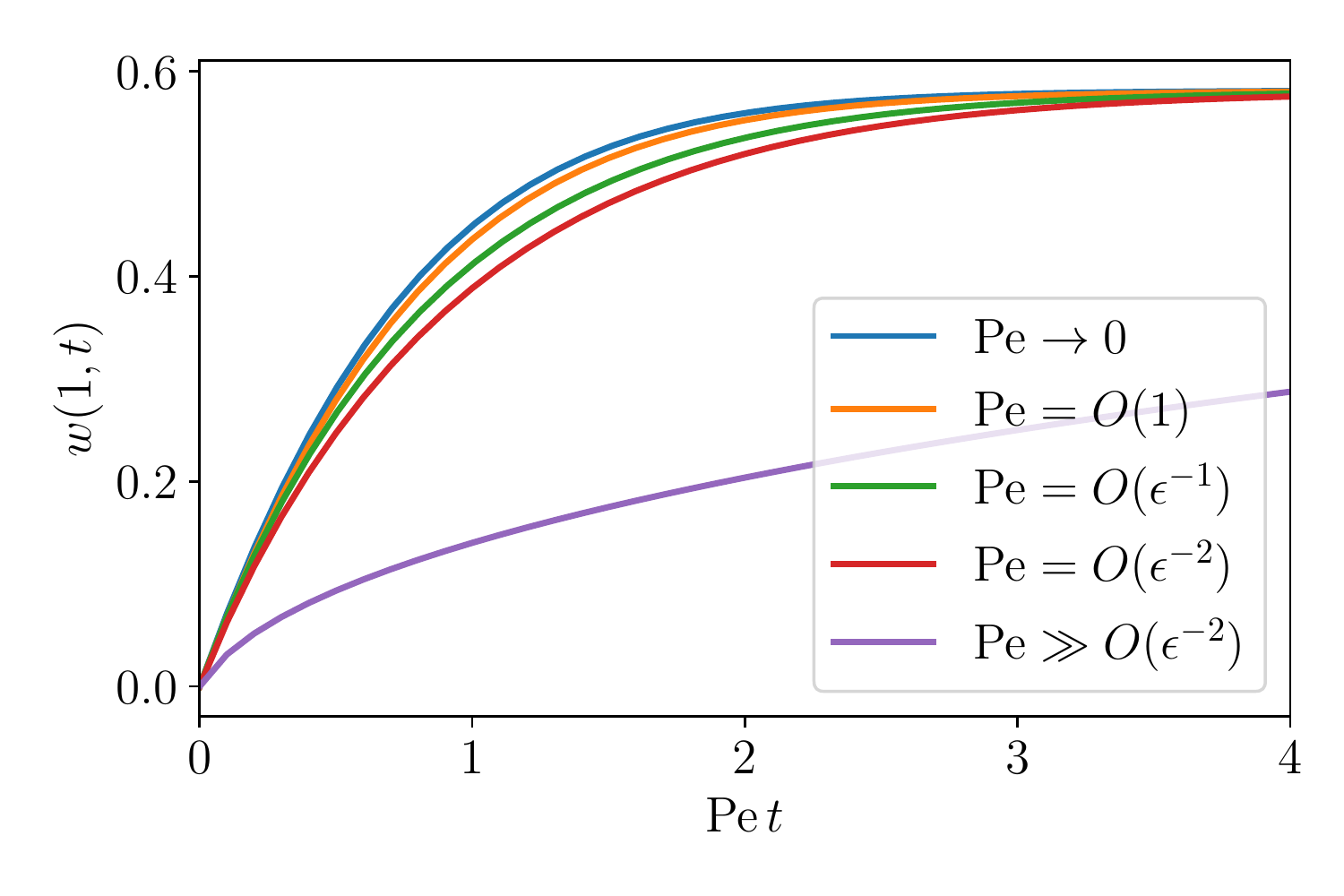}}
  \subfigure[]{\includegraphics[width=0.48\textwidth]{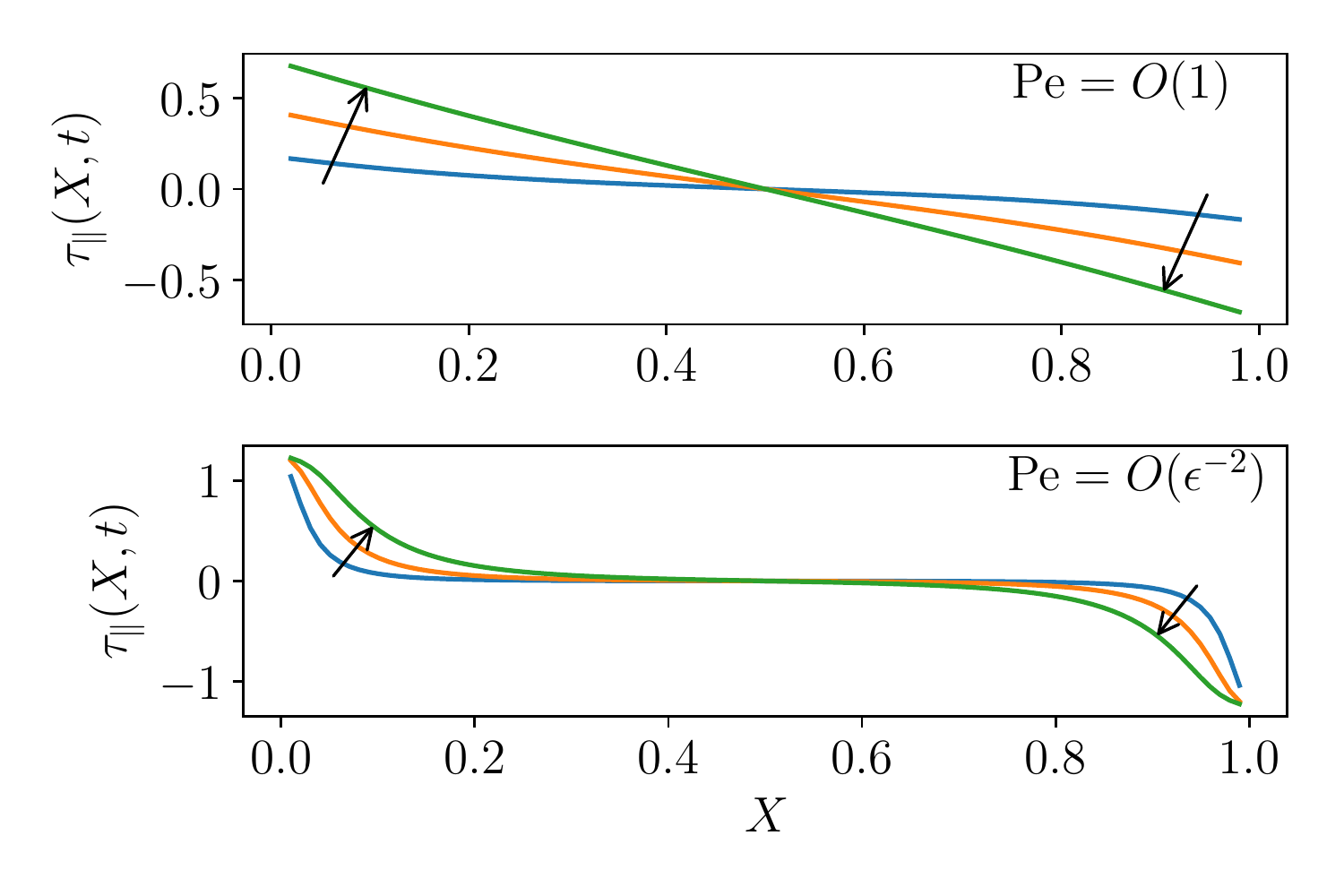}}
  \caption{(a) Asymptotic solutions for plate deflection
    plotted in terms of $\Pe\, t$ for five different drying regimes.
    (b) Asymptotic solutions for the in-plane traction when $\Pe = 1$ (top)
    and $\Pe = \epsilon^{-2}$ (bottom). The solutions are shown when
    $\Pe\, t = 0.1$, $0.3$, and $0.6$.  The arrows show the direction of
    increasing time. See
    text for full details.}
\label{fig:Pe_compare}
\end{figure}

By plotting the deflection at the end of the beam as a function of
$\Pe\, t$, we find that the curves obtained from different values of
$\Pe$ nearly overlap; see Fig.~\ref{fig:Pe_compare}~(a).
Thus, once the time scale of evaporation is
accounted for, the evolution of the beam deflection is relatively
insensitive to the evaporation rate for this range of P\'eclet numbers.
The small differences 
between the curves shown in Fig.~\ref{fig:Pe_compare}~(a)
could be attributed to two physical mechanisms. 
The first is that larger P\'eclet numbers
lead to non-uniform in-plane tractions that become increasingly localised
at the contact line.
The second is that a larger P\'eclet number will lead to a more rapid depletion
of fluid from the free surface, which, in turn,
will reduce the evaporative flux $\mathcal{V}(\phif)$ and
hence the rate of deflection. The first mechanism can be ruled out by noticing
that the curves in Fig.~\ref{fig:Pe_compare}~(a) do overlap for $\Pe\, t < 0.5$.
However, the in-plane tractions during this time are very different, as
shown in Fig.~\ref{fig:Pe_compare}~(b)
when $\Pe = 1$ and $\Pe = \epsilon^{-2}$. Thus, the
differences in the curves can be attributed to the depletion
of fluid near the film surface as $\Pe$ increases.

To further demonstrate the impact of fluid depletion at the free surface on
the rate of deflection,
we consider an extreme case and
numerically solve \eqref{ad:c3:Phi} when $\Pe_1 = 100$, corresponding to
$\Pe \gg O(\epsilon^{-2})$. In this case, a compositional boundary layer
of width $O(\Pe_1^{-1/2})$ develops at the free surface, where the Eulerian
fluid
fraction decreases to approximately $\phif_\infty$. As a result, the evaporation
flux $\mathcal{V}(\phif)$ undergoes a considerable decrease. 
When the plate deflection is obtained from
\eqref{nd:p:w_ode_thick} and plotted in terms of $\Pe\,t$, the corresponding
curve is markedly different from those obtained from smaller P\'eclet numbers;
see Fig.~\ref{fig:Pe_compare}~(a).
When $\Pe = O(\epsilon^{-2})$ or smaller, the deflection
initially grows linearly with time.
However, when $\Pe \gg O(\epsilon^{-2})$, the kinetics
are qualitatively different and the deflection grows
approximately with $t^{1/2}$.

\section{Discussion and conclusion}
\label{sec:conclusion}

Stoney's equation relating the in-plane film stress, $\Sxxx$, to the radius of curvature
of the cantilever beam, $R$, can be written in dimensional form as
\begin{align}
\Sxxx = \frac{E_p H_p^2}{6 H_f^0 R},
\label{eqn:stoney}
\end{align}
where $H_f^0$ denotes the constant film thickness.
To fairly compare against the one-dimensional models derived 
in Sec.~\ref{sec:fem}, we must consider the
thin-film limit $\epsilon \ll \delta$.  By re-dimensionalising
\eqref{nd:p:w_ode_thick} and using $\p^2 w / \p X^2 \sim 1 / R$, we obtain
\begin{align}
\langle \Sxxx \rangle = \frac{\bar{E}_p H_p^2}{6 H_f R},
\label{eqn:us}
\end{align}
where $\bar{E}_p = E_p / (1 - \nub^2)$.  Thus, the equations presented here
are consistent with Stoney's.  The differing factor of $1 - \nub^2$ can be
attributed to Stoney working under the assumption of plane stress, whereas
\eqref{nd:p:w_ode_thick} has been obtained under the assumption of plane
strain.  The agreement between Stoney's result and our own suggests that
Stoney's choice of neutral axis is not incorrect after all, as was suggested by
Chiu~\cite{chiu1990}. That being said, Stoney's derivation is based on the
assumption that bending is driven by the in-plane film traction, whereas the
asymptotic analysis reveals that the vertical traction can be just as relevant
when the film is thin and sufficiently stiff. 
However, when reducing the plate
model to a beam model, the terms associated with the vertical traction
cancel out. Although Stoney did not recognise the importance of the vertical
traction, their formula remains correct due to this serendipitous
cancellation.  A direct comparison of \eqref{eqn:stoney}
and \eqref{eqn:us} shows that the in-plane stress in Stoney's 
formula should be interpreted as the mean stress
across the film thickness.  However, given that significant in-plane
stress gradients can also arise during drying, 
it would be more accurate to interpret 
Stoney's $\Sxxx$ as the mean stress across and along the film.

Tomar \etal\cite{tomar2020} conducted cantilever experiments using
drops of polymer solutions.  By analysing their data using an equation 
that is similar
to \eqref{eqn:stoney}, they found that the film stress increases
like $t$ for thin drops and like $t^{1/2}$ for thicker drops.  
These observations are consistent with the results in 
Sec.~\ref{sec:Pe_compare}, in which the small-time evolution of the
beam deflection is seen to change from $t$ to $t^{1/2}$ as the P\'eclet number,
$\Pe = V_e^0 \hf \mu_f / (\epsilon^2  k_0 E_f)$ increases.  
For a given polymer solution, the aspect ratio $\epsilon$ of the drop
will be fixed and determined by the equilibrium contact angle.
Thus, increasing the drop thickness is equivalent to increasing
the P\'eclet number.

Using asymptotic methods, we have systematically derived a hierarchy
of simplified models of the cantilever experiment.  Our models
extend Stoney-like formulae by accounting for the time dependence
of the drying process and the generation of non-uniform in-plane
and transverse stresses in the film.  Fitting the time-dependent 
solutions to experimental data could lead to new insights into the
solid mechanics of drying films and enable difficult-to-measure
quantities, such as the Young's modulus of the film, to be inferred.  

Due to the complex rheology of drying films, a number of extensions
to this work are possible.  For instance, the Young's modulus of
a colloidal mixture can increase by several orders of
magnitude during drying~\cite{bouchaudy2019}.  
As a result, the system will pass through many of the asymptotic
regimes that we have identified.  Matched asymptotic expansions
could be used to obtain a reduced drying model that spans
all of regimes and captures the evolution of the material 
properties of the film.  The asymptotic analysis could also be
adapted to films with viscoelastic or elasto-viscoplastic
rheologies and applied to other contexts in which the film undergoes
thermal expansion, growth, or swelling.  Finally, the
asymptotic framework that we have laid out could also be applied
to drying-induced delamination of thin films.

\section*{Acknowledgements}

We thank Ludovic Pauchard for many insightful discussions about
drying colloidal films and the cantilever experiment.

\begin{appendix}

\section{Derivation of the plate model}
\label{app:plate_model}

The purpose of this appendix is to derive (i) the modified FvK equations
for a thin plate subject to a non-uniform in-plane traction on its upper
surface and (ii) the corresponding
boundary conditions at the free edges of the plate.
We consider
rectangular plates of length $L$, width $W = O(L)$, and height $H_p$.
In the reference state, the domain of the plate can be written in terms
of Lagrangian coordinates as
$0 \leq X_1 \leq L$, $-W/2 \leq X_2 \leq W/2$, and
$-H_p/2 \leq Z \leq H_p/2$. The plate is assumed to be clamped at the $X_1 = 0$
boundary and free at the $X_1 = L$ and $X_2 = \pm W / 2$ boundaries.

\subsection{Derivation of the modified FvK equations}
\label{app:fvk}

The derivation of the modified FvK equations
begins by considering the equations of nonlinear elasticity
in the reference configuration, which we summarise here. 
Conservation of linear momentum is given by
\begin{align}
	\N \cdot \tens{S} &= \vec{0}, \label{fvk:lin}
\end{align}
where $\tens{S}$ is the first Piola--Kirchhoff (PK1) stress tensor.
For convenience, we define the in-plane and shear components of the
PK1 stress tensor as 
$\Sxx = {\sf S}_{\alpha \beta} \vec{e}_\alpha \otimes \vec{e}_\beta$
and $\Sxz = \Saz \vec{e}_\alpha$, respectively. 
Conservation of angular momentum implies that the
Cauchy stress tensor $\tens{T} = J^{-1} \tens{S} \tens{F}^T$
is symmetric, which leads to the relation
\begin{align}
  \tens{S} \tens{F}^T &= \tens{F}\tens{S}^T. \label{fvk:ang}
\end{align}
The deformation gradient tensor
is related to the
displacement $\vec{u}(\vec{X},t) = \vec{x}(\vec{X},t) - \vec{X}$ by
\begin{align}
  \tens{F} = \tens{I} + \N \vec{u}.
  \label{fvk:F}
\end{align}
The mechanical response of the plate is described using the
Saint Venant--Kirchhoff constitutive relation given by
\begin{align}
  \tens{S} = \tens{F} \tens{P}, \qquad
  \tens{P} = \frac{\nub E_p}{(1+\nub)(1-2\nub)}\tr( \tens{E}) \tens{I} + \frac{E_p}{1+\nub}\tens{E},
  \label{fvk:SFP}
\end{align}
where $\nub$ and $E_p$ are the Poisson's ratio and Young's modulus
of the plate, respectively;
$\tens{P}$ is the second Piola--Kirchhoff
stress tensor; and $\tens{E} = (1/2) (\tens{F}^T \tens{F} - \tens{I})$ is
the strain tensor.

We assume that the upper surface of the plate, located at $Z = H_p / 2$, 
experiences a traction given by $\TT = \TTx + \TTz \ez$. The
bottom surface of the plate, located at $Z = -H_p / 2$
is assumed to be stress free. Therefore,
the following boundary conditions are imposed:
\subeq{
  \label{fvk:bc:surf}
\begin{alignat}{2}
  \tens{S}\cdot \vec{e}_z &= \TT, &\quad Z &= H_p / 2; \\
  \tens{S}\cdot \vec{e}_z &= 0, &\quad Z &= -H_p / 2.
\end{alignat}}
The boundary conditions at the edges of the plate will be discussed in
Appendix~\ref{app:fvk_bc}; they are not required in the derivation
of the bulk equations.

The equations are non-dimensionalised following Howell \etal\cite{howell2009}.
We let $\XX \sim L$, $Z \sim H_p$, and define $\delta = H_p / L \ll 1$.
The displacements, components of the (PK1) stress tensor, and
traction vector are scaled according to
$\ubx \sim \delta H_p$, $\ubz \sim H_p$, 
$\Sxx \sim \delta^2 E_p$,
$\Sxz \sim \delta^3 E_p$, $\Sza \sim \delta^3 E_p$,
$\Szz \sim \delta^4 E_p$,
$\TTx \sim \delta^3 E_p$ and $\TTz \sim \delta^4 E_p$. Although this scaling
differs from that of Sec.~\ref{sec:plate_scaling},
it can be derived in the same way,
and it simplifies the notation used in the subsequent calculations.

The displacements and (PK1) stress tensor are expanded in powers of $\delta$ as
\subeq{
\begin{align}
  \ubx &= \ubx^{(0)}(\XX, Z, t) + \delta \ubx^{(1)}(\XX, Z, t) + O(\delta^2), \\
  \ubz &= \wb(\XX,t) + \delta \ubz^{(1)}(\XX,t) + \delta^2 \ubz^{(2)}(\XX, Z, t) + O(\delta^3),
         \\
  \tens{S} &= \tens{S}^{(0)}(\XX,Z,t) + \delta \tens{S}^{(1)}(\XX, Z, t) +
             \delta^2 \tens{S}^{(2)}(\XX, Z, t)
             + O(\delta^3). \label{fvk:S_exp}
\end{align}}
To streamline the derivation and reduce the algebra,
the first two contributions to the vertical
displacement $\ubz$ are taken to be independent of the vertical coordinate $Z$.
Using the steps outlined below, it is straightforward to retain the $Z$ dependence
and then show that $\pdf{\wb}{Z} = \pdf{\ubz^{(1)}}{Z} = 0$.
By taking into consideration how the components of the stress tensor have been
scaled, we can deduce that
\subeq{
\begin{align}
  \tens{S}^{(0)} &= \Sxx^{(0)}, \\
  \tens{S}^{(1)} &= \Sxx^{(1)} + \Sxz^{(1)}\otimes\ez + \Sza^{(1)} \ez \otimes \vec{e}_\alpha,
                   \label{fvk:S1}
  \\
  \tens{S}^{(2)} &= \Sxx^{(2)} + \Sxz^{(2)}\otimes\ez + \Sza^{(2)} \ez \otimes \vec{e}_\alpha + \Szz^{(2)} \ez \otimes \ez.
                   \label{fvk:S2}
\end{align}}
The boundary conditions on the upper and lower surfaces of the plate
\eqref{fvk:bc:surf} become 
\subeq{\label{fvk:bc:nd:surf}
\begin{alignat}{2}
  \Sxz^{(1)} &= \TTx, &\quad Z &= 1 / 2; \label{fvk:bc:nd:Sxz_top} \\
  \Szz^{(2)} &= \TTz, &\quad Z &= 1 / 2; \label{fvk:bc:nd:Szz_top}\\
  \Sxz^{(1)} &= 0, &\quad Z &= -1 / 2; \label{fvk:bc:nd:Sxz_bot} \\
  \Szz^{(2)} &= 0, &\quad Z &= -1 / 2. \label{fvk:bc:nd:Szz_bot} 
\end{alignat}}
The deformation gradient tensor \eqref{fvk:F} can be written as
$\tens{F} = \tens{F}^{(0)} + \delta \tens{F}^{(1)} + \delta^2 \tens{F}^{(2)} + O(\delta^3)$, where
\subeq{
  \begin{align}
    \tens{F}^{(0)} &= \tens{I},
                     \label{fvk:F0}
    \\
    \tens{F}^{(1)} &= \pd{\ubx^{(0)}}{Z} \otimes \ez + \ez \otimes \Nx \wb,
                     \label{fvk:F1}
    \\
    \tens{F}^{(2)} &= \Nx \ubx^{(0)} + \pd{\ubx^{(1)}}{Z} \otimes \ez + \ez \otimes \Nx \ubz^{(1)} + \pd{\ubz^{(2)}}{Z} \ez \otimes \ez.
                     \label{fvk:F2}
  \end{align}}
Using these expansions in \eqref{fvk:ang} gives
\begin{align}
  \Sxx^{(0)} + \delta \tens{S}^{(1)} + \delta \Sxx^{(0)} (\tens{F}^{(1)})^T
  + O(\delta^2) = 
  (\Sxx^{(0)})^T + \delta (\tens{S}^{(1)})^T + \delta \tens{F}^{(1)} (\Sxx^{(0)})^T
  + O(\delta^2),
\end{align}
which has been simplified using \eqref{fvk:F0}.
The $O(1)$ contributions imply that
the in-plane components of the stress tensor are symmetric, i.e.
$\Sxx^{(0)} = (\Sxx^{(0)})^T$. The $O(\delta)$ contributions lead to
the equation
\begin{align}
(\tens{S}^{(1)})^T - \tens{S}^{(1)} = \Sxx^{(0)} (\tens{F}^{(1)})^T
  - \tens{F}^{(1)}\Sxx^{(0)}.
\end{align}
By right-multiplying with the basis vector $\ez$ and using
\eqref{fvk:S1} and \eqref{fvk:F1}, we find that
\begin{align}
  \Sza^{(1)} \vec{e}_\alpha - \Sxz^{(1)} = \Sxx^{(0)} \Nx \wb.
  \label{fvk:Sza}
\end{align}

We now examine the leading-order contributions to the balance of
linear momentum \eqref{fvk:lin}. After making use of \eqref{fvk:Sza}, these
can be written as
\subeq{\label{fvk:mom_comp}
  \begin{align}
    \Nx \cdot \Sxx^{(0)} + \pd{\Sxz^{(1)}}{Z} = 0, \label{fvk:mom_x} \\
    \Nx \cdot \Sxz^{(1)} + \Nx \cdot\left(\Sxx^{(0)} \Nx \wb\right)
    + \pd{\Szz^{(2)}}{Z} = 0.
    \label{fvk:mom_z}
  \end{align}}
By integrating
both equations in \eqref{fvk:mom_comp} from $Z = -1/2$ to $Z = 1/2$ and using
the boundary conditions \eqref{fvk:bc:nd:surf}, we obtain
\subeq{
\begin{align}
  \Nx \cdot \Sxxbar^{(0)} + \TTx = 0, \label{fvk:mom_x_mean} \\
  \Nx \cdot \bar{\tens{S}}^{(1)}_\perp + \Nx \cdot \left(\Sxxbar^{(0)} \Nx \wb\right)
  + \TTz = 0, \label{fvk:mom_z_mean}
\end{align}}
where the mean in-plane and transverse shear stresses are given by
\begin{align}
  \Sxxbar^{(0)} = \int_{-1/2}^{1/2} \Sxx^{(0)}\, \d Z, \qquad
  \Sxzbar^{(1)} = \int_{-1/2}^{1/2} \Sxz^{(1)}\, \d Z.
\end{align}

In order to complete the derivation,
expressions for $\Sxx^{(0)}$ and $\Sxz^{(1)}$ are
required. These are found by expanding the stress-strain relation
given by \eqref{fvk:SFP}.
Before doing so, it is useful to examine
the strain tensor, which has the asymptotic form
$\tens{E} = \delta \tens{E}^{(1)} + \delta^2 \tens{E}^{(2)} + O(\delta^3)$, where
$\tens{E}^{(1)} = \sym (\tens{F}^{(1)})$, $\tens{E}^{(2)} = \sym(\tens{F}^{(2)}) + (1/2)(\tens{F}^{(1)})^T \tens{F}^{(1)}$, and $\sym(\cdot)$ denotes the symmetric
part of a tensor, e.g.\ $\sym(\tens{F}) = (1/2)(\tens{F} + \tens{F}^T)$.
By using \eqref{fvk:F1} and \eqref{fvk:F2}, we can write the contributions
to the strain tensor as
\begin{align}
  \tens{E}^{(1)} &= \sym\left(\pd{\ubx^{(0)}}{Z}\otimes \ez + \ez \otimes \Nx \wb\right),
                   \label{fvk:E1}
  \\
  \tens{E}^{(2)} &= \sym(\tens{F}^{(2)}) + \frac{1}{2}\left(\pd{\ubx^{(0)}}{Z}
                   \cdot \pd{\ubx^{(0)}}{Z} \ez \otimes \ez + \Nx \wb \otimes \Nx \wb\right).
                   \label{fvk:E2}
\end{align}
Since $\tens{P}$ is linear in the strain tensor $\tens{E}$, it follows
that $\tens{P}$ can be expanded as
$\tens{P} = \delta \tens{P}^{(1)} + \delta^2 \tens{P}^{(2)}+ O(\delta^3)$.
Therefore, from \eqref{fvk:SFP}, we have that
\begin{align}
  \delta^2 \Sxx^{(0)} + O(\delta^3)  = (\tens{I} + \delta \tens{F}^{(1)} + O(\delta^2))(\delta \tens{P}^{(1)} + \delta^2 \tens{P}^{(2)} + O(\delta^3)).
  \label{fvk:SFP_asy}
\end{align}
The $O(\delta)$ contributions to \eqref{fvk:SFP_asy}
imply that $\tens{P}^{(1)} = 0$ and hence
$\tens{E}^{(1)} = 0$. Thus, setting $\tens{E}^{(1)} = 0$ in \eqref{fvk:E1}
leads to a differential equation for $\ubx^{(0)}$, which can be solved to
find
\begin{align}
  \ubx^{(0)}(\XX, Z, t) = \ubar(\XX, t) - Z \Nx \wb,
  \label{fvk:ubar}
\end{align}
where $\ubar$ is a `constant' of integration. This constant is chosen to
coincide with the mean horizontal displacement in the beam, defined by
\begin{align}
  \ubar = \int_{-1/2}^{1/2} \ubx^{(0)}\,\d Z.
\end{align}
Substituting \eqref{fvk:ubar}
into \eqref{fvk:E2} and using \eqref{fvk:F2} leads to
\begin{align}
  \tens{E}^{(2)} = \frac{1}{2}\left(\Nx \ubar + (\Nx \ubar)^T\right) - Z \tens{H} &+ \sym\left(\pd{\ubx^{(1)}}{Z}\otimes \ez + \ez \otimes \Nx \ubz^{(1)}\right) \nonumber
  + \pd{\ubz^{(2)}}{Z}\ez \otimes \ez  \\ &+ \frac{1}{2}\left(|\Nx \wb|^2 \ez \otimes \ez + \Nx \wb \otimes \Nx \wb\right),
  \label{fvk:E2_2}
\end{align}
where $\tens{H} = \Nx(\Nx \wb)$ is a symmetric tensor corresponding to the
Hessian of $\wb$. The trace of \eqref{fvk:E2_2} is given by
\begin{align}
  \tr(\tens{E}^{(2)}) = \Nx \cdot \ubar - Z \Nx \cdot (\Nx \wb) + \pd{\ubz^{(2)}}{z} + |\Nx \wb|^2.
  \label{fvk:tr_E2}
\end{align}
The $O(\delta^2)$ contributions to \eqref{fvk:SFP_asy} imply that
\begin{align}
  \Sxx^{(0)} = \tens{P}^{(2)} = \frac{\nub}{(1+\nub)(1-2\nub)}\tr (\tens{E}^{(2)})
  \tens{I} + \frac{1}{1+\nub} \tens{E}^{(2)}.
  \label{fvk:S0}
\end{align}
It is now possible to  eliminate $\ubz^{(2)}$ from the problem
by left- and right-multiplying with $\ez$ and using $\ez \cdot \Sxx^{(0)} \cdot \ez = 0$, $\ez \cdot \tens{E}^{(2)} \cdot \ez = \pdf{\ubz^{(2)}}{z} + (1/2) |\Nx \wb|^2$, and \eqref{fvk:tr_E2} to obtain
\begin{align}
  \pd{\ubz^{(2)}}{z} = -\frac{1}{2(1 - \nub)}\left(|\Nx \wb|^2 + 2 \nub\left[\Nx \cdot \ubar - Z \Nx \cdot (\Nx \wb)\right]\right),
  \label{fvk:ubz2}
\end{align}
By substituting \eqref{fvk:ubz2} into \eqref{fvk:E2_2} and simplifying,
the in-plane strain is found to be given by
\begin{align}
  \tens{E}^{(2)}_{\parallel} &= \frac{1}{2}\left[\Nx \ubar + (\Nx \ubar)^T
                               + \Nx \wb \otimes \Nx \wb\right] - Z \tens{H}.
                               \label{fvk:E2_3}
\end{align}
Moreover, after some algebra it can be shown that
\begin{align}
  \tr(\tens{E}^{(2)}_{\parallel}) &= \left(\frac{1 - \nub}{1 - 2\nub}\right) \tr(\tens{E}^{(2)}).
\end{align}
Hence, the in-plane component of \eqref{fvk:S0} can be written as
\begin{align}
  \Sxx^{(0)} = \frac{1}{1-\nub^2}\left[\nub \tr( \tens{E}^{(2)}_{\parallel})
  \Ixx + (1-\nub) \tens{E}^{(2)}_{\parallel}\right],
  \label{fvk:Sxx_0}
\end{align}
where the in-plane strain is given by \eqref{fvk:E2_3}. By
integrating over the thickness of the plate, the mean in-plane stress
and strain are given by
\subeq{\label{fvk:mean_SE}
\begin{align}
  \Sxxbar^{(0)} &= \frac{1}{1-\nub^2}\left[\nub \tr (\bar{\tens{E}}^{(2)}_{\parallel})
  \Ixx + (1-\nub) \bar{\tens{E}}^{(2)}_{\parallel}\right], \\
  \bar{\tens{E}}^{(2)}_{\parallel} &= \frac{1}{2}\left[\Nx \ubar + (\Nx \ubar)^T
                                 + \Nx \wb \otimes \Nx \wb\right],
\end{align}}
which may be combined with \eqref{fvk:mom_x_mean} to determine a system of equations
for the mean in-plane displacements $\ubar$.

We are now in a position to compute $\Sxz^{(1)}$ and hence determine
 a problem for $\wb$. We first note that the in-plane stress tensor
\eqref{fvk:Sxx_0}
can be written as
\begin{align}
  \Sxx^{(0)} = \Sxxbar^{(0)} + Z \tens{A}, \qquad
  \tens{A} = -\frac{1}{1-\nub^2}\left[\nub \tr(\tens{H}) \Ixx + (1- \nub) \tens{H}\right],
  \label{fvk:Sxx_A}
\end{align}
where $\tens{A}$ is a symmetric tensor that is proportional to the bending
moments about the plane $Z = 0$.
Substituting \eqref{fvk:Sxx_A}
into \eqref{fvk:mom_x} and using \eqref{fvk:mom_x_mean} leads to
\begin{align}
  \pd{\Sxz^{(1)}}{Z} = \TTx - Z \Nx \cdot \tens{A}.
\end{align}
Integrating and imposing $\Sxz^{(1)} = 0$ at $Z = -1/2$ gives
\begin{align}
  \Sxz^{(1)} = \left(Z + \frac{1}{2}\right) \TTx + \frac{1}{2}\left(Z^2 - \frac{1}{4}\right)\Nx \cdot \tens{A},
\end{align}
from which it follows that the mean transverse shear stress is
\begin{align}
  \Sxzbar^{(1)} = \frac{1}{2}\TTx + \frac{1}{12} \Nx \cdot \tens{A}.
  \label{fvk:Sxzbar}
\end{align}
Substituting \eqref{fvk:Sxzbar} into \eqref{fvk:mom_z_mean} and
simplifying leads to
\begin{align}
  -\frac{1}{12(1-\nub^2)}\nx^4 \wb
  + \Nx \cdot \left(\Sxxbar^{(0)}\Nx \wb\right) = -\frac{1}{2}\Nx \cdot \TTx
  - \TTz, \label{fvk:fvk}
\end{align}
which completes the main steps of the derivation.
Dropping the superscripts $(0)$ and $(2)$ in 
\eqref{fvk:mom_x_mean},
\eqref{fvk:ubar},
\eqref{fvk:mean_SE}, and 
\eqref{fvk:fvk} and then
re-dimensionalising leads to modified FvK equations presented 
in Sec.~\ref{sec:plate_scaling}.

\subsection{Derivation of the boundary conditions at a free edge}
\label{app:fvk_bc}

The boundary conditions at the free edges of the plate can be determined
from a boundary-layer analysis of the three-dimensional
equations of nonlinear elasticity, i.e.\ \eqref{fvk:lin}--\eqref{fvk:bc:surf}.
We use the non-dimensionalisation
discussed in Appendix~\ref{app:fvk} so that the domain
of the plate is defined by $0 \leq X_1 \leq 1$, $-\mathcal{W}/2 \leq X_2 \leq \mathcal{W}/2$, and
$-1/2 \leq Z \leq 1/2$, where $\mathcal{W} = W / L = O(1)$
represents the ratio of the width
to the length of the plate. We will
analyse the boundary layer at $X_1 = 1$ in detail and then generalise the
results to the boundaries at $X_2 = \pm \mathcal{W} / 2$.
The derivation presented
here is based on Howell \etal\cite[Chap.~6.4]{howell2009} but is extended
to the case of nonlinear elasticity and accounts for in-plane
(longitudinal) tractions on the upper surface of the plate.

The analysis begins by writing $X_1 = 1 + \delta \xi$ and rescaling
the (non-dimensionalised) transverse shear and normal components of the stress
as
$\Saz = \delta^{-1}\tSaz$, $\Sza = \delta^{-1} \tSza$, and
$\Szz = \delta^{-2} \tSzz$. Tildes are used to denote dependent variables
in the boundary layer. This rescaling means that all components of the
stress tensor $\tilde{\tens{S}}$ have the same order of magnitude in the
boundary layer, in contrast to
the bulk. With this rescaling, the conservation of linear momentum
\eqref{fvk:lin} becomes
\subeq{\label{fvk:bl:lin}
\begin{align}
  \pd{\tS_{11}}{\xi} + \delta \pd{\tS_{12}}{X_2} + \pd{\tS_{1z}}{Z} = 0,\\
  \pd{\tS_{21}}{\xi} + \delta \pd{\tS_{22}}{X_2} + \pd{\tS_{2z}}{Z} = 0,\\
  \pd{\tS_{z1}}{\xi} + \delta \pd{\tS_{z2}}{X_2} + \pd{\tS_{zz}}{Z} = 0.
  \label{fvk:bl:lin_z}
\end{align}}
Conservation of angular momentum \eqref{fvk:ang} implies that
\begin{align}
  \tilde{\tens{S}} \tilde{\tens{F}}^T = \tilde{\tens{F}} \tilde{\tens{S}}^T.
  \label{fvk:bl:ang}
\end{align}
The stress-strain relation \eqref{fvk:SFP} can be written as
\begin{align}
  \delta^2 \tilde{\tens{S}} = \tilde{\tens{F}}\tilde{\tens{P}},
  \qquad
  \tilde{\tens{P}} = \frac{\nub}{(1+\nub)(1-2\nub)}
  \tr(\tilde{\tens{E}})
  \tens{I} + \frac{1}{1+\nub} \tilde{\tens{E}}.
  \label{fvk:bc:SFP}
\end{align}
The boundary conditions along the upper and lower surfaces of the plate
\eqref{fvk:bc:surf} are given by
\subeq{
\begin{alignat}{2}
  \tS_{\alpha z} &= \delta \TTi{\alpha}, &\quad Z &= 1/2; \\
  \tS_{zz} &= \delta^2 \TTi{z}, &\quad Z &= 1/2; \\
  \tS_{\alpha z} &= 0, &\quad Z &= -1/2; \\
  \tS_{zz} &= 0, &\quad Z &= -1/2.
\end{alignat}}
The edge of the plate is taken to be stress free; therefore, we impose
\subeq{
\begin{align}
  \tS_{\alpha 1} = 0, \quad \xi = 0; \\
  \tS_{z1} = 0, \quad \xi = 0.
\end{align}}

The displacements and the stress tensor are asymptotically
expanded in powers of $\delta$ as
\subeq{
\begin{align}
  \tubx &= \tubx^{(0)}(\xi, X_2, Z, t) + O(\delta), \\
  \tubz &= \tubz^{(0)}(\xi, X_2, Z, t) + O(\delta), \\
  \tilde{\tens{S}} &= \tilde{\tens{S}}^{(0)}(\xi, X_2, Z, t) +
                     \delta \tilde{\tens{S}}^{(1)}(\xi, X_2, Z, t)
                     + O(\delta^2).
\end{align}}
The deformation gradient tensor has the asymptotic form
$\tilde{\tens{F}} = \tilde{\tens{F}}^{(0)} + O(\delta)$, where
\begin{align}
  \tilde{\tens{F}}^{(0)} = \tens{I} + \pd{\tubz^{(0)}}{Z} \ez \otimes \ez
  + \pd{\tubz^{(0)}}{\xi} \ez \otimes \vec{e}_1 + O(\delta).
\end{align}
From the $O(1)$ contributions to \eqref{fvk:bc:SFP}, we can deduce that
$\tilde{\tens{E}}^{(0)} = (1/2)[(\tilde{\tens{F}}^{(0)})^T\tilde{\tens{F}}^{(0)} - \tens{I}] = \tens{0}$ and hence $\tilde{\tens{F}}^{(0)} = \tens{I}$.
Consequently, the leading-order contribution to the vertical displacement
is independent of $\xi$ and $Z$; by matching to the outer solution as $\xi \to -\infty$, we obtain $\tubz^{(0)} = \wb(1,X_2,t)$. Importantly, the $O(1)$
contributions to \eqref{fvk:bl:ang} imply that the stress tensor is symmetric to
leading order
\begin{align}
  \tilde{\tens{S}}^{(0)} = (\tilde{\tens{S}}^{(0)})^T,
\end{align}
a result that will be of considerable use in the subsequent analysis.
The boundary conditions at the upper and lower surfaces can be expanded as
\subeq{
\begin{alignat}{3}
  \tS_{\alpha z}^{(0)} &= 0, &\quad \tS_{\alpha z}^{(1)} &= \TTi{\alpha}, &\quad Z &= 1/2; \label{fvk:bl:bc:Saz_up}
  \\
  \tS_{zz}^{(0)} &= 0, &\quad \tS_{zz}^{(1)} &= 0, &\quad Z &= 1/2;
  \label{fvk:bl:bc:Szz_up}
  \\
  \tS_{\alpha z}^{(0)} &= 0, &\quad \tS_{\alpha z}^{(1)} &= 0, &\quad Z &= -1/2;
  \label{fvk:bl:bc:Saz_low}
  \\
  \tS_{zz}^{(0)} &= 0, &\quad \tS_{zz}^{(1)} &= 0, &\quad Z &= -1/2.
  \label{fvk:bl:bc:Szz_low}
\end{alignat}}
The stress-free conditions at the edge of the plate are
\subeq{
\begin{align}
  \tS_{\alpha 1}^{(0)} = 0, \quad \xi = 0,
  \label{fvk:bl:bc:Sx}
  \\
  \tS_{z1}^{(0)} = 0, \quad \xi = 0.
  \label{fvk:bl:bc:Sz}
\end{align}}

Having simplified the kinematics and the form of the stress tensor, we
are now in a position to derive the boundary conditions for the modified
FvK equations. We start by
considering the $O(1)$ contributions to \eqref{fvk:bl:lin}, which
can be written as
\subeq{
\begin{align}
  \pd{\tS_{11}^{(0)}}{\xi} + \pd{\tS_{1z}^{(0)}}{Z} = 0, \label{fvk:bl:lin_10}\\
  \pd{\tS_{21}^{(0)}}{\xi} + \pd{\tS_{2z}^{(0)}}{Z} = 0, \label{fvk:bl:lin_20}\\
  \pd{\tS_{z1}^{(0)}}{\xi} + \pd{\tS_{zz}^{(0)}}{Z} = 0. \label{fvk:bl:lin_30}
\end{align}}
Integrating \eqref{fvk:bl:lin_10} and \eqref{fvk:bl:lin_20}
from $Z = -1/2$ to $Z = 1/2$ and using the boundary conditions
\eqref{fvk:bl:bc:Saz_up} and \eqref{fvk:bl:bc:Saz_low} yields
\begin{align}
  \pd{}{\xi} \int_{-1/2}^{1/2} \tS_{\alpha 1}^{(0)}\,\d Z = 0.
\end{align}
By integrating with respect to $\xi$ and using the boundary condition
\eqref{fvk:bl:bc:Sx}, we deduce that
\begin{align}
  \int_{-1/2}^{1/2} \tS_{\alpha 1}^{(0)}\,\d Z = 0
\end{align}
for all $\xi$. Thus, taking the limit as $\xi \to -\infty$ and using
the matching condition
\begin{align}
  \lim_{\xi \to -\infty} \tS_{\alpha 1}^{(0)}\, \d Z = \lim_{X_1 \to 1} \mathsf{S}_{\alpha 1}^{(0)}\, \d Z,
  \label{fvk:bl:match_Sa1}
\end{align}
furnishes two stress-free boundary conditions for \eqref{fvk:mom_x_mean} given by
\begin{align}
  \bar{\mathsf{S}}_{\alpha 1}^{(0)}(1,X_2,t) = 0.
\end{align}
By applying a similar procedure to \eqref{fvk:bl:lin_30} and using the
symmetry of the stress tensor, we find that
\begin{align}
  \int_{-1/2}^{1/2} \tS_{z 1}^{(0)}\,\d Z = \int_{-1/2}^{1/2} \tS_{1z}^{(0)}\,\d Z = 0.
  \label{fvk:bl:S1z_mean}
\end{align}

We now multiply \eqref{fvk:bl:lin_10} by $Z$ and integrate across
the thickness of the
plate to obtain, after using \eqref{fvk:bl:bc:Saz_up}, \eqref{fvk:bl:bc:Saz_low},
and \eqref{fvk:bl:S1z_mean},
\begin{align}
  \pd{}{\xi} \int_{-1/2}^{1/2} \tS_{11}^{(0)}Z \,\d Z = 0.
  \label{fvk:bl:dx_S11}
\end{align}
By integrating \eqref{fvk:bl:dx_S11} over the boundary layer and using the matching
condition \eqref{fvk:bl:match_Sa1} along with \eqref{fvk:Sxx_A}
and the stress-free condition \eqref{fvk:bl:bc:Sx}, we find that
\begin{align}
  \frac{1}{12} \mathsf{A}_{11}(1,X_2,t) = 0,
\end{align}
which provides one of the two boundary conditions for \eqref{fvk:fvk}. By multiplying
\eqref{fvk:bl:lin_20} by $Z$ and repeating the process, we find that
\begin{align}
  \int_{-1/2}^{1/2}\tS_{2z}^{(0)}\,\d Z = \pd{}{\xi}\int_{-1/2}^{1/2}\tS_{21}^{(0)} Z\,\d Z.
  \label{fvk:bc:S2z_mean}
\end{align}

To obtain the second boundary condition for \eqref{fvk:fvk}, we
consider the $O(\delta)$ contributions to \eqref{fvk:bl:lin_z}:
\begin{align}
  \pd{\tS_{z1}^{(1)}}{\xi} + \pd{\tS_{z2}^{(0)}}{X_2} + \pd{\tS_{zz}^{(1)}}{Z} = 0.
  \label{fvk:bl:lin_31}
\end{align}
Integrating \eqref{fvk:bl:lin_31}
across the thickness of the plate and over the boundary
layer, and using the boundary conditions \eqref{fvk:bl:bc:Szz_up},
\eqref{fvk:bl:bc:Szz_low}, and \eqref{fvk:bl:bc:Sx}
gives
\begin{align}
  \lim_{\xi \to -\infty} \int_{-1/2}^{1/2}\tS_{z1}^{(1)}\,\d Z = \pd{}{X_2}\int_{-\infty}^{0} \int_{-1/2}^{1/2} \tS_{z2}^{(0)}\,\d Z \d \xi.
  \label{fvk:bl:Sz1}
\end{align}
To proceed, the integrands on both sides of \eqref{fvk:bl:Sz1} are
rewritten using
the symmetry of the stress tensor, $\tS_{\alpha z}^{(0)} = \tS_{z \alpha}^{(0)}$.
Then, \eqref{fvk:bc:S2z_mean} is substituted into the right-hand side to obtain
\begin{align}
  \lim_{\xi \to -\infty} \int_{-1/2}^{1/2}\tS_{1z}^{(1)}\,\d Z = \pd{}{X_2}\left[\int_{-1/2}^{1/2}\tS_{21}^{(0)} Z\,\d Z\right]^{\xi = 0}_{\xi = -\infty}.
  \label{fvk:bl:S1z_2}
\end{align}
The boundary term evaluated at $\xi = 0$ can be set to zero using the stress-free
condition \eqref{fvk:bl:bc:Sx}. The remaining terms associated with the
far field can be matched with the outer solution using the matching conditions
\subeq{
\begin{align}
  &\lim_{\xi \to -\infty} \int_{-1/2}^{1/2}\tS_{1z}^{(1)}\, \d Z =
  \lim_{X_1 \to 1} \bar{\mathsf{S}}_{1z}^{(0)}
  =  \lim_{X_1 \to 1}  \left(\frac{1}{2}\TTi{1} + \frac{1}{12}\pd{\mathsf{A}_{1\alpha}}{X_\alpha}\right),
  \\
 &\lim_{\xi \to -\infty} \int_{-1/2}^{1/2}\tS_{21}^{(0)}\, \d Z = \lim_{X_1 \to 1} \bar{\mathsf{S}}_{21}^{(0)} = \lim_{X_1 \to 1} \frac{1}{12} \mathsf{A}_{12}.
\end{align}}
Thus, \eqref{fvk:bl:S1z_2} can be written as 
\begin{align}
  \frac{1}{2}\TTi{1} + \frac{1}{12}\pd{\mathsf{A}_{1\alpha}}{X_\alpha} = -\frac{1}{12} \pd{\mathsf{A}_{12}}{X_2},
  \quad X_1 = 1,
\end{align}
which provides the second and final boundary condition for \eqref{fvk:fvk}.

To summarise, the non-dimensional boundary conditions at the free edge
located at $X_1 = 1$ are given by
\subeq{
  \begin{alignat}{2}
    \bar{\mathsf{S}}_{\alpha 1}^{(0)} &= 0, &\quad X_1 &= 1; \\
    \mathsf{A}_{11} &= 0, &\quad X_1 &= 1;
    \\
    \frac{1}{2}\TTi{1} + \frac{1}{12}\pd{\mathsf{A}_{1\alpha}}{X_\alpha} &= -\frac{1}{12} \pd{\mathsf{A}_{12}}{X_2}, &\quad X_1 &= 1.
  \end{alignat}}
The final two boundary conditions can be written in terms of the vertical
displacement $\wb$ using the definition of $\tens{A}$ in
\eqref{fvk:Sxx_A} to obtain
\subeq{
\begin{alignat}{2}
  \pdd{\wb}{X_1} + \nub \pdd{\wb}{X_2} &= 0, &\quad X_1 &= 1; \\
  \frac{1}{12(1-\nub^2)}\left[\frac{\p^3\wb}{\p X_1^3} + (2-\nub)\frac{\p^3\wb}{\p X_1 \p X_2^2}\right] &= \frac{1}{2}\TTi{1}, &\quad X_1 &= 1.
\end{alignat}}
By generalising the results, the boundary conditions at the free edges
at $X_2 = \pm \mathcal{W}/2$ are given by
\subeq{
  \begin{alignat}{2}
    \bar{\mathsf{S}}_{\alpha 2}^{(0)} &= 0, &\quad X_2 &= \pm \mathcal{W}/2; \\
    \nub \pdd{\wb}{X_1} + \pdd{\wb}{X_2} &= 0, &\quad X_2 &= \pm \mathcal{W}/2; \\
  \frac{1}{12(1-\nub^2)}\left[(2-\nub)\frac{\p^3\wb}{\p X_1^2 \p X_2} + \frac{\p^3\wb}{\p X_2^3}\right] &= \frac{1}{2}\TTi{2}, &\quad X_2 &= \pm \mathcal{W}/2.
\end{alignat}}
The boundary conditions given in Sec.~\ref{sec:beam_bc} can be obtained from
these by dropping the $(0)$ superscript and re-dimensionalisation.

\end{appendix}

\bibliographystyle{ieeetr}
\bibliography{refs}

\end{document}